\documentclass[a4paper,fleqn,usenatbib]{mnras}

\usepackage{newtxtext,newtxmath}
\usepackage[T1]{fontenc}
\usepackage{ae,aecompl}
\usepackage{epsfig}
\usepackage{epstopdf}

\usepackage{graphicx}	% Including figure files
\usepackage{amsmath}	% Advanced maths commands
\usepackage{amssymb}	% Extra maths symbols
\usepackage{wasysym}
\usepackage{latekpackages/booktabs/booktabs}
\usepackage{csvsimple}
\usepackage{subfig}
\usepackage{tablefootnote}
\usepackage{csquotes}
\usepackage{textcomp}
\usepackage[english]{babel}
\usepackage{lscape}
\usepackage{longtable}
\usepackage{xcolor}
\usepackage{soul}
\newcommand{\Rp}{R_{\textnormal{p}}}
\newcommand{\Rs}{R_{\star}}
\newcommand{\RpRs}{\Rp/\Rs}

\title[Exoplanet Transmission Spectra Library]{A library of \texttt{ATMO} forward model transmission spectra for hot Jupiter exoplanets}

\author[Goyal et al.]{Jayesh M. Goyal$^{1}$\thanks{E-mail: jgoyal@astro.ex.ac.uk},
Nathan Mayne$^{1}$, David K. Sing$^{1}$, Benjamin Drummond$^{1}$, 
\newauthor Pascal Tremblin$^{2}$, David S. Amundsen$^{3,4}$, Thomas Evans$^{1}$, Aarynn L. Carter$^{1}$,
\newauthor Jessica Spake$^{1}$, Isabelle Baraffe$^{1}$, Nikolay Nikolov$^{1}$, James Manners$^{1,5}$ 
\newauthor Gilles Chabrier$^{1,6}$ and Eric Hebrard$^{1}$
\\
$^{1}$Astrophysics Group, School of Physics and Astronomy, University of Exeter, Exeter EX4 4QL, UK \\
$^{2}$Malson de la Simulation, CEA-CNRS-INRIA-UPS-UVSQ, USR 3441, Centre detude de Saclay, France \\
$^{3}$NASA Goddard Institute for Space Studies, New York, NY 10025, USA\\
$^{4}$Department of Applied Physics and Applied Mathematics, Columbia University, New York, NY 10025, USA\\
$^{5}$Met Office, FitzRoy Road, Exeter, EX1 3PB, UK\\
$^{6}$Ecole Normale Supérieure de Lyon, CRAL, UMR CNRS 5574, F-69364 Lyon Cedex 07, France\\
}

\date{Accepted XXX. Received YYY; in original form ZZZ}

\pubyear{2017}

\begin{document}
\label{firstpage}
\pagerange{\pageref{firstpage}--\pageref{lastpage}}
\maketitle

\begin{abstract}
We present a grid of forward model transmission spectra, adopting an isothermal temperature-pressure profile, alongside corresponding equilibrium chemical abundances for 117 observationally significant hot exoplanets (Equilibrium Temperatures of 547-2710\,K). This model grid has been developed using a 1D radiative-convective-chemical equilibrium model termed \texttt{ATMO}, with up-to-date high temperature opacities. We present an interpretation of observations of ten exoplanets, including best fit parameters and $\chi^{2}$ maps. In agreement with previous works, we find a continuum from clear to hazy/cloudy atmospheres for this sample of hot Jupiters. The data for all the 10 planets are consistent with sub-solar to solar C/O ratio, 0.005 to 10 times solar metallicity and water rather than a methane dominated infrared spectra.  We then explore the range of simulated atmospheric spectra for different exoplanets, based on characteristics such as temperature, metallicity, C/O-ratio, haziness and cloudiness. We find a transition value for the metallicity between 10 and 50 times solar, which leads to substantial changes in the transmission spectra. We also find a transition value of C/O ratio, from water to carbon species dominated infrared spectra, as found by previous works, revealing a temperature dependence of this transition point ranging from $\sim$0.56 to $\sim$1-1.3 for equilibrium temperatures from $\sim$900 to $\sim$2600 K. We highlight the potential of the spectral features of HCN and C$_2$H$_2$ to constrain the metallicities and C/O ratios of planets, using JWST observations. Finally, our entire grid ($\sim$460,000 simulations) is publicly available and can be used directly with the JWST simulator \texttt{PandExo} for planning observations.

\end{abstract}


%%%%%%%%%%%%%%%%%%%%%%%%%%%%%%%%%%%%%%%%%%%%%%%%%%

%%%%%%%%%%%%%%%%% BODY OF PAPER %%%%%%%%%%%%%%%%%%

\section{Introduction}
\label{section:Introduction}

The number of exoplanets that have been discovered has now reached a staggering 3529\footnote{Source: NASA Exoplanet Archive as of 5th Oct. 2017}. This count of planets will increase dramatically with the launch of the CHEOPS \citep{Broeg2013} and TESS missions \citep{Ricker2014} in 2018. However, accurate atmospheric characterisation has been performed for only a small subset of these exoplanets, primarily due to technological limitations, but also due to complications in modelling their atmospheres. Alkali metal elements like sodium and potassium have been detected in the atmospheres of various exoplanets, for example HD~209458b \citep{Charbonneau2002, Sing2008, Redfield2008, Snellen2008, Sing2016} and XO-2b \citep{Sing2011}. Water has also been detected in many hot Jupiter atmospheres, \citep[e.g][]{Deming2013, Wakeford2013, Stevenson2014}. Additionally, \citet{Kreidberg2014} and \citet{Sing2016} highlighted the possibility of clouds and hazes in some of them. Most of these works used transmission spectra derived from observations made using the Hubble Space Telescope (HST). 

Although HST observations have led to the detection of several species \citep{Sing2013, Sing2016}, it is limited by its wavelength coverage (0.2 to 1.7 \textmu m). The launch of the James Webb Space Telescope (JWST) in 2018 will enable probing exoplanet atmospheres from wavelengths of 0.6 all the way up-to 28 \textmu m \citep{Beichman2014, Greene2016}. Therefore, it will be extremely valuable for the detection of species with signatures in the mid to near-infrared, which can provide constraints on various physical parameters such as the temperature, C/O ratio and metallicity. In this paper we present a grid of forward model transmission spectra for 117 exoplanets that are scientifically important targets for characterisation. The grid for each target consists of a range of variables; atmospheric temperature, metallicity, C/O ratio, haziness and cloudiness (described in Section \ref{section:Parameter Space Selection}). This grid is publicly available online\footnote{\url{https://bd-server.astro.ex.ac.uk/exoplanets/}} , and we encourage the community to use it as a tool to assist them in planning future observations, such as with JWST, HST and various ground based telescopes, along-with interpreting existing datasets. It can provide a useful complement for interpretation, alongside atmospheric retrieval analysis.

The efforts in modelling atmospheres of exoplanets began just after the discovery of the first exoplanet orbiting a sun like star in 1995 \citep{Mayor1995}. \citet{Burrows1997} provided a theoretical basis for understanding the spectral features of exoplanet and brown dwarf atmospheres. \citet{Seager2000, Brown2001} and \citet{Hubbard2001} all predicted forward model transmission spectra for HD~209458b which later led to the detection of a sodium feature in its atmosphere \citep{Charbonneau2002}. \citet{Sudarsky2003} presented a systematic exploration of model spectra. \citet{Fortney2010} provided a detailed analysis of the effect of temperature, surface gravity and metallicity on transmission spectra for various hot Jupiter planets. More recently, \citet{Molliere2015} developed a generalised grid of forward model emission spectra for a range of planetary gravity values and other planetary characteristics without focusing on specific planets. Finally, \citet{Molliere2016} presented a grid of emission and transmission spectra for 18 important JWST targets, with a sophisticated cloud scheme included in their model. 

The outline of this paper is as follows. In Section \ref{section:Model Details} we explain the input physics of the model including our treatment of radiative transfer, opacity sources and their implementation, and chemistry. In Section \ref{section:Grid Setup} we describe the basis for the selection of planets in the current grid and the model setup, along-with the  description and justification of the chosen parameter space. In Section \ref{section:Model Validation} we present a comparison between spectra derived from isothermal and radiative-convective equilibrium pressure-temperature profiles (hereafter termed \enquote{isothermal $P$-$T$ profiles} and \enquote{consistent $P$-$T$ profiles}, respectively). In Section \ref{section:Interpretation of Observations} we present an interpretation of the observations from \citet{Sing2016}, and the inferred best fit characteristics. In Section \ref{section:Transmission Spectra : variation with parameters} we provide the analysis of the model simulations over the entire parameter space for a subset of planets. In Section \ref{Simulating JWST observations with ATMO} we demonstrate the application of the grid to plan observations, by using one of our simulations as an input to the JWST simulator \texttt{PandExo} \citep{Batalha2017}. Finally, we conclude in Section \ref{section:Conclusions}. The Appendix \ref{app:chi-maps of all planets}, \ref{subsection:Molecular Features}, \ref{app:Pressure Broadening Sources} and \ref{app:Planets and their parameters in Grid}, show the $\chi^{2}$ maps, transmission spectral features of individual molecules, pressure broadening sources and the table of selected planets, respectively. 

\section{Model Details}
\label{section:Model Details}

\subsection{Model General Structure}
\label{section:Model General Structure}

\texttt{ATMO} is a 1-D radiative-convective equilibrium model for planetary atmospheres \citep{Tremblin2015, Tremblin2016, Amundsen2014, Drummond2016}. It has been applied to interpret observations of several exoplanets both as a forward and retrieval model \citep{Evans2016, Wakeford2017,Evans2017}. It solves the radiative transfer equation for a given set of opacities, $P$-$T$ profile and chemical abundances. The code also solves for the $P$-$T$ profile that satisfies hydrostatic equilibrium and conservation of energy. It can compute equilibrium and non-equilibrium chemical abundances described in detail in Section \ref{subsection:Chemistry}.

\texttt{ATMO} adopts an optical depth grid with plane parallel geometry. However, spherical geometry is considered while computing transmission spectra, as the radiation travels through the limb of the planetary atmosphere. Minimum and maximum optical depth limits, at a particular reference wavelength specified during the model initialisation, which are also a function of opacity and chemistry,  govern the pressure domain of the model atmosphere. The details of energy flux balance to compute radiative-convective equilibrium $P$-$T$ profiles can be found in \citet{Drummond2016}. We note that transmission spectra computed using $P$-$T$ profiles in radiative-convective equilibrium are used only in Section \ref{section:Model Validation}. Elsewhere in this paper, all the transmission spectra are computed using isothermal $P$-$T$ profiles. 

\texttt{ATMO} can solve the radiative transfer (RT) equation using the line by line (LBL) or correlated-\textit{k} approach. LBL implies very high spectral resolution (\texttt{ATMO} normally uses 0.001 $\textup{cm}^{-1}$ evenly spaced in wavenumber), but  is computationally very expensive and not practical for generating consistent radiative-convective equilibrium $P$-$T$ profiles and the corresponding spectra. To overcome this problem, the correlated-\textit{k} approximation \citep{Lacis1991, Amundsen2014} is used to solve radiative transfer while achieving the required accuracy (see Section \ref{subsection:Correlated-k Methodology} for more details). 

\subsection{Opacity Database}
\label{subsection:Line Lists}
The opacity database and its treatment are two of the most important aspects of any atmosphere model. These opacity computations require absorption coefficients for the spectrally significant gaseous species. These absorption coefficients are computed using line list databases from various sources. The HITRAN (High Resolution TRANsmission) database \citep{Rothman2013} is the most widely used opacity source in various atmospheric models. However, this database is established at a reference temperature of $296$\,K \citep{Rothman2010}, with HITEMP \citep{Rothman2010} its high temperature version available only for certain molecules. Expected temperatures on hot Jupiter exoplanets can be substantially higher than $296$\,K  , for example WASP-107b has an equilibrium temperature (T$_{\rm{eq}}$) of  $770$\,K while that of WASP-12b is $2580$\,K. In such conditions, the HITRAN low temperature line lists can underestimate the absorption of radiation by several orders of magnitude.

The line lists available from various sources primarily consist of Einstein coefficients or oscillator strengths. These quantities are used, along with the effect of line broadening to calculate the absorption cross-sections for each molecule as a function of wavelength/wavenumber, to be used in the radiative transfer equation. The details of these computations can be found in \citet{Amundsen2014}. \texttt{ATMO} considers Doppler broadening and pressure broadening for these computations. Doppler broadening becomes significant in low pressure and high temperature environments. Therefore, it is important in the high altitude region of hot Jupiter atmospheres, probed by transmission spectra.

Hot Jupiter exoplanets are expected to be H$_2$ and He dominated. Therefore, pressure broadening due to H$_2$ and He species has to be taken into account for each radiatively important gaseous species in the model \citep[see][for details]{Amundsen2014, Hedges2016}. However, HITRAN only provides air and self pressure broadened line widths for various gases which will not be accurate for hot Jupiter exoplanets. We include an up-to-date set of opacities for high temperature exoplanet atmospheres in \texttt{ATMO} primarily from ExoMol \citep{Tennyson2016}, with H$_2$ and He broadening taken into account for the species where data are available, otherwise we use the air broadening parameters from HITRAN. Table \ref{table:Linelist} shows the updated source of line lists compared to \citet{Amundsen2014}, for various molecules and the corresponding partition functions used in \texttt{ATMO}.  Updated pressure broadening parameters for each molecule are also documented in Appendix \ref{app:Pressure Broadening Sources}. We note that we exclude CrH opacities in the current model simulations, due to non availability of thermochemical constants to compute equilibrium chemical abundances of CrH.

At higher metallicities, atmospheric abundance of species other than H$_2$ and He such as  CO, H$_2$O, CO$_2$, H$_2$S, etc., become significant. In such conditions, the effect of broadening due to all major species on all the other radiatively important species should be taken into account. It is difficult to accurately comment on the effect of pressure broadening at high metallicities since no study has been done in that area according to our knowledge, although the need for laboratory measurements in this region of the parameter space has been highlighted in \citet{Fortney2016}. However, equilibrium chemistry calculations show that even at 200 times solar metallicity the composition remains H$_2$ and He dominated, allowing us to perform simulations up-to this upper limit of metallicity. Since absorption coefficient calculations are sensitive to atmospheric composition, one of the future goals of our research is to generate an opacity database for a larger range of compositions.

We have considered only those opacities making a significant contribution to the derived spectra in our analysis. For example, C$_2$H$_2$ and C$_2$H$_4$ have almost overlapping absorption peaks throughout the spectrum except between 10 to 12 \textmu m. However, C$_2$H$_2$ opacity dominates over C$_2$H$_4$ opacity. Also equilibrium chemistry dictates that if C$_2$H$_4$ is present in the atmosphere C$_2$H$_2$ will also be present \citep{Moses2011} with almost equal or higher concentrations, even at high C/O ratios. Therefore, we have included only C$_2$H$_2$ in our current analysis, since C$_2$H$_2$ will effectively mask the features of C$_2$H$_4$.

\begin{table*}
	\centering	
	\begin{tabular}{|p{2cm}|p{4cm}|p{4cm}l}    
		\hline
		Molecule & Line list & Partition Function \\
		\hline
		H$_2$O & \citet{Barber2006} & \citet{Barber2006} \\
		CO$_2$ & \citet{Tashkun2011} & \citet{Rothman2009} \\
		CO & \citet{Rothman2010} & \citet{Rothman2009}\\
		CH$_4$ & \citet{Yurchenko2014} & \citet{Yurchenko2014} \\
		NH$_3$ & \citet{Yurchenko2011} & \citet{Sauval1984} \\
		Na & VALD3$^1$ & \citet{Sauval1984} \\ 
 		K & VALD3$^1$ & \citet{Sauval1984} \\
		Li & VALD3$^1$ & \citet{Sauval1984} \\
		Rb & VALD3$^1$ & \citet{Sauval1984} \\
		Cs & VALD3$^1$ & \citet{Sauval1984} \\
		TiO & \citet{Plez1998} & \citet{Sauval1984} \\
		VO & \citet{Volinelist2016} & \citet{Sauval1984} \\
		FeH & \citet{Fehlinelist2010} & \citet{Fehlinelist2010} \\
		CrH$^2$  & \citet{Tennyson2012} & \citet{Burrows2002} \\
	        PH$_3$ & \citet{Ph3linelist2014} & \citet{Ph3linelist2014} \\
	        HCN & \citet{Harris2006} & \citet{Harris2006} \\
		        & \citet{Barber2014} & \citet{Barber2014} \\
		C$_2$H$_2$ & \citet{Rothman2013} & \citet{Rothman2013} \\
		H$_2$S & \citet{Rothman2013} & \citet{Rothman2013} \\
		SO$_2$ & \citet{Underwood2016} & \citet{Underwood2016} \\
		H$_2$-H$_2$ CIA & \citet{Ciahitranpaper2012} & N/A \\
		H$_2$-He CIA & \citet{Ciahitranpaper2012} & N/A \\
		\hline
	\end{tabular}
	\centering
	\caption{Molecular line lists used in \texttt{ATMO} and their sources. Pressure broadening sources are shown in Table  \ref{table:Broadening Source} in Appendix \ref{app:Pressure Broadening Sources}. \newline $^1$\citet{Heiter2008} (\url{http://vald.astro.uu.se/~vald/php/vald.php}). \newline $^2$Note : CrH opacities are not included in the grid (see Section \ref{subsection:Line Lists}).} 
	\label{table:Linelist}
\end{table*}

\subsection{Correlated-\textit{k} Methodology}
\label{subsection:Correlated-k Methodology}

The correlated-\textit{k} approximation is a standard approach used in many Earth based atmospheric models, both 1D and 3D \citep{Goody1989, Lacis1991, Socrates1996} and also many of the forward models developed for exoplanet atmospheres \citep{Fortney2010, Molliere2015, Malik2017}. \citet{Amundsen2014} created a correlated-\textit{k} opacity database for \texttt{ATMO} which has been updated for this analysis with more species, all of them listed in Table \ref{table:Linelist}. This database is on a pressure and temperature grid which extends from $70$\,K to $3000$\,K and $10^{-4}$ to $10^8$ Pa ($10^{-9}$ to $10^3$ bar) with 20 and 40 points, respectively, giving a total of 800 points for each species and each band, covering the complete range of temperatures and pressures expected in exo-planetary atmospheres. These correlated-\textit{k} opacity files are at 3 different spectral resolutions, the lowest resolution with 32 bands, medium resolution with 500 bands and highest resolution with 5000 bands. The 500 and 5000 bands are evenly spaced in wavenumber between 1 $\textup{cm}^{-1}$ and 50000 $\textup{cm}^{-1}$. The lowest resolution 32 band files are used for generating consistent radiative-convective equilibrium $P$-$T$ profiles, since the model has to iterate numerous times between radiative transfer and chemistry at each level, making it computationally expensive. 500 and 5000 band files are used to generate transmission and emission spectra of a planet. All the spectra in this paper have been calculated using 5000 bands, which corresponds to R$\sim$5000 at 0.2 \textmu m while decreasing to R$\sim$100 at 10 \textmu m.

The correlated-\textit{k} methodology used in \texttt{ATMO} \citep{Amundsen2014} is based on the methodology adopted within the Met Office SOCRATES radiative transfer model \citep{Socrates1996}. \texttt{ATMO} has been validated against SOCRATES, by comparing outputs from both models for hot Jupiter environments  \citep{Amundsen2014}. \texttt{ATMO} in correlated-\textit{k} mode is also routinely validated against the LBL methodology by comparing fluxes and heating rates.

As described earlier, \textit{k}-coefficients for each gaseous species included in the model are computed, for a range of temperatures and pressures. Depending on the chemical composition of the atmosphere these opacities are combined together, to obtain a total opacity. Chemical composition will be different for different planets, and will also change with parameters such as  temperature, metallicity and C/O ratio. Using pre-mixed opacities is not flexible, and is accurate only for a particular atmospheric composition \citep{Amundsen2017}. Therefore, combining  \textit{k}-coefficients of different gases to obtain the total opacity of the atmosphere is crucial for flexibility and accuracy. \texttt{ATMO} adopts the random overlap method with resorting and rebinning \citep{Lacis1991, Amundsen2017} to combine \textit{k}-coefficients \enquote{on the fly} depending on the chemical composition, temperature and pressure at each atmospheric level, for each spectral band, during each iteration. Therefore, using the technique of random overlap allows us to simulate atmospheres for a certain range of temperatures, metallicities and C/O ratio. This \enquote{on the fly} combination of \textit{k}-coefficients using the random overlap technique also makes the model physically consistent, which means that the opacities, and thereby the $P$-$T$ structure, are consistent with the chemical composition of the atmosphere at any given iteration. 

\subsection{Radiative Transfer}
\label{subsection:Radiative Transfer}
Radiative transfer in \texttt{ATMO} is solved numerically using the discrete ordinate method with isotropic scattering, but used only for calculating consistent $P$-$T$ profiles and emission spectra, while transmission spectra is computed as shown in Section \ref{section:Transmission Spectra}. The details of the radiative transfer computation implemented in \texttt{ATMO} can be found in \citet{Drummond2016}. We here discuss some of the recent new additions to \texttt{ATMO} used in this analysis. 

\subsubsection{Multi-gas Rayleigh Scattering}
\label{subsubsection:Multi-gas Rayleigh Scattering}
Rayleigh scattering is one of the most important process affecting the radiation budget and the albedo of the planetary atmosphere from ultraviolet to visible wavelengths. Rayleigh scattering due to any of the species present in the atmosphere is given by \citep{Liou1980}

\begin{equation}
   \sigma_{\textup{n}}^{\textup{RAY}}(\lambda) = \frac{32\pi^{3}(m_\textup{r} - 1)^2}{3\lambda^{4}n^2}f(\rho_\textup{n}),
   \label{eq:rayleigh1} 
\end{equation}

where $\lambda$ is wavelength in $\textup{cm}$, $\sigma_{n}^{\textup{RAY}}(\lambda)$ is Rayleigh scattering cross section in $\textup{cm}^2$, $m_\textup{r}$ is the (real) refractive index for that particular gas, $n$ is the number density in $\textup{cm}^{-3}$. To consider the anisotropy of scattering particles, a correction factor $f(\rho_n)$  is applied given by

\begin{equation}
   f(\rho_\textup{n}) = \frac{6 + 3\rho_\textup{n}}{6 - 7\rho_\textup{n}},
   \label{eq:rayleigh2} 
\end{equation}

where $\rho_n$ is the depolarisation factor. Additionally, being additive in nature, the total Rayleigh scattering in the atmosphere will be the sum of the scattering due to individual species. 

In H$_2$ and He dominated atmospheres with solar metallicity, it is only the Rayleigh scattering due to H$_2$ and He that is significant. However, with an increasing metallicity of the planetary atmosphere, the abundance of other gases such as CO$_2$, CO, H$_2$O and CH$_4$ start increasing substantially (although the atmosphere remains H$_2$ and He dominated for metallicities $\leq$ 200$\times$ solar) \citep{Moses2013b}. In such cases, Rayleigh scattering due to these other species also become significant. Therefore, we have included multi-gas Rayleigh scattering, due to the species CO, N$_2$, CH$_4$, NH$_3$, H$_2$O, CO$_2$, H$_2$S and SO$_2$,  in addition to H$_2$ and He in \texttt{ATMO} model. The H$_2$ refractive index is adopted from 
\citet{Leonard1974} and that of He from \citet{Mansfield1969}. Depolarisation factors for both are taken from \citet{Rayleigh1919} and \citet{Penndorf1957}. The source of refractive index and depolarisation factor for CO, N$_2$, CH$_4$ and CO$_2$ is \citet{Sneep2005}, for NH$_3$ and H$_2$O is \citet{Allen2000} and for H$_2$S and SO$_2$ 
is National Physical Laboratory (NPL\footnote{\url{http://www.kayelaby.npl.co.uk/general\_physics/2\_5/2\_5\_7.html}}) database. The wavelength dependance of the refractive index is neglected in our calculations.

\subsubsection{Haze and Cloud Treatment}
\label{subsubsection:Haze and Cloud Treatment}

In \texttt{ATMO} the opacity of haze, small scattering aerosol particles suspended in the atmosphere, is implemented as a parameterised enhanced Rayleigh scattering. This can be represented by $\sigma(\lambda) = \alpha_\textup{haze}\sigma_{0}$ where $\sigma(\lambda)$ is the total scattering cross-section with haze, $\alpha_\textup{haze}$ is the haze enhancement factor and $\sigma_{0}(\lambda)$ is the scattering cross-section due to all other gases (since \texttt{ATMO} considers multi-gas scattering), and is computed using Equation \ref{eq:rayleigh1} and \ref{eq:rayleigh2}. 

Clouds are treated as large particles with grey opacity.  Therefore, we use a simple treatment of clouds similar to \citet{Benneke2012} and \citet{Sing2016}. In this treatment, clouds are primarily scattering in nature thus decreasing the amount of radiation received by the observer at Earth when the exo-planetary limb is being observed in transmission. The result of significant cloud opacity on transmission spectra is, obscured or muted molecular absorption features depending on the cloud strength, which is governed by the particle size, chemical and radiative properties of the particles. Since, at this stage it is extremely difficult to constrain the type of aerosol particles in exoplanetary atmospheres \citep{Wakeford2015}, we simply tune the strength of grey scattering to represent clouds. Therefore, the size of absorption features is a function of the strength of grey scattering, representing the cloud deck. This can be represented by $\kappa(\lambda)_c = \kappa(\lambda) + \alpha_\textup{cloud}\kappa_{H_2}$, where $\kappa(\lambda)_c$ is the total scattering opacity in $\textup{cm}^{2}/\textup{g}$, $\kappa(\lambda)$ is the scattering opacity due to nominal Rayleigh scattering in similar units, $\alpha_\textup{cloud}$ is the variable cloudiness factor governing the strength of grey scattering and $\kappa_{H_2}$ is the scattering opacity due to H$_2$ at 350 nm which is $ \sim 2.5 \times 10^{-3} \textup{cm}^{2}/\textup{g}$.   This value is calculated using Equations \ref{eq:rayleigh1}  and \ref{eq:rayleigh2} for the scattering cross-section in $\textup{cm}^2$ and divided by the mass of the H$_2$ molecule in $grams$ to obtain scattering opacity in $\textup{cm}^{2}/\textup{g}$, assuming a completely H$_2$ atmosphere (that is H$_2$ mole fraction abundance of 1).

\subsubsection{Transmission Spectra}
\label{section:Transmission Spectra}

\begin{figure}
\includegraphics[width=\columnwidth]{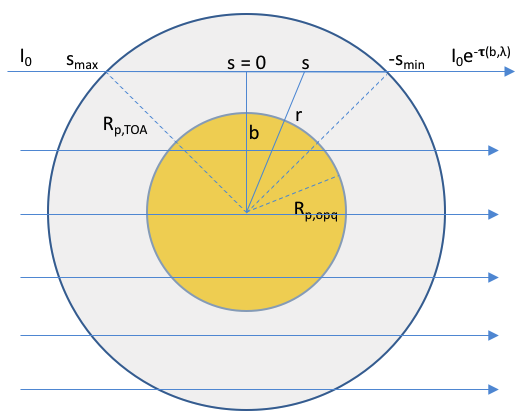}
 \caption{Geometry of transmission spectrum computation. $I_0$ is the incoming stellar radiation, $R_\textup{p,eff}(\lambda)$ is the wavelength dependent effective planetary radius including the atmosphere, $R_\textup{p,opq}(\lambda)$ is the radius below which the atmosphere is optically thick, $R_\textup{p,TOA}$ is the radius at the top of the atmosphere, $b$ is the impact parameter, $\tilde{\tau}(b,\lambda)$ is the atmospheric optical depth, $s$ is the ray path coordinate.}
    \label{fig:schematic_transspectra}
\end{figure}

Transmission spectra in \texttt{ATMO} is computed as shown in schematic Figure \ref{fig:schematic_transspectra} using the equation

\begin{equation} 
R^{2}_\textup{p,eff}(\lambda) = R^{2}_\textup{p,opq}(\lambda) + 2\int_{R_\textup{p,opq}}^{R_\textup{p,TOA}}b\,\textup{d}b(1 - e^{-\tilde{\tau}(b,\lambda)}), \\
\label{eq:transmission spectra}
\end{equation}

where $R_\textup{p,eff}(\lambda)$ is the wavelength dependent effective planetary radius including the atmosphere, $R_\textup{p,opq}(\lambda)$ is the radius below which the atmosphere is optically thick, $R_\textup{p,TOA}$ is the radius at the top of the atmosphere, $b$ is the impact parameter, all in $\textup{cm}$ and $\tilde{\tau}(b,\lambda)$ is the atmospheric optical depth which for a 1D $P$-$T$ profile is spherically symmetric and independent of $\phi$ but is a function of impact parameter $b$ and given by

\begin{equation}
\tilde{\tau}(b,\lambda) = \int_{-s_\textup{min}}^{s_\textup{max}}\textup{d}s\kappa_{\rho}(\lambda,s)\rho(s),  \\
\end{equation}

where $s_\textup{min}$ is the minimum path coordinate of the ray as it leaves the atmosphere as illustrated in Figure \ref{fig:schematic_transspectra}, while $s_\textup{max}$ is the maximum path coordinate where the ray enters the atmosphere both in $\textup{cm}$, $\rho(s)$ is the density in $\textup{g}/\textup{cm}^3$ at path $s$ given by  $\sqrt{r^2 - b^2}$  in $\textup{cm}$ and $\kappa_{\rho}(\lambda,s)$ is the opacity as a function of wavenumber and path $s$ in $\textup{cm}^2/\textup{g}$. This gives the effective radius of the planet as a function of wavelength which represents the model transmission spectra of the planet. It is worth noting that the chemical and the thermodynamic structure of the atmosphere imprint their signature in the transmission spectra via $\rho$ and $\kappa$ variables. We note that we assume single scattering and neglect refraction while computing our transmission spectra.

\subsection{Chemistry}
\label{subsection:Chemistry}
 
 \texttt{ATMO} has two chemistry schemes, a Gibbs energy minimisation scheme following \citet{Gordon1994} used for equilibrium chemistry calculations and a chemical kinetics scheme that currently adopts the chemical network of \citet{Venot2012}. The exact methodology and implementation details of both schemes are explained in \citet{Drummond2016}. The chemical kinetics scheme can also be used to simulate non-equilibrium physical processes like vertical mixing and photochemistry. The coupling of the radiative-convective scheme with the chemistry (equilibrium and non-equilibrium) scheme also allows fully consistent modelling, where both the $P$-$T$ profile and the chemical abundances are solved for simultaneously. Therefore, it provides a final $P$-$T$ and chemical abundances profiles which are physically consistent with each other. However, as a grid for a range of planets requires extensive computational resources, we restrict ourselves to equilibrium chemistry for this work.
		
The Gibbs energy minimisation scheme follows the method of \citet{Gordon1994}, with the thermochemical data for each species taken from \citet{McBride1993,McBride2002}. For this particular analysis, a total of 258 chemical species comprising of both gaseous and condensate species were included. The 23 elements included in the model to form these 258 species are H, He, C, N, O, Na, K, Si, Ar, Ti, V, S, Cl, Mg, Al, Ca, Fe, Cr, Li, Cs, Rb, F and P. Local chemical equilibrium abundances are computed by minimising the Gibbs energy independently on each model level. This scheme has been validated by reproducing the results of the TECA chemical equilibrium code \citep{Venot2012}, as well as the analytical solutions to chemical equilibrium by \citet{Burrows1999} and \citet{Heng2016}. For more details on the chemistry schemes see \citet{Drummond2016}. 
	
\texttt{ATMO} considers three options when calculating the chemical equilibrium abundances: \\
1) Gas-phase only - Only gas phase species are included and condensed phase species assumed to be negligible. \\
2) Local condensation - Condensed species are allowed to form, depleting the gas-phase abundance of the elements locally but each model level is independent and has the same elemental abundance. \\
3) Rainout condensation - Condensed species are allowed to form and the elemental abundance of the elements within those condensed species are progressively depleted along the profile. \\

The gas-phase only approach (1) is likely to be valid for very hot atmospheres where the temperature is above the condensation temperature of most condensate species. The local condensation option (2) assumes that the formation of condensates in one model level does not effect the availability of elements in other model levels. Finally, the rainout condensation approach (3) assumes that once condensates are formed the particles sink in the atmosphere and the elements that comprise that condensate are depleted stoichiometrically from the layers above \citep[e.g][]{Barshay1978, Burrows1999}. All our model simulations in the grid are performed using the equilibrium chemistry scheme and including condensation with rainout. We have adopted condensation with rainout mechanism in this paper since it is the most common assumption in planetary atmospheric models \citep{Burrows1999, Lodders2006, Fortney2008, Mbarek2016}
	
The solar elemental abundances are adopted from \citet{Caffau2011}. In a particular simulation, the elemental abundances are then adjusted for the set metallicity and C/O ratio parameters. The metallicity is taken into account by multiplying the abundances of the elements (except H, He and O) by the appropriate factor, and then re-normalising such that the sum of the fractional abundances is equal to unity. We note that the oxygen abundance is set via the carbon abundance and the prescribed C/O ratio following \citet{Moses2011}, and the C/O ratio refers to total elemental abundance across gas and condensate phase.

\begin{figure*}
\begin{center}
 \subfloat[]{\includegraphics[width=\columnwidth]{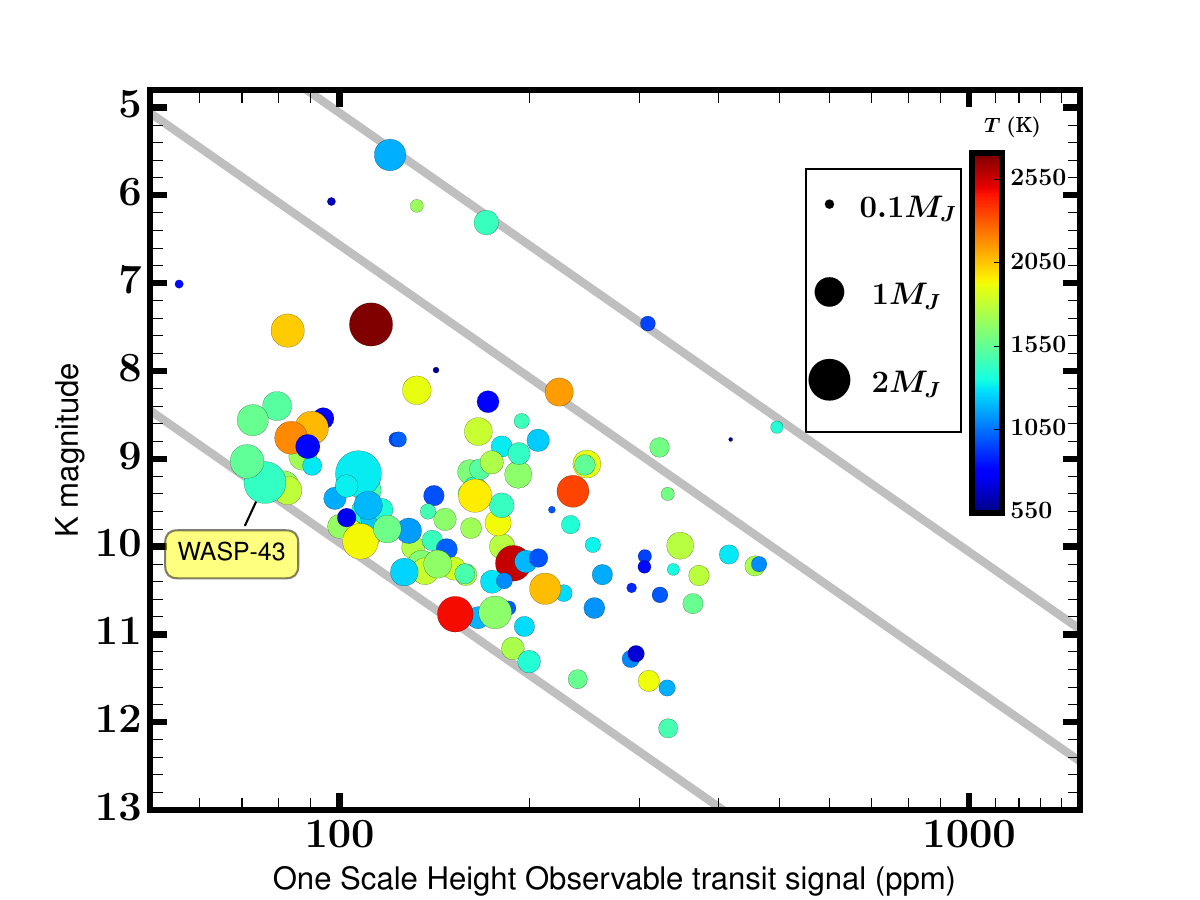}}
 \subfloat[]{\includegraphics[width=\columnwidth]{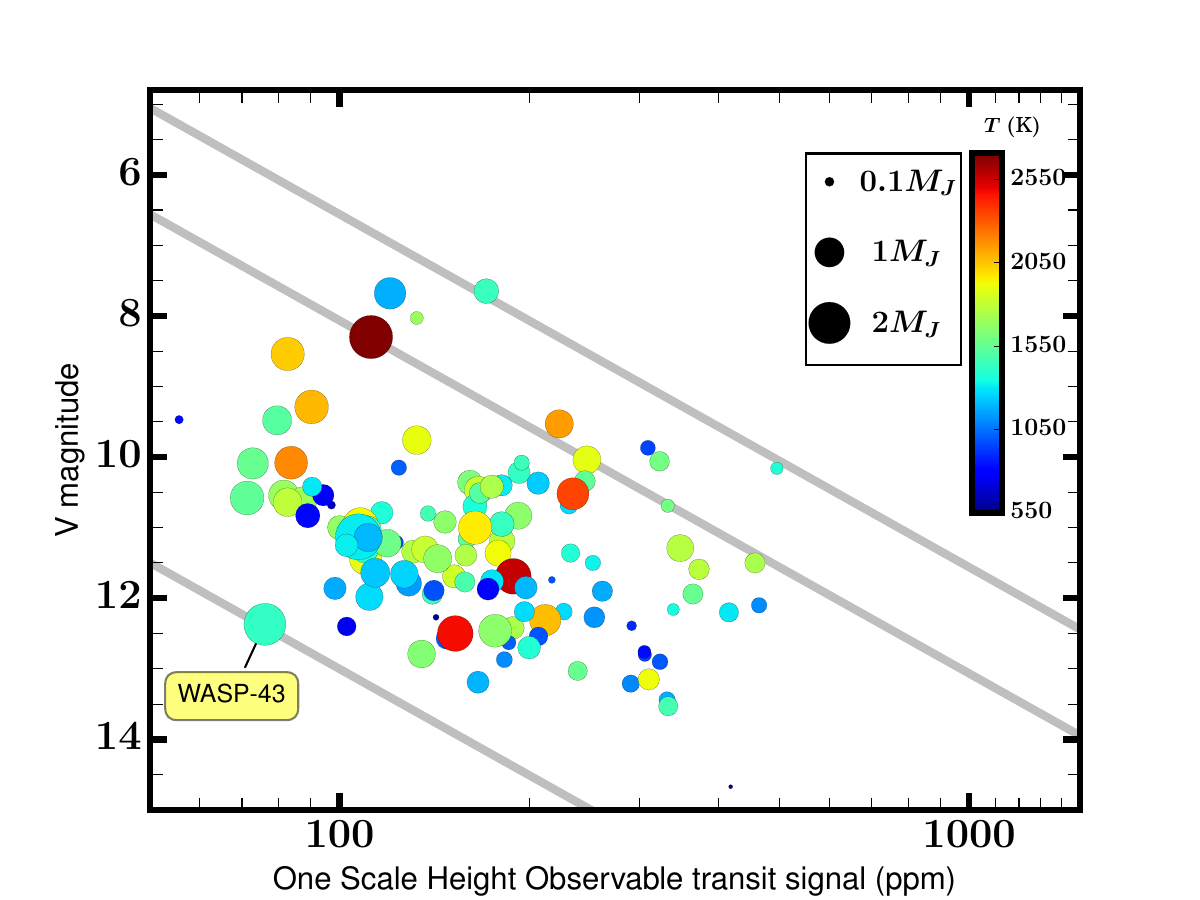}}
\end{center}
 \caption{\textbf{(a)} - Figure showing host star K magnitude vs. one atmospheric scale height observable transit signal in parts per million (ppm) for all the planets selected in the grid. Colours indicate the temperature of the planet based on the scale, and the size shows the planet mass relative to the mass of the Jupiter. Grey lines indicate contours at theoretical relative signal to noise (SNR) ratio values of 20, 10 and 2.1 (for WASP-43b) from top to bottom. \textbf{(b)} same as (a) but for V magnitude and SNR contours of 20, 10 and 1 (for WASP-43b) from top to bottom.}
    \label{fig:Planet_Selection1}
\end{figure*}

\section{Grid Setup}
\label{section:Grid Setup}

We use 50 vertical model levels with minimum and maximum optical depths of $10^{-7}$ and $10$ at 1 \textmu m, respectively. This covers the atmospheric region that is characterised via transmission spectra ($\sim$0.1-100 millibar), with reasonable computational time for each model run. However, when we compute $P$-$T$ profiles as in model validation Section \ref{section:Model Validation}, our maximum optical depth limit is $2\times10^5$ at 1 \textmu m, since we need to compute temperatures even in the higher pressure region ($\sim10^3$ bar).  Also to standardise the comparison of transmission spectra for a range of variables, we set the pressure at which the radius of the planet is defined at 1 millibar \citep{Lecavelier2008}. We note that there exists a  degeneracy between reference transit radius and associated reference pressure as highlighted by \citet{Lecavelier2008, Heng2017}.
Transmission spectra probes the atmospheric region around $\sim$0.1-100 millibar, therefore we restrict our upper atmosphere model pressure to $10^{-6}$ bars. The input stellar spectra for each planetary model grid are taken from the BT-Settl\footnote{\url{https://phoenix.ens-lyon.fr/Grids/BT-Settl/AGSS2009/SPECTRA/}} models \citep{Allard2012, Rajpurohit2013}. These stellar spectra are selected according to closest obtained host star temperature, gravity and metallicity from the TEPCAT database \citep{Tepcat2011}. All the parameters required for model initialisation like stellar radius, planetary radius, planetary equilibrium temperature, surface gravity and semi-major axis are also adopted from TEPCAT\footnote{\url{http://www.astro.keele.ac.uk/jkt/tepcat/allplanets-ascii.txt}} database, along with observational parameters like stellar $V_{mag}$  and $K_{mag}$ for target selection (see Appendix \ref{app:Planets and their parameters in Grid}).

\subsection{Target Selection}
\label{section:Target Selection}

An order of magnitude estimate of the observable transit signal can be calculated using basic geometry \citep{Winn2010}, by using planet parameters and taking the ratio of the annular area of the planetary atmosphere to that of the stellar surface area given by 

\begin{equation} 
OTS = \frac{2 R_\textup{p} H}{R^2_{*}} {10^6},
\label{eq:One Scale Height Transit Signal} 
\end{equation}

where $OTS$ is the observable transit signal for one scale height of the atmosphere in parts per million (ppm),  $H$ is scale height given by  $KT/ \mu g$, where $K$ is the Boltzmann constant, $T$ is the planetary equilibrium temperature, $\mu$ is the mean molecular weight of the planetary atmosphere which in this case is for a H$_2$ and He dominated atmosphere, $g$ is the planetary surface gravity, $R_p$ is the planetary radius within which the planet is optically thick at all wavelengths and $R_*$ is the stellar radius, all in CGS units. The $OTS$ for each planet and their host star  $V_{mag}$ and $K_{mag}$ are plotted in Figure \ref{fig:Planet_Selection1}, along with contours, at a particular relative theoretical signal to noise ratio (SNR) given by 

\begin{equation} 
SNR_\textup{c} = SNR_\textup{ref}\frac{OTS_\textup{c}}{OTS_\textup{ref}}10^{\frac{-(V_\textup{c} - V_\textup{ref})}{5}},
\label{eq:SNR} 
\end{equation}

where $SNR_c$ is the theoretical relative signal to noise ratio of the contour, $SNR_\textup{ref}$ is the same for the reference planet, $OTS_{c}$ are the range of one scale height observable transit signal plotted in the contours,  while $OTS_\textup{ref}$ is the $OTS$ for the reference planet, $V_{c}$ and $V_\textup{ref}$ are the V magnitudes in the contours and reference planet host star, respectively. In our case we have taken WASP-12b as our reference to plot SNR contours in Figure \ref{fig:Planet_Selection1}. This SNR for WASP-12b is calculated with five scale height transit depth value and the average noise calculated from \citet{Mandell2013} for one transit. These contours are used to select observationally significant atmospheres of exoplanets as shown in Figure \ref{fig:Planet_Selection1}.

We select the planets with theoretical relative SNR greater than that of WASP-43b in $V_{mag}$ and $K_{mag}$ as shown in Figure \ref{fig:Planet_Selection1}. We have deliberately chosen to make this grid planet specific, rather than exploring the huge parameter space of mass, radius, gravity etc., which would have increased the size of the grid substantially. Making it planet specific is also very helpful to directly use it for observational proposals and interpretation without interpolation. All the planets with their parameters and references, selected in our current grid of model simulations from TEPCAT database \citep{Tepcat2011}, are shown in Appendix \ref{app:Planets and their parameters in Grid}.

\begin{table*} 
	\centering	
	\begin{tabular}{|p{3cm}|p{2cm}|p{2cm}|p{2cm}|p{2.2cm}|}    
		\hline
		Temperature & Metallicity & C/O-ratio & Haze & Grey \\
		 (K) & (x solar) &   & enhancement factor ($\alpha_\textup{haze}$) & cloudiness factor ($\alpha_\textup{cloud}$) \\ 
		\hline
		T$_{\rm{eq}}$ - 300 & 0.005 & 0.15 & 1 (No Haze) & 0 (No Cloud)\\
                 T$_{\rm{eq}}$ - 150 & 0.1 & 0.35 & 10 & 0.06\\
                 T$_{\rm{eq}}$  & 1 & 0.56 & 150 & 0.2\\
                 T$_{\rm{eq}}$+ 150 & 10 & 0.70 & 1100 & 1\\
                 T$_{\rm{eq}}$ + 300 & 50 & 0.75 &   &  \\
                   & 100 & 1 &   &  \\
                   & 200 & 1.5  &   &  \\
		\hline
	\end{tabular}%}
	\caption{Table showing the entire parameter space of the grid. The temperature is with respect to the planetary equilibrium temperature (T$_{\rm{eq}}$). The C/O ratio of 0.56 is solar value. The haze enhancement factor is with respect to gaseous Rayleigh scattering. The grey cloudiness factor is with respect to H$_2$ scattering cross-section at 350 nm.} 
	\label{tab:grid_param}
\end{table*}

\subsection{Parameter Space Selection}
\label{section:Parameter Space Selection}
This section describes in detail, the parameter space of the grid for which model transmission spectra are generated. These parameters have been selected based on the most important physical parameters affecting the transmission spectra and the computational feasibility of running the simulations for a range of planets. For each planet, five major parameters are varied and are listed in Table \ref{tab:grid_param}. The first parameter is the temperature of the planet, which is not a well constrained parameter observationally, since it is dependent on various other properties of the atmosphere. However, it has a profound effect on the transmission spectra of a planet \citep{Fortney2010}. The zeroth order T$_{\rm{eq}}$ calculated based on the distance of the planet from the host star is the only known parameter. When computing transmission spectra we are concerned with the temperature approximately around the 1 millibar pressure region of the atmosphere. Therefore, T$_{\rm{eq}}$ is used as a first guess. We vary the temperature of the planetary atmosphere in increments of $150$\,K to a maximum of $\pm300$\,K, with respect to the T$_{\rm{eq}}$ of the planet, giving a total of 5 temperature grid points per planet as shown in Table  \ref{tab:grid_param}. The selection of $150$\,K increment is based on the typical temperature uncertainty in the observational transmission spectra \citep{Lecavelier2008}. The selection of maximum variation of $\pm300$\,K is based on a compromise between computational feasibility and accuracy required to capture major spectral features. The metallicity of a planet is a parameter which indirectly determines the chemical composition of its atmosphere, thereby affecting its observable signatures in the transmission spectra. The metallicity is varied from sub-solar to super-solar values: 0.005, 0.5, 1, 10, 50, 100 and 200 times solar.

\citet{Oberg2011} and \citet{Madhusudhan2016} provided evidence of utilising C/O ratios to constrain the location of planetary formation in the debris disk. Its effect on the exoplanet atmospheric chemistry has been studied extensively by \citet{Seager2005, Kopparapu2012, Madhusudhan2012, Moses2013}. In particular, \citet{Molliere2015} developed a very extensive grid for various C/O ratios and analysed its effect on the emission spectrum. The C/O ratios are selected here based on the current important transition values guided by previous studies. Our selection of lower C/O ratios (0.15 and 0.35) was guided by model fitting to observations, since some of the observations were consistent with very low C/O ratio (see Section \ref{section:Interpretation of Observations}). Therefore, our parameter space contains C/O ratios of 0.15, 0.35, 0.56, 0.7, 0.75, 1 and 1.5. The solar C/O ratio is $\sim$0.56 \citep{Caffau2011}. 

\citet{Sing2016} presented a comparative planetology of various exoplanets that highlighted the importance of haze and clouds in understanding and characterising  exoplanet atmospheres using transmission spectra. \citet{Lecavelier2008} and \citet{Sing2015, Sing2016} highlighted the effect of haze in muting the spectral features in transmission spectra. Therefore, haze in the form of Rayleigh scattering having variable strengths with respect to the nominal multi-gas Rayleigh scattering has been included in the grid as a fourth parameter.  $\alpha_\textup{haze}$, the haze enhancement factor, explained in Section \ref{subsubsection:Haze and Cloud Treatment} is varied in the grid in steps such that it leads to approximately one scale height change in the transmission spectrum, where the Rayleigh scattering dominates, which leads to multiplication factors 1, 10, 150, 1100 times nominal multi-gas Rayleigh scattering in the grid.

 A grey scattering opacity representing clouds of different scattering cross-sections is used as a fifth parameter. We use a grey cloud strength factor ($\alpha_\textup{cloud}$) (see Section \ref{subsubsection:Haze and Cloud Treatment}) of 0.06, 0.2 and 1 corresponding to scattering opacity ($\kappa$) of $\sim$ $1.5 \times 10^{-4}$, $5 \times 10^{-4}$ and $2.5 \times 10^{-3}$ $\textup{cm}^{2}/\textup{g}$, respectively in the grid. $\alpha_\textup{cloud}=0$ corresponds to clear sky scenario. These factors were chosen based on the change in the 1.4 \textmu m H$_2$O spectral feature due to addition of grey clouds, particularly for the test case of HD~189733b \citep{McCullough2014, Sing2016, Heng2017}. However, since these factors correspond to fixed values of scattering opacity, they are independent of planetary parameters. The factors 0.06, 0.2 and 1 led to the transit radius ratio of this 1.4 \textmu m feature being reduced to $\sim$$66\%$, $33\%$ and $15\%$, respectively, compared to clear atmosphere case (see Figure \ref{fig:cloud_hd189733b} discussed in Section \ref{subsection:Effect of Haze and Clouds}). In \texttt{ATMO}, clouds can be specified at any level in the atmosphere. However, we specify clouds throughout the atmosphere (all 50 levels), while changing its scattering strength to represent the degree of cloudiness. 

\begin{figure*}
\begin{center}
 \subfloat[]{\includegraphics[width=\columnwidth]{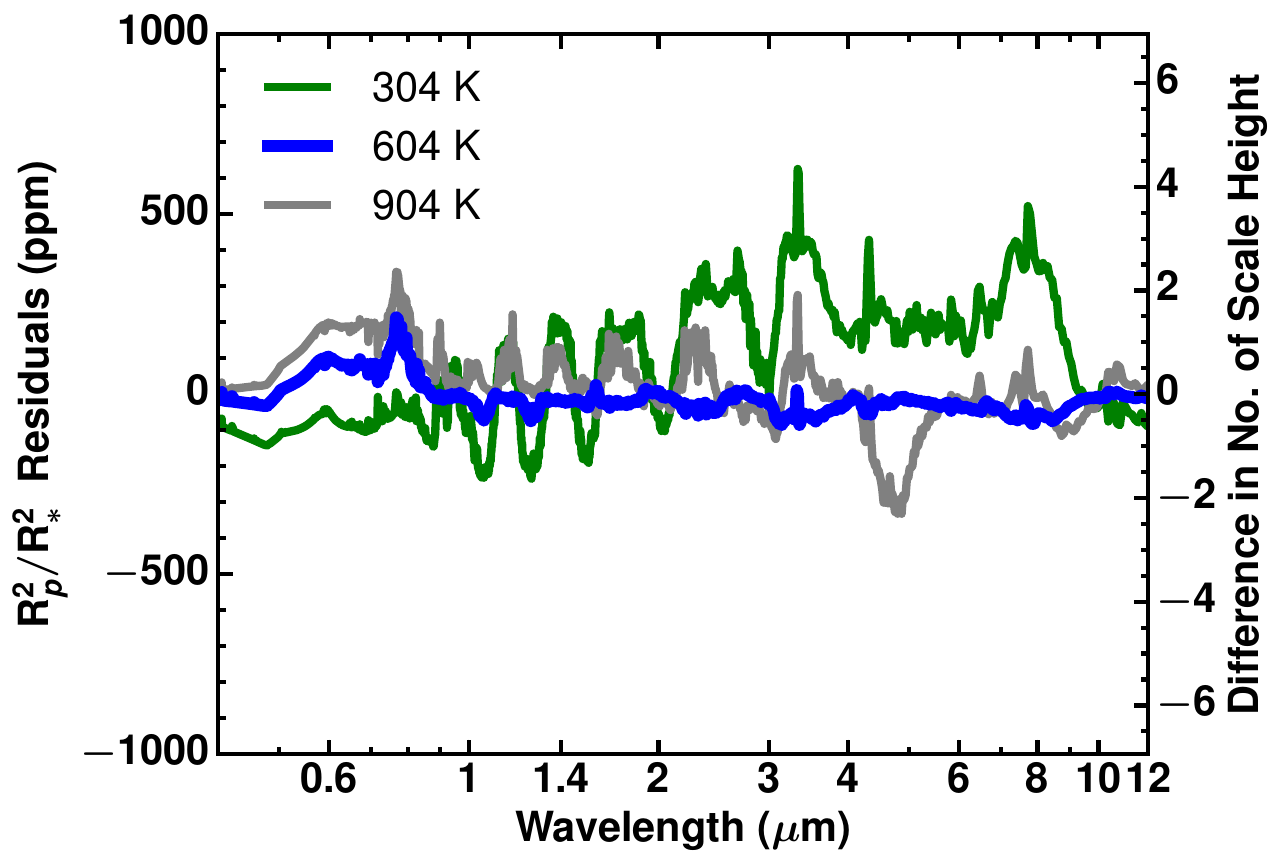}\label{fig:Iso_eq_diff_spectra_gj3470}}
 \subfloat[]{\includegraphics[width=\columnwidth]{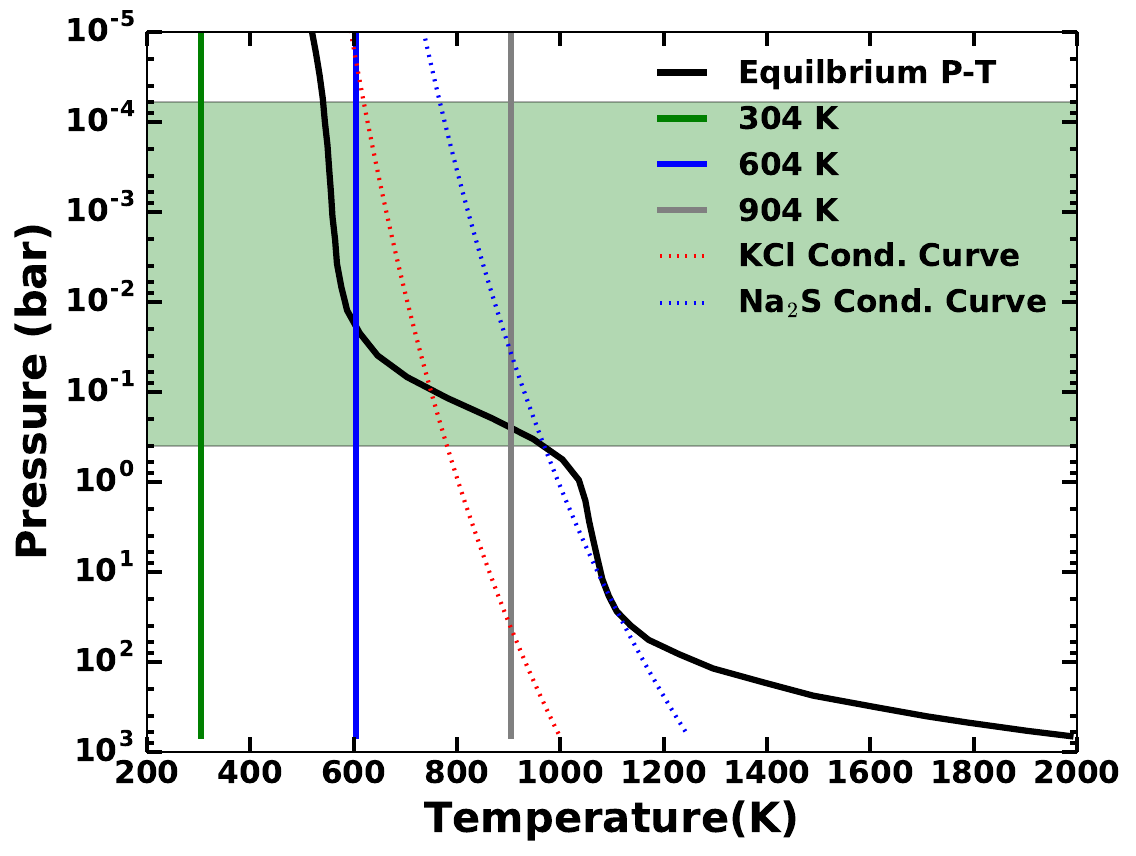}\label{fig:Iso_eq_pt_gj3470}}
\end{center}
 \caption{\textbf{(a)} Figure showing differences (residuals) in transit depth ($\Rp^2/\Rs^2$) generated using radiative-convective equilibrium $P$-$T$ profiles and isothermal $P$-$T$ profile (in the sense consistent minus isothermal) for the isothermal temperatures in our parameter space for GJ~3470b which are $304$\,K (green), $604$\,K (blue) and $904$\,K (grey). Thicker line in blue for $604$\,K shows minimum residuals and green line for $304$\,K shows maximum residuals. Spectra with equilibrium $P$-$T$ profile is using the recirculation factor of 0.5 (see Section \ref{section:Model Validation}). Residuals are shown both in transit depth in parts per million (ppm) on left  and number of scale heights on right Y-axis. X-axis shows wavelength in \textmu m. \textbf{(b)} Figure showing radiative-convective equilibrium $P$-$T$ profiles for a recirculation factor of 0.5 (black), and isothermal $P$-$T$ profiles in our parameter space for GJ~3470b which are $304$\,K (green), $604$\,K (blue) and $904$\,K (grey). The condensation curves for KCl and Na$_2$S are also shown with dotted lines in red and blue respectively. Shaded green region highlights the atmospheric pressures (altitude) probed using the transmission spectra. X-axis shows temperature in Kelvin and Y-axis shows pressure in bar.}
\end{figure*}

\begin{figure*}
\begin{center}
 \subfloat[]{\includegraphics[width=\columnwidth]{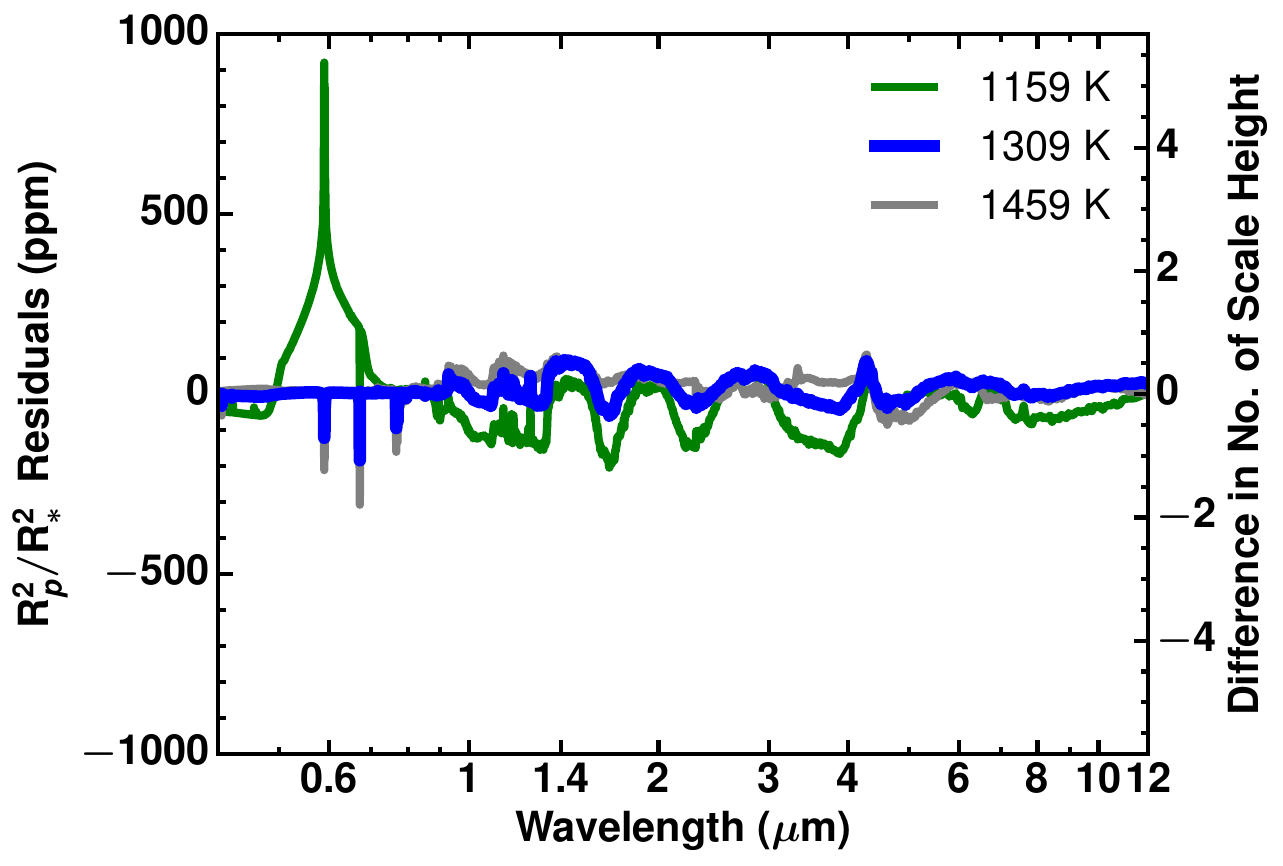}\label{fig:Iso_eq_diff_spectra_hd209}}
 \subfloat[]{\includegraphics[width=\columnwidth]{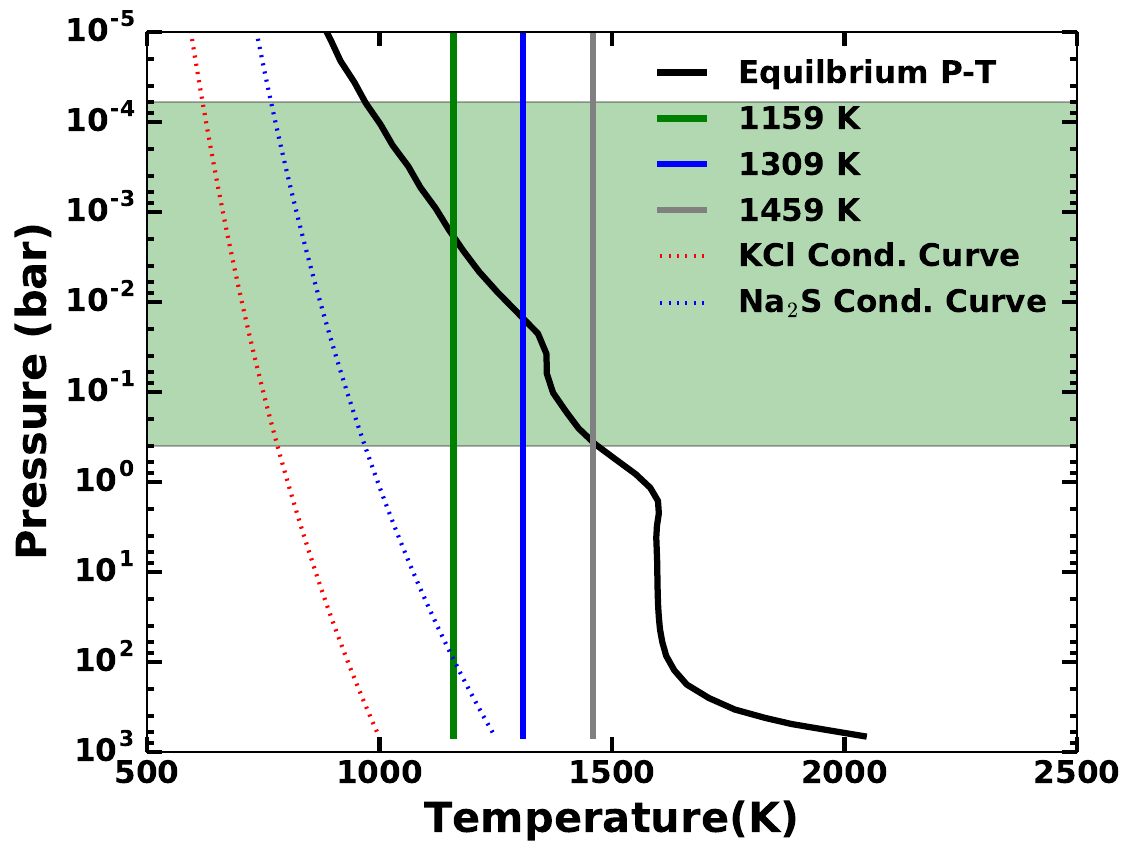}\label{fig:Iso_eq_pt_hd209}}
\end{center}
 \caption{\textbf{(a)} Figure showing residuals similar to Figure \ref{fig:Iso_eq_diff_spectra_gj3470}, but for hotter planet, HD~209458b (T$_{\rm{eq}}$ = $1459$\,K) at $1159$\,K (green), $1309$\,K (blue) and $1459$\,K (grey). Thicker line in blue for $1309$\,K shows minimum residuals and green line for $1159$\,K shows maximum residuals. \textbf{(b)} Figure similar to Figure \ref{fig:Iso_eq_pt_gj3470}, but for HD~209458b showing radiative-convective equilibrium $P$-$T$ profiles for a recirculation factor of 0.5 (black), and isothermal $P$-$T$ profile at $1159$\,K (green), $1309$\,K (blue) and $1459$\,K (grey).}
\end{figure*}

\begin{figure*}
\begin{center}
 \subfloat[]{\includegraphics[width=\columnwidth]{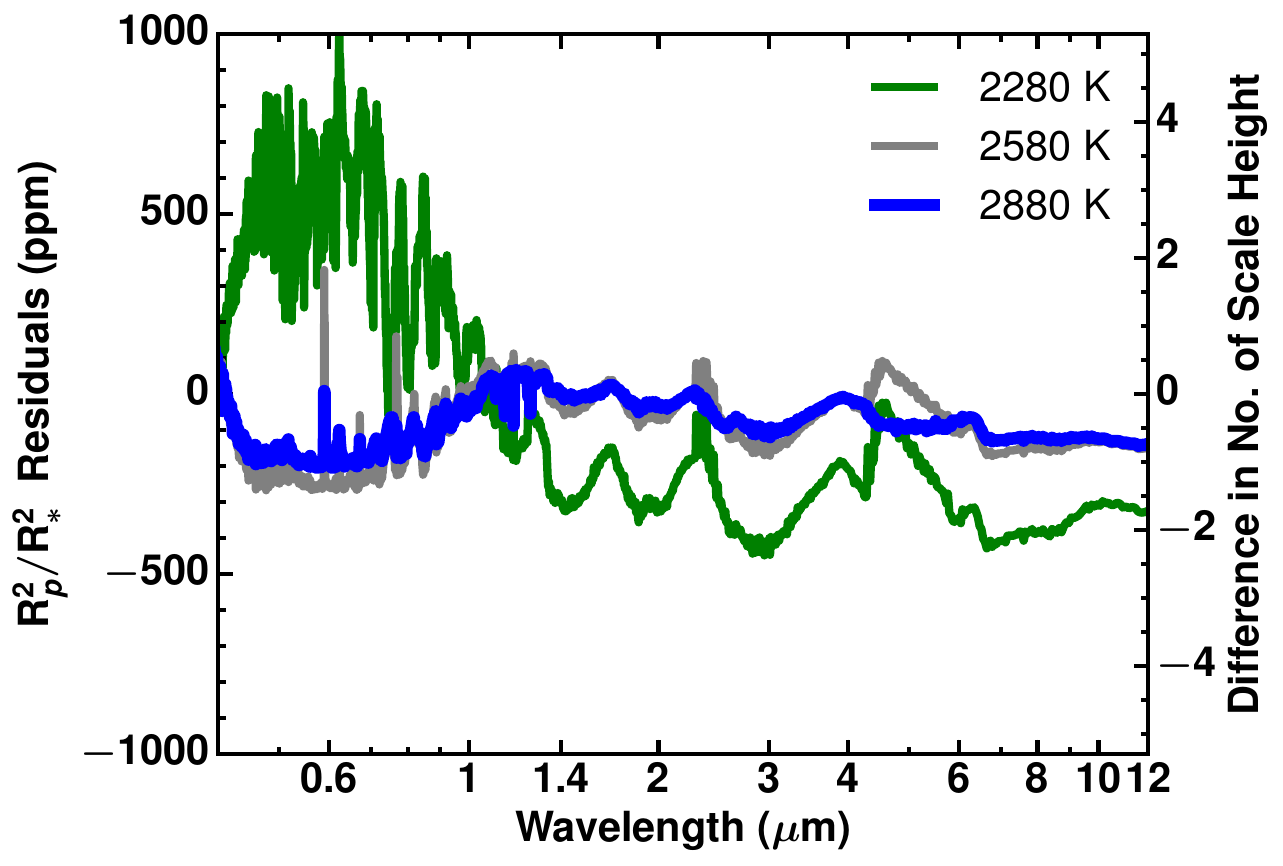}\label{fig:Iso_eq_diff_spectra_wasp12}}
 \subfloat[]{\includegraphics[width=\columnwidth]{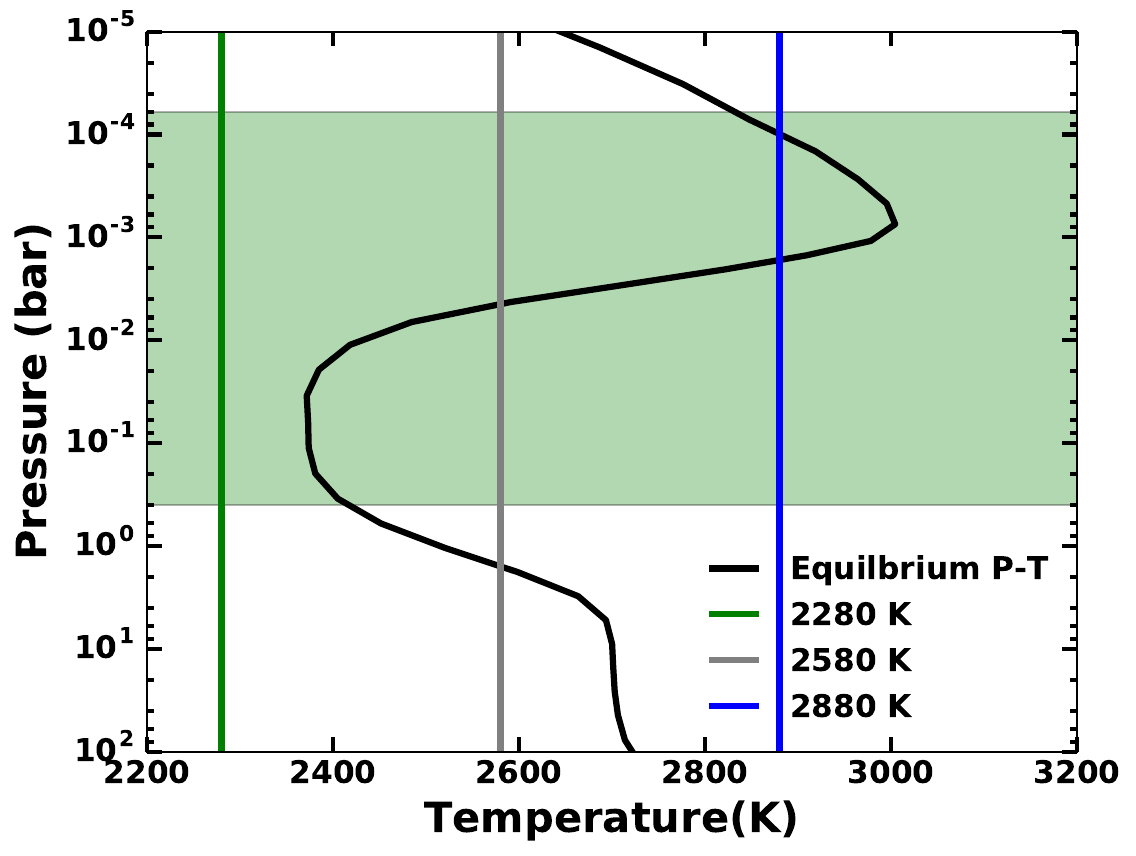}\label{fig:Iso_eq_pt_wasp12}}
\end{center}
 \caption{\textbf{(a)} Figure showing residuals similar to Figure \ref{fig:Iso_eq_diff_spectra_gj3470}, but for a hotter planet WASP-12b (T$_{\rm{eq}}$ = $2580$\,K) at $2580$\,K (green), $2730$\,K (grey) and $2880$\,K (blue). Thicker line in blue for $2880$\,K shows minimum residuals and green line for $2280$\,K shows maximum residuals. \textbf{(b)} Figure similar to Figure \ref{fig:Iso_eq_pt_gj3470} but for WASP-12 showing radiative-convective equilibrium $P$-$T$ profiles for a recirculation factor of 0.5 (black), and isothermal $P$-$T$ profile at $2280$\,K (green), $2580$\,K (grey) and $2880$\,K (blue).}
\end{figure*}
		
\section{Model Validation with Equilibrium $P$-$T$ Transmission Spectra}
\label{section:Model Validation}

We currently calculate transmission spectra adopting isothermal $P$-$T$ profiles. However, this will be extended to include $P$-$T$ profiles in radiative-convective equilibrium (consistent $P$-$T$ profiles) in our upcoming work. In this section we explain the differences between transmission spectra computed from isothermal $P$-$T$ profiles and those from consistent $P$-$T$ profiles. We note that model transmission spectra with isothermal $P$-$T$ profiles have been shown sufficient to explain the observations \citep[see][]{Fortney2005, Heng2017}.

To quantify the effect of assuming an isothermal $P$-$T$ profile as opposed to calculating a $P$-$T$ profile consistent with radiative-convective equilibrium, we compared the two approaches for planets spanning a wide range of T$_{\rm{eq}}$. Namely, GJ~3470b (T$_{\rm{eq}}$ = $604$\,K), HD~209458b (T$_{\rm{eq}}$ = $1459$\,K) and WASP-12b (T$_{\rm{eq}}$ = $2580$\,K). Computing a $P$-$T$ profile consistent with radiative-convective equilibrium requires adoption of an angle of incidence ($\theta$) for the radiative flux, and a \enquote{recirculation factor} \citep[treated as a reduction in incoming flux as in][]{Fortney2007}. The recirculation factor simulates the redistribution of input stellar energy in the planetary atmosphere, by the dynamics, where a value of $1$ equates to no redistribution, while $0.5$ represents efficient redistribution. Simulations adopting solar metallicity, solar C/O ratio, without cloud or haze, were then performed for the consistent case, adopting $\theta = 60^{\circ}$ (equating to the dayside average)  and a recirculation factor of 0.5, and compared to each of the counterpart different temperature isothermal simulations in our grid. 

The differences (residuals) between simulated spectra derived from the consistent simulations and their isothermal counterparts at three different temperatures adopted in the grid are shown for GJ~3470b, HD~209458b and WASP-12b in Figures \ref{fig:Iso_eq_diff_spectra_gj3470}, \ref{fig:Iso_eq_diff_spectra_hd209} and \ref{fig:Iso_eq_diff_spectra_wasp12}, respectively. Figures \ref{fig:Iso_eq_pt_gj3470}, \ref{fig:Iso_eq_pt_hd209} and \ref{fig:Iso_eq_pt_wasp12} show the derived consistent (equilibrium)  $P$-$T$ profiles and the adopted isothermal profiles for these simulations, alongside the condensation curves of KCl and Na$_2$S. Figures \ref{fig:Iso_eq_diff_spectra_gj3470}, \ref{fig:Iso_eq_diff_spectra_hd209} and \ref{fig:Iso_eq_diff_spectra_wasp12}  include both examples of the largest and smallest residuals, and reveal that the differences are all less than $\sim$1 scale height for the closest matching isothermal spectrum at all the wavelengths. Therefore, very high precision measurements (e.g $\sim$150 ppm for HD~209458b) would be needed to detect temperature variations via altitude-dependent scale height differences in the transmission spectra probed region.

In the case of GJ~3470b, residuals within $\sim$\,1 scale height are seen for the isothermal temperature of 604\,K in Figure \ref{fig:Iso_eq_diff_spectra_gj3470} since this temperature is closest to the consistent $P$-$T$ profile in the transmission spectra probed region, which is almost isothermal, as shown in Figure \ref{fig:Iso_eq_pt_gj3470}. The residuals are largest for the coolest isothermal simulation at 304\,K, since it is substantially different from the consistent $P$-$T$ profile. For HD~209458b and Wasp-12b, the residuals of the closest matching isothermal spectrum are again within $\sim$\,1 scale height, despite the $P$--$T$ profile being far from isothermal. For optical wavelengths large residuals can be seen, for the coolest isothermal temperature, at the core of the strong Na lines for HD~209458b and TiO/VO lines for WASP-12b. For HD~209458b this large difference is caused by the condensation of Na$_2$S which occurs, as shown in Figure \ref{fig:Iso_eq_pt_hd209}, in the coolest isothermal simulation, at pressures above $10^2$\,Pa (where the Na$_2$S condensation curve intersects the temperature of 1159\,K), but not in the consistent version. In our model we assume efficient settling of condensates i.e. \enquote{rainout}, which as described in Section \ref{subsection:Chemistry}, depletes the atmosphere above the condensation point of the constituent species. This leads to the absence of Na features in the spectrum derived from the coolest isothermal simulation, and thus, large residuals when compared to the radiative-convective equilibrium version. A similar effect is found for WASP-12b, but due to condensation of TiO/VO bearing species in the coolest isothermal simulation as shown in Figure \ref{fig:Iso_eq_diff_spectra_wasp12}. However, additionally, there is substantial deviation of coolest isothermal $P$-$T$ profile from that of consistent profile, as seen in Figure \ref{fig:Iso_eq_pt_wasp12}, increasing the residuals. It is important to note that the residuals found between the spectra derived from the isothermal and consistent $P$-$T$ profile simulations are also a function of the recirculation factor adopted in the latter. As the recirculation factor is an unconstrained parameter this introduces uncertainties into the consistent calculation thereby effecting the match with the isothermal spectra.

In summary, for all the test case planets from our grid shown in Figures \ref{fig:Iso_eq_diff_spectra_gj3470}, \ref{fig:Iso_eq_diff_spectra_hd209} and \ref{fig:Iso_eq_diff_spectra_wasp12}, the assumption of an isothermal atmosphere leads to observationally negligible differences for the closest matching, most appropriate isothermal temperature,  except where the temperatures are cool enough for condensation and subsequent rainout to occur (as is the case for lowest isothermal temperature for these planets). In practice, different isothermal temperatures can be used for different altitudes, as was done for for the Na line in HD~189733b \citep{Huitson2012}, which would avoid this issue.

As described in Section \ref{subsection:Chemistry} we assume efficient settling of condensed species (rainout) while computing the equilibrium chemistry in our current simulations, which is a widely adopted assumption in the literature \citep{Burrows1999, Lodders2006, Fortney2008, Mbarek2016}. Without a sophisticated cloud model, calculating whether a given condensate will be present in the atmosphere or settle is not possible, so the best we can provide are the two limiting cases of efficient settling (rainout) and efficient vertical lofting (local condensation or no-rainout). This concern has prompted us to also provide a matching grid of isothermal simulations via the website in the near future, adopting the opposite assumption, i.e. efficient lofting of condensed species (no-rainout). However, we note that all the simulations in this paper are performed under the assumption of condensation with rainout.

\begin{figure*}
\includegraphics[width=\textwidth, height=622 pt]{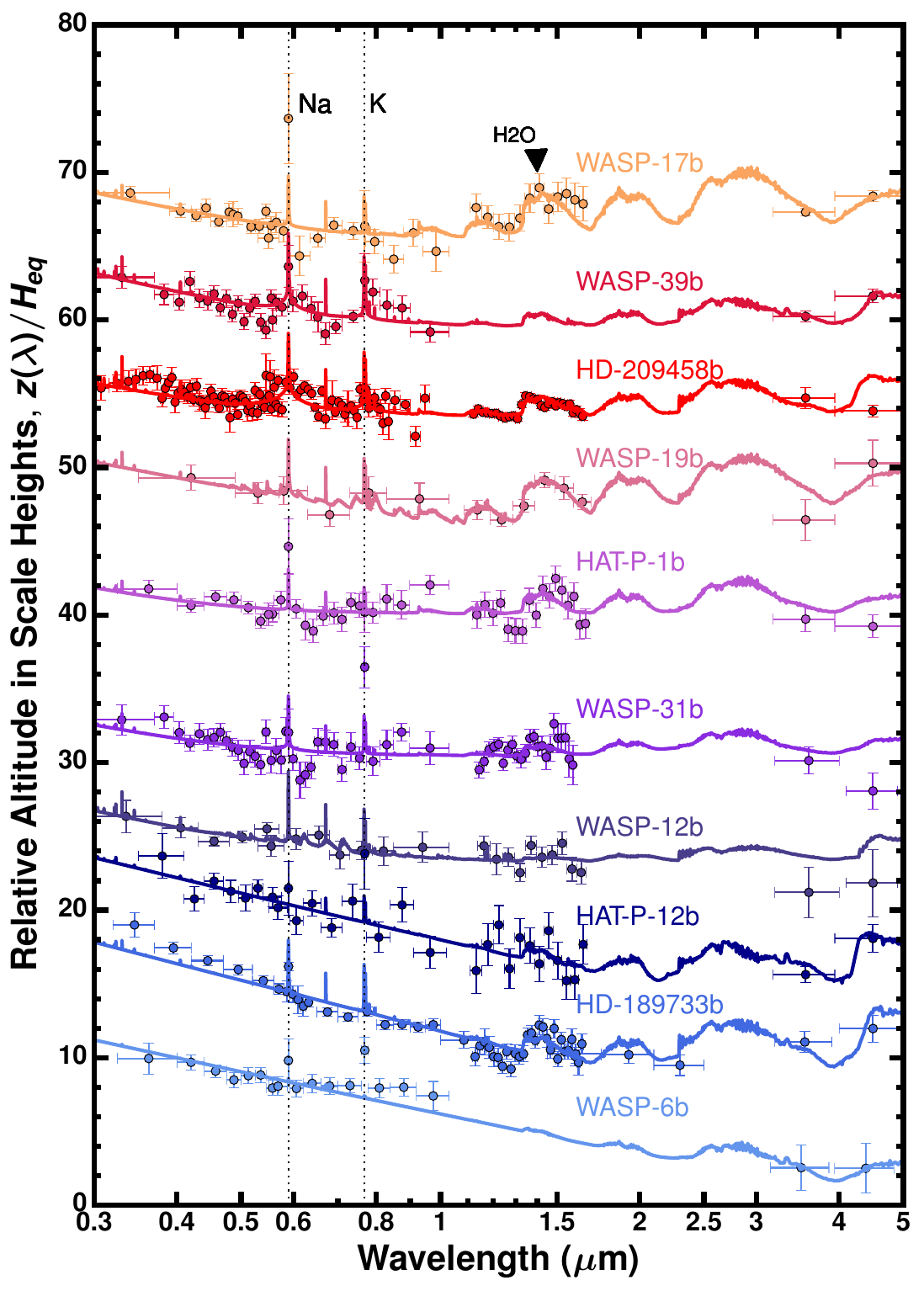}
 \caption{Figure showing the ATMO forward model grid applied to observations of 10 exoplanets from \citet{Sing2016}. The Y-axis shows relative altitude in scale height. Solid lines show best fit forward models and filled circular markers show HST observations with error-bars. Planet names are placed above their respective spectra. Dashed lines indicate expected Na and K features. Comparatively clear atmospheres at the top have strong H$_2$O and alkali features. The strength of these features decreases from top to bottom as planets become more hazy and cloudy.}
    \label{fig:allplanets}
\end{figure*}

\begin{table*}  
	\centering	
	\begin{tabular}{|p{1.75cm}|p{0.75cm}|p{1cm}|p{1.25cm}|p{0.75cm}|p{1cm}|p{1cm}|p{1cm}|p{0.5cm}|p{1.75cm}|p{3cm}|}     
		\hline
		Planet & T$_{\textup{eq}}$ & T$_{\textup{bestfit}}$ & Metallicity & C/O & Haze & Cloud & $\chi^{2}$ & DOF & Reduced $\chi^{2}$ & Data Source\\
		  & (K) & (K) & (x solar) &   & ($\alpha_\textup{haze}$)  & ($\alpha_\textup{cloud}$)  &  &  & &\\ 
		\hline
		WASP-17b & 1755 & 1755 & 0.1 & 0.15 & 10 & 0.2 & 29.67 & 38 & 0.7807 & \citet{Sing2016}\\
		WASP-39b & 1116 & 1266 & 1 & 0.56 & 10 & 0.2 & 41.84 & 34 & 1.23 & \citet{Fischer2016, Sing2016}\\
		HD-209458b & 1459 & 1459 & 10 & 0.56 & 10 & 0.5 & 230.61 & 123 & 1.874 & \citet{Sing2016}\\
		WASP-19b & 2077 & 1927 & 0.1 & 0.35 & 10 & 0 & 7.21 & 13 & 0.555 & \citet{Huitson2013, Sing2016}\\
		HAT-P-1b & 1322 & 1322 & 0.1 & 0.15 & 10 & 1.0 & 50.06 & 41 & 1.22 & \citet{Wakeford2013, Nikolov:2014aa}\\
		WASP-31b & 1575 & 1575 & 0.005 & 0.35 & 1 & 0.06 & 82.48 & 60 & 1.37 & \citet{Sing2015, Sing2016}\\
		WASP-12b & 2580 & 2880 & 0.1 & 0.56 & 150 & 1 & 21.53 & 23 & 0.936 & \citet{Sing2013, Sing2016}\\
		HAT-P-12b & 960 & 1110 & 10 & 0.56 & 1100 & 0 & 27.72 & 30 & 0.924 & \citet{Sing2016}\\
		HD-189733b & 1191 & 1491 & 1 & 0.56 & 150 & 0 & 90.69 & 52 & 1.744 & \citet{Pont2013, McCullough2014, Sing2016}\\
		WASP-6b & 1184 & 1184 & 0.005 & 0.15 & 1100 & 0 & 29.55 & 18 & 1.641 &  \citet{Nikolov2015, Sing2016}\\
		\hline
	\end{tabular}%}
	\caption{Table showing best fit planetary characteristics for all the observed exoplanets from \citet{Sing2016}. The C/O ratio of 0.56 is solar value. The haze enhancement factor is with respect to gaseous Rayleigh scattering. The grey cloudiness factor is with respect to H$_2$ scattering cross-section at 350 nm. DOF refers to degrees of freedom applied to best fit.} 
	\label{tab:grid best fit}
\end{table*}

\begin{figure}
\includegraphics[width=\columnwidth]{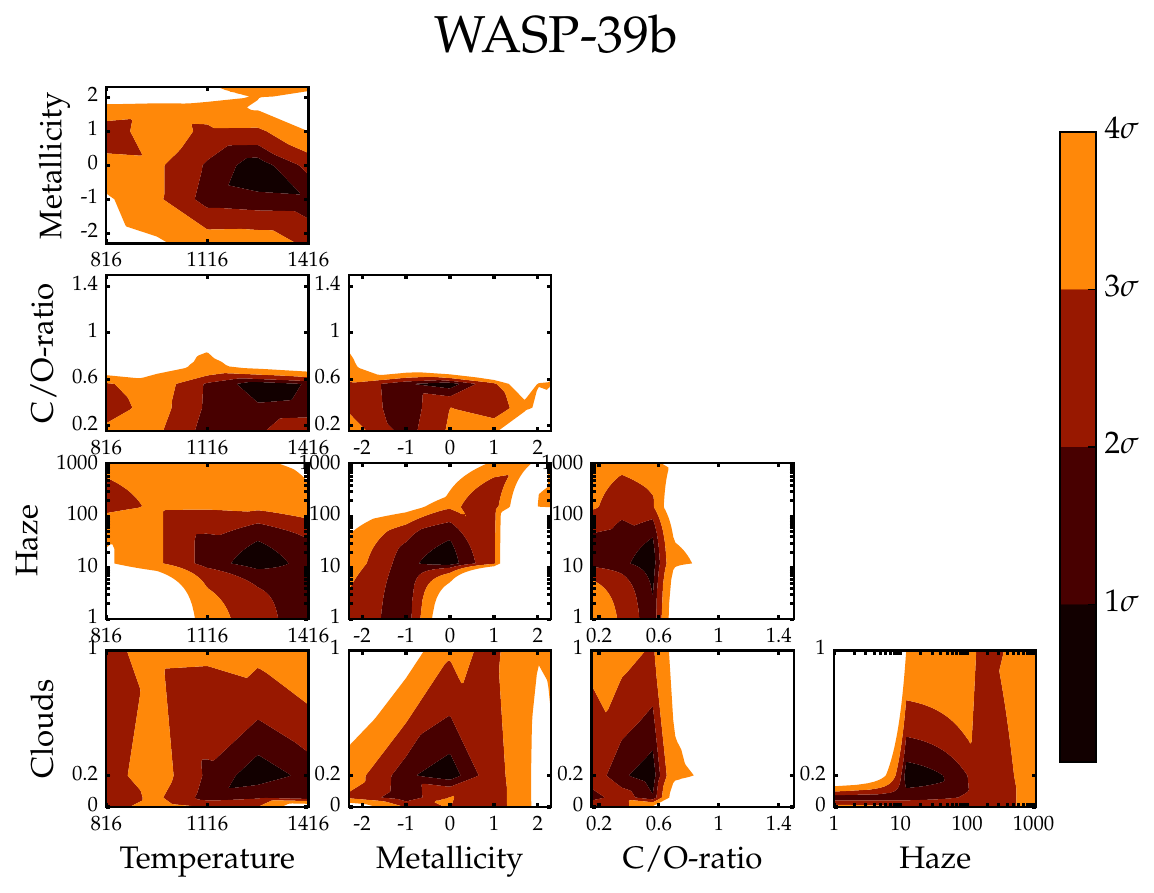}
 \caption{Figure showing $\chi^{2}$ map for WASP-39b. Contours of $\chi^{2}$ are shown for all the combinations of grid parameters. Axis for cloud and haze factors are log-scaled. Metallicity is also log-scaled, 0 being solar metallicity and 2 being 100 times solar metallicity. Colours indicate confidence intervals as shown in colormap to the right (see Section \ref{section:Interpretation of Observations} for details).}
    \label{fig:chimap_wasp39b}
\end{figure}

\section{Interpretation of Observations}
\label{section:Interpretation of Observations}

In this section we interpret the observations of ten hot Jupiter exoplanets from \citet{Sing2016}. The best fitting planetary characteristics are determined for each planet, using transmission spectra in chemical equilibrium and the standard technique of $\chi^{2}$ minimisation, where the only free parameter is a vertical offset between the data and model. We have also compared our physical interpretations with previous studies.

Figure \ref{fig:allplanets}  shows the best fitting model spectra with observations for all the planets. This figure can be directly compared to Figure 1 of  \citet{Sing2016}. However, in this paper the best fit transmission spectra come from a homogenous set of forward models from our grid,  compared to a combination of different models shown in \citet{Sing2016}. The best fitting planetary characteristics along with their $\chi^{2}$ values are shown in Table \ref{tab:grid best fit}. We also present the $\chi^{2}$ maps \citep{Madhusudhan2009}, to demonstrate how the physical parameters are constrained for each planet, in Figure \ref{fig:chimap_wasp39b} for WASP-39b and in Figures \ref{fig:chimap_wasp17b} to \ref{fig:chimap_wasp6b} in Appendix \ref{app:chi-maps of all planets} for all the other planets. For this,  $\chi^{2}$ is computed for each model simulation. Then we fix values of a pair of parameters whilst allowing all others to be free. This is repeated for all pair combinations and all the possible combinations of grid parameters. The resulting $\chi^{2}$  space is mapped along with confidence intervals, which are obtained under the assumption of a $\chi^{2}$  distribution with two degrees of freedom, since there are two unconstrained variables for each plot on the map.

\subsection {WASP-17b}
\label {subsection:WASP-17b}
The best fit WASP-17b forward model (topmost) in Figure \ref{fig:allplanets} shows that the data are consistent with sub-solar metallicities and sub-solar C/O ratios as shown in  Table \ref{tab:grid best fit}. The best fitting model gives a reduced chi square value of 0.82 which can be considered to be an excellent fit for a purely forward model. It also shows signature of haze due to Rayleigh scattering of the order of  $\alpha_\textup{cloud}=10$, in agreement with the retrieval analysis of \citet{Barstow2016}. The data are also consistent with cloudiness factor of $\alpha_\textup{cloud}=0.2$. However, the Na feature in our best fit model is not as strong as in the observations. The reason for this is unclear and retrieval models are also not able to explain this strong feature as shown in  \citet{Barstow2016}. The  $\chi^{2}$ map for WASP-17b is shown in  Figure \ref{fig:chimap_wasp17b} in Appendix \ref{app:chi-maps of all planets}. It shows that the data are consistent with the lowest possible C/O ratio in our parameter space. Therefore, current observations do not show clear features indicative of carbon bearing species. This finding was also one of the initial motivations to expand our parameter space to C/O ratio as low as 0.15. The best fit model shows that H$_2$O features dominate the infrared spectra.

\subsection {WASP-39b}
\label {subsection:WASP-39b}
For WASP-39b shown in Figure \ref{fig:allplanets}, the data are consistent with $\alpha_\textup{haze}=10$ and $\alpha_\textup{cloud}=0.2$ with solar metallicity and solar C/O ratio. They show one of the clearest atmosphere of the set as concluded by \citet{Fischer2016}, but our results also show weak haziness and cloudiness.  Figure \ref{fig:chimap_wasp39b} shows the $\chi^{2}$ map for WASP-39b, revealing that except the temperature all other values are very well constrained for this planet. The $1\sigma$ credible range for the temperature of the planet is higher than the upper limit of our parameter space. The metallicity of the planet is well constrained between solar and slightly sub-solar values, while the C/O ratio is well constrained near a solar C/O ratio. The data are also consistent with the presence of Na and K, albeit at a low significance.

\subsection {HD~209458b}
\label {subsection:HD209458b}
HD~209458b is the best observationally constrained planet in our sample. Our four cloudiness factors were insufficient to constrain the HD~209458b observed spectra, therefore we added two more cloudiness factors ($\alpha_\textup{cloud}= 0.5$ and $10$) specifically for HD~209458b. We find that the data are consistent with ten times solar metallicity and a combination of Rayleigh haze and grey clouds, with muted H$_2$O, Na and K features. However, a rise in the $\RpRs$ values between 0.3 and 0.4 \textmu m in the Rayleigh slope part of the spectrum is still not explained by the model. This may be due to thermospheric effects or missing opacity. The $\chi^{2}$ map of HD~209458b is shown in Figure \ref{fig:chimap_hd209458b} in  Appendix \ref{app:chi-maps of all planets}. It demonstrates that the best fit values of HD~209458b are very well constrained with 4$\sigma$ confidence. This is mainly due to the high wavelength resolution in observations compared to any other planet. However, it is interesting to see that some of the parameters like metallicity have a bi-modal structure in the maps. Therefore, although observations are consistent with a 10 times solar metallicity model, a 0.1 times solar metallicity model can also explain the observations, with the corresponding decrease in haziness, cloudiness and temperature.

\subsection {WASP-19b}
\label {subsection:WASP-19b}
WASP-19b is the planet with least observational data points, posing a challenge to accurately constrain its characteristics. This can be seen in the $\chi^{2}$ map plot for WASP-19b in Figure \ref{fig:chimap_wasp19b}  in Appendix \ref{app:chi-maps of all planets}. Temperature, clouds and haze are the least constrained, while the C/O ratio is constrained to values less than solar, and metallicity between 0.1 to 1 times solar, considering 1$\sigma$ confidence intervals. However, data are consistent with sub-solar metallicity and sub-solar C/O ratio with haze, but no grey clouds. The H$_2$O features are clearly visible in the model and observations, muted by haze and consistent with a low C/O ratio atmosphere, all in agreement with \citet{Huitson2013}. The best fit forward model also suggests a weak narrow Na feature for WASP-19b, which has not been detected in the observations due to lack of sufficient data points. Importantly, the lack of TiO/VO features also suggest a lower than equilibrium temperature for this planet. 

\subsection {HAT-P-1b}
\label {subsection:HAT-P-1b}
The HAT-P-1b data are consistent with 0.1 times solar metallicity, sub-solar C/O ratio of 0.15, $\alpha_\textup{haze}=10$ and substantial grey cloudiness factor of  $\alpha_\textup{cloud}=1$. They are consistent with H$_2$O features similar to \citet{Wakeford2013} but are strongly muted, which can be attributed to the extreme cloudiness. One of the most important discrepancies between the data and the model, is that the best fit forward model in chemical equilibrium predicts a very weak, narrow Na feature, compared to a larger feature in the observations implying that it might have enhanced (non-solar or disequilibrium) Na concentration in agreement with \citet{Nikolov:2014aa}. However the strength of the Na feature in the best-fit no-rainout chemistry  scenario is similar to observations (not shown here), potentially implying some missing physical process is preventing it from raining out. The $\chi^{2}$ map for HAT-P-1b is shown in Figure \ref{fig:chimap_hatp01} in Appendix \ref{app:chi-maps of all planets}. It demonstrates that similar to WASP-17b, the HAT-P-1b data are consistent with the lowest considered C/O ratio in our parameter space, i.e. C/O$=$0.15. Interpreted characteristics of this planet are very similar to that of WASP-17b, but with more cloudiness. 

\subsection {WASP-31b}
\label {subsection:WASP-31b}
The WASP-31b data are consistent with a 0.005 times solar metallicity and the C/O ratio of 0.35 with no enhanced Rayleigh scattering, but a grey cloud enhancement factor $\alpha_\textup{cloud}=0.06$. One of the important discrepancies between the data and the model,  is that the observations suggest a possible K feature without any Na feature, which none of the forward models in our parameter space for this planet are able to reproduce. Na and K have very similar condensation curves so they are both expected in the spectrum in chemical equilibrium conditions (see \ref{subsection:HAT-P-12b} for the exception). \citet{Sing2015} interpreted a strong haze and cloud deck with K feature, but our best fit forward model in chemical equilibrium suggests a more clear atmosphere, with very weak K feature and extremely sub-solar metallicity. This discrepancy points towards a sub-solar Na/K abundance in agreement with \citet{Sing2015}. It also highlights the degeneracy existing between the effect of metallicity and clouds/haze on spectral features and can be seen in $\chi^{2}$ map for WASP-31b in Figure \ref{fig:chimap_wasp31b}. 

\subsection {WASP-12b}
\label {subsection:WASP-12b}
WASP-12b has the highest equilibrium temperature among our observed targets. \citet{Madhusudhan2011a} concluded a high C/O ratio and weak thermal inversion for this planet based on Spitzer infrared measurements. However, HST WFC3 optical observations from \citet{Sing2016} show a completely flat spectra with just a Rayleigh scattering slope. The data from \citet{Sing2016} are consistent with an extremely hazy and cloudy atmosphere. They show evidence for aerosols and absence of TiO/VO. The best fit values reach the upper edge of our parameter space for clouds ($\alpha_{cloud=1}$) along with haziness factor of  $\alpha_\textup{haze}=150$, sub-solar metallicity and solar C/O ratios. The $\chi^{2}$ map of WASP-12b is shown in Figure \ref{fig:chimap_wasp12b} in Appendix \ref{app:chi-maps of all planets}.  It shows that many of the parameters are unconstrained, which is mainly due to the feature-less spectrum and also since the slope of scattering due to haze deviates from $\lambda^{-4}$, the standard Rayleigh scattering slope. However, \citet{Kreidberg2015} obtained more precise data between 0.8 and 1.6 \textmu m with 6 HST transits along with detection of a H$_2$O feature. When data from \citet{Kreidberg2015} are used along with the data from \citet{Sing2016}, they are consistent with an equilibrium temperature of $2280$\,K, solar metallicity, C/O ratio of 0.7, haze factor of  $\alpha_\textup{haze}=150$ and cloud factor of  $\alpha_\textup{cloud}=1$. It is also consistent with the 1.4 \textmu m  H$_2$O feature. Since our best fit model suggests a C/O ratio of 0.7, ruling out carbon-rich spectra, it is in agreement with retrieval results of \citet{Kreidberg2015}, within the 1$\sigma$ uncertainties. 

\subsection {HAT-P-12b}
\label {subsection:HAT-P-12b}
The data for HAT-P-12b are consistent with a strong enhanced Rayleigh scattering, $\alpha_\textup{haze}=1100$, reaching the upper limit of parameter space, but without any grey clouds. The best-fit model shows evidence for K, but not Na. Interestingly, this particular scenario, where a K feature is present but Na is not, is produced in our model simulations for HAT-P-12b shown in Figure \ref{fig:temp_hatp12b}. However, this scenario is not in agreement with other spectral features, therefore, is not selected as best fit model. However, the temperature required to obtain K features without any Na features lie within 2 $\sigma$ uncertainties of best fit temperature values. Figure \ref{fig:chimap_hatp12b} in Appendix \ref{app:chi-maps of all planets} shows $\chi^{2}$ map for HAT-P-12b, which also suggests extremely high haziness for this planet is well constrained. 

\subsection {HD~189733b}
\label {subsection:HD189733b}
HD~189733b is the planet with the second highest number of observations of our targets and has one of the strongest enhanced Rayleigh scattering signatures in agreement with \citet{Pont2013}. The data shows H$_2$O and Na feature as found in \citet{McCullough2014, Sing2016}. They are consistent with $\alpha_\textup{haze}=150$, solar metallicity and solar C/O ratio, shown in Figure \ref{fig:allplanets} and Table \ref{tab:grid best fit}. However, the forward model also predicts a Na feature which is not seen in the observations. The $\chi^{2}$ maps show that most of the model parameters are well constrained as seen in Figure \ref{fig:chimap_hd189733b} in  Appendix \ref{app:chi-maps of all planets}. Only the temperature of the planet  tends to hit the upper edge of our parameter space. 

\subsection {WASP-6b}
\label {subsection:WASP-6b}
WASP-6b has very few observations similar to WASP-19b making it very difficult to constrain its physical parameters. There is a strong signature of haze with $\alpha_\textup{haze}=1100$ also in agreement with \citet{Nikolov2015}, but no Na or K signature, tentatively seen in observations. 

\bigskip

With the best fit estimates of all the planets using forward models, we see a continuum from clear to cloudy/hazy atmospheres as found by \citet{Sing2016}. The data for all the 10 planets are consistent with sub-solar to solar C/O ratio. This is also in agreement with the retrieval analysis of \citet{Benneke2015}, where they concluded C/O ratios of HD~209458b, WASP-19b, WASP-12b, HAT-P-1b, HD~189733b and WASP-17b to be less than 0.9. Therefore, current observations do not show clear features, indicative of carbon bearing species. The metallicity for these planetary atmospheres are also consistent with extremely sub-solar value of 0.005 to that of 10 times solar metallicity and tend to favour a H$_2$O dominated, rather than a CH$_4$ (carbon species) dominated atmosphere. As seen in Table \ref{tab:grid best fit}, good fits are obtained in 6 out of 10 planets ($\chi^{2}_{r}$ near 1) with a wide parameter space exploration using a forward model.  

\section{Transmission Spectra : variation with parameters}
\label{section:Transmission Spectra : variation with parameters}

The major spectral features of various species in the transmission spectrum of exoplanets are described in Appendix \ref{subsection:Molecular Features}. In this section, we explore the parameter space for a subset of planets, to demonstrate their effect on the transmission spectra. For brevity, we select three planets across different equilibrium temperature regimes. The effect on the transmission spectra of these planets over the entire parameter space is investigated, along with their physical interpretation. The three planets for which we present the analysis are, HAT-P-12b, WASP-17b and WASP-12b,  with equilibrium temperatures of $960$\,K, $1755$\,K  and  $2580$\,K, respectively. These three different planets cover the full range of currently observed hot jupiter planets. 

\begin{figure}
\includegraphics[width=\columnwidth]{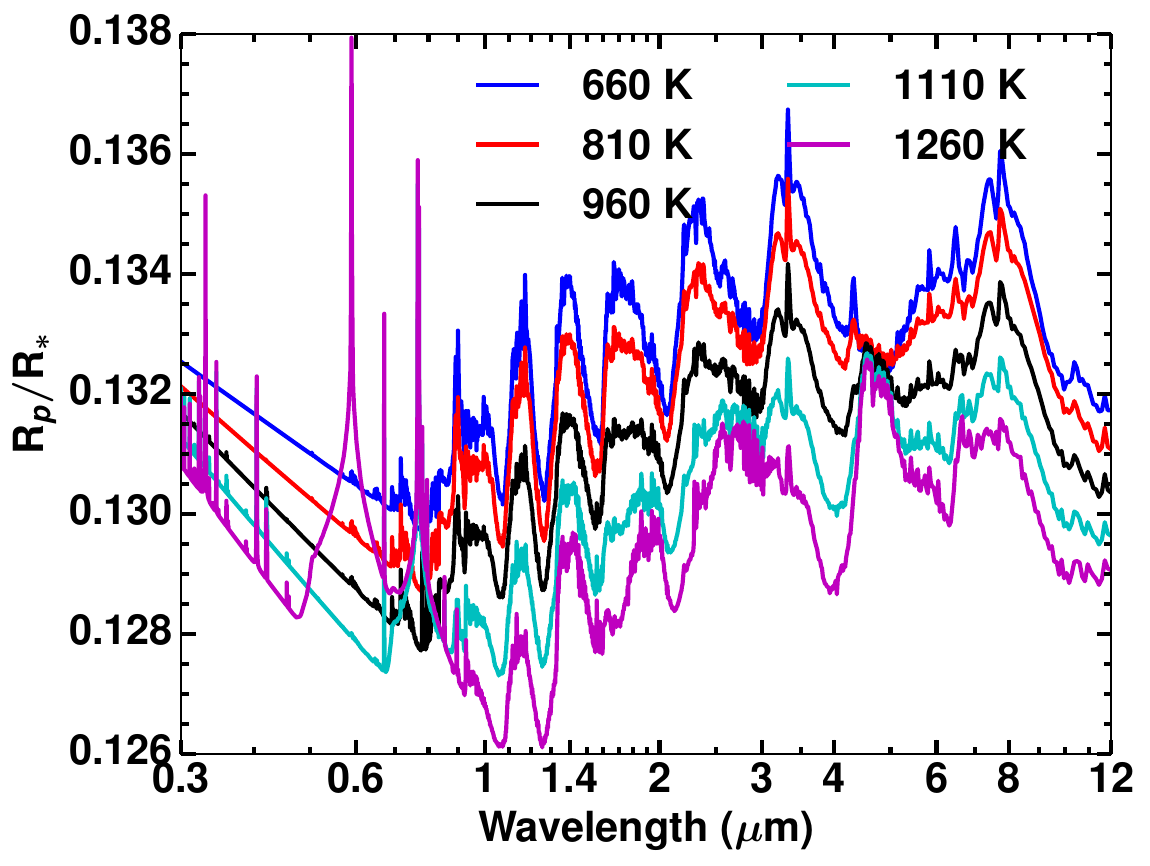}
 \caption{Figure showing HAT-P-12b transmission spectra for a range of temperatures (in Kelvin) at solar metallicity, solar C/O ratio and clear atmosphere. X-axis is wavelength in \textmu m and Y-axis transit radius ratio ($\RpRs$).}
    \label{fig:temp_hatp12b}
\end{figure}

\begin{figure*}
\begin{center}
 \subfloat[]{\includegraphics[width=\columnwidth]{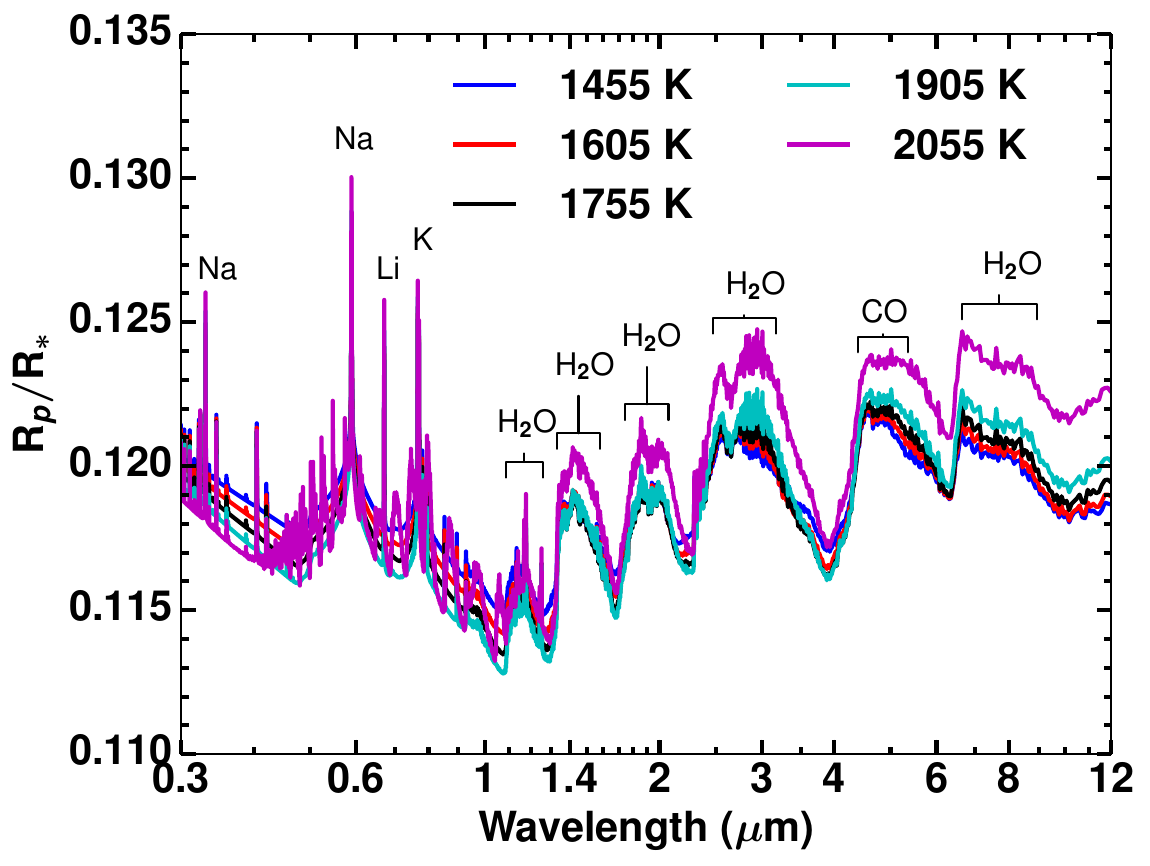}\label{fig:temp_wasp17b}}
 \subfloat[]{\includegraphics[width=\columnwidth]{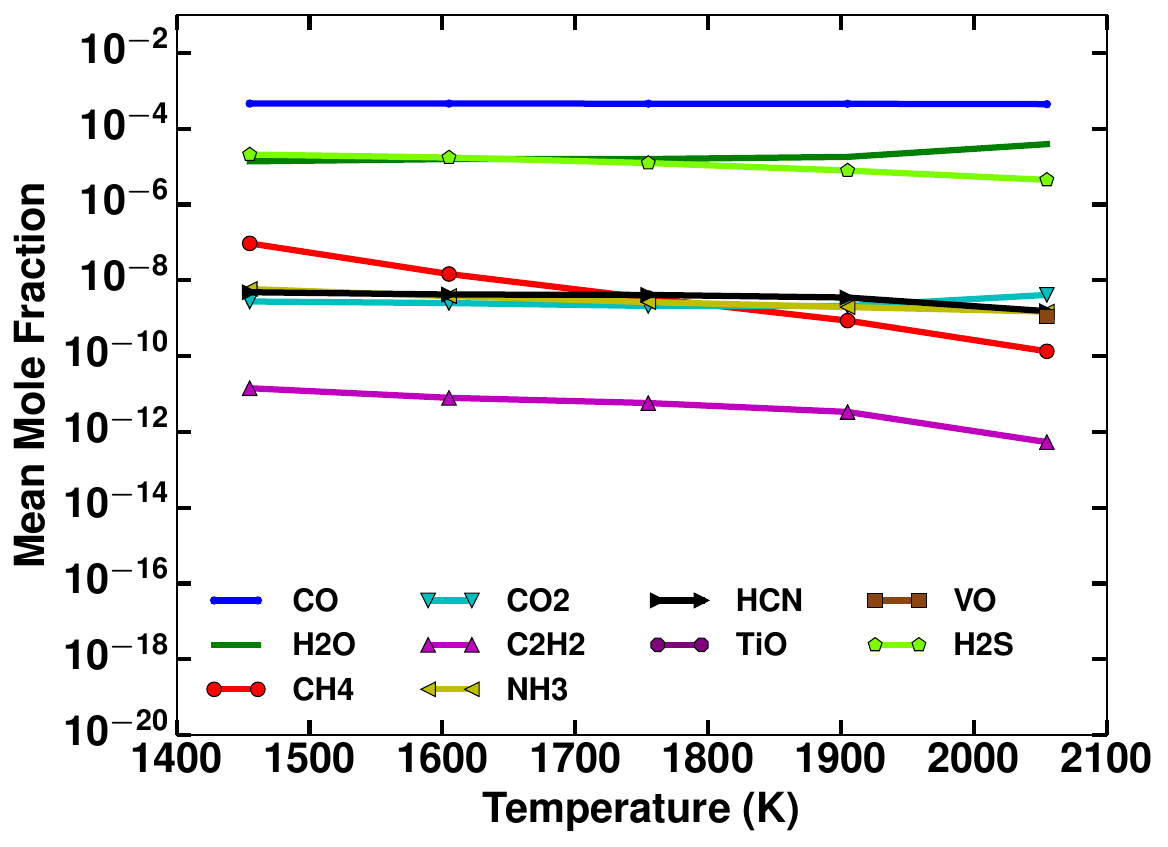}\label{fig:chem_temp_wasp17b}}
\end{center}
 \caption{\textbf{(a)} Figure showing WASP-17b transmission spectra for a range of temperatures, similar to Figure \ref{fig:temp_hatp12b}, with major molecular features shown at equilibrium temperature ($1755$\,K). \textbf{(b)} Figure showing change in mean chemical abundances between 0.1 and 100 millibar for various molecules, with change in temperature for WASP-17b at solar metallicity, solar C/O ratio and clear atmosphere. X-axis is temperature in Kelvin while Y-axis shows mean abundances in units of mole fraction. TiO/VO shown in legends have been rained out for this planet.}
\end{figure*}

\begin{figure}
\includegraphics[width=\columnwidth]{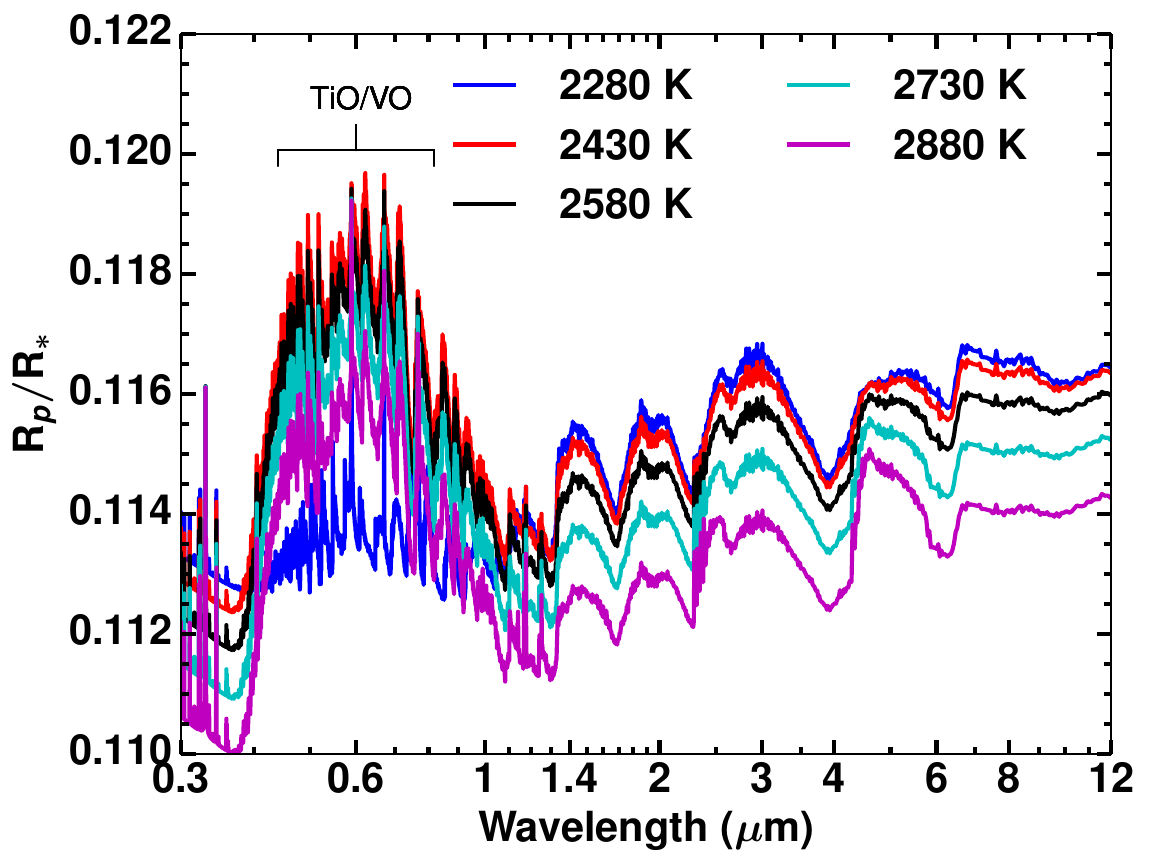}
 \caption{Figure showing WASP-12b transmission spectra for a range of temperatures, similar to Figure \ref{fig:temp_hatp12b}}
    \label{fig:temp_wasp12b}
\end{figure}

\subsection {Effect of temperature}
\label {subsection:Effect of temperature}
The temperature structure of the planet governs the most important physical and chemical mechanisms in a planetary atmosphere. The isothermal temperature we adopt, is indicative of the temperature at the $\sim$1 millibar pressure level. The metallicity and C/O ratio were fixed to solar values to explore the effect of temperature, without any changes due to other parameters.
 
Changes in the transmission spectral features for a range of temperatures for HAT-P-12b with an equilibrium temperature of $960$\,K are shown in Figure \ref{fig:temp_hatp12b}. As the temperature increases from $660$\,K to $1260$\,K, the dominant spectral features vary substantially. For temperatures from $660$\,K to $960$\,K, the spectra are dominated by CH$_4$ without any Na, K or other alkali metal features. However, at a temperature of $1110$\,K we observe a K feature at 0.74 \textmu m and most importantly H$_2$O features start to dominate over those of CH$_4$. The alkali metal features of Li, Rb and Cs also start showing their signatures at this temperature. When the temperature is increased to $1260$\,K, Na features become visible which were absent at $1110$\,K, implying a threshold value of temperature, below which it rains-out Na and above which it is sustained in the gas phase. Moreover, spectral features of CO also appear at T>960 K, at 4.5 \textmu m

The change in transmission spectral features for a range of temperatures for WASP-17b with an equilibrium temperature of $1755$\,K are shown in Figure \ref{fig:temp_wasp17b}. The features become stronger with the increase in temperature, because the scale height increases,  implying hot planets are the best targets for transmission spectroscopy as found in previous studies \citep{Fortney2010}. Additionally, at temperatures greater than $\sim$2000\,K, features of VO can be seen near the K absorption band. The temperature of $\sim$2055\,K (based on grid resolution) is where it becomes possible for VO to be in the gas phase, thus we see its very weak features. However, with the increase in temperature there is a gradual increase in the concentration of both TiO and VO and their features become increasingly significant which can be seen in higher equilibrium temperature planets. 

Figure \ref{fig:temp_wasp12b} shows spectra for WASP-12b with an equilibrium temperature of $2580$\,K. Here we clearly see the gradual increase in the TiO/VO features as we increase the temperature from $2280$ to $2880$\,K, with the spectra substantially dominated by TiO/VO with extremely large effective $\RpRs$ in the optical. This shows that the presence of TiO/VO in these high temperature planetary atmospheres leads to a substantial signature in transmission spectra in agreement with previous studies \citep{Burrows1999,Fortney2010}. Importantly, it shows that VO is sustained in the atmosphere for temperatures greater than $\sim$2050\,K, but TiO is sustained only after $\sim$2350\,K with abundances greater than VO. Therefore, it is TiO, not VO that leads to these huge spectral features at optical wavelengths for high temperature planets in chemical equilibrium, which can be seen in Figure \ref{fig:temp_wasp12b}. It is also important to note that TiO/VO features develop around Na and K features, thereby masking them. We would expect Na and K to be ionised as such high temperatures, but we currently don't include ionisation in our model. We also see that the ratio of optical and near-infrared bands $\RpRs$ change with changing temperature. We note that TiO/VO features are seen above $1500$\,K if we simulate assuming gas phase chemistry as concluded by \citep{Molliere2016}. However, if we consider condensation with rainout which is a more physical representation of a planetary atmosphere, we see TiO/VO features only after $\sim$ $2000$\,K. We note that using consistent $P$-$T$ profiles might change this result. 

To understand the change in equilibrium chemical abundances with temperature, we calculate the simple linear mean abundances for some spectrally important species in the transmission spectra probed region (0.1-100 millibar).  Figure \ref{fig:chem_temp_wasp17b} shows these mean abundances for WASP-17b. CO is the most abundant chemical species after H, H$_2$ and He (not shown here). Surprisingly, H$_2$S is also as abundant as H$_2$O, but with a weak spectral signature, therefore it has not yet been detected in any exoplanet atmosphere. However, H$_2$S abundances decrease with increasing temperature while that of H$_2$O increases, especially after $1900$\,K. The drop in CH$_4$ abundances with increase in temperature is substantial, going from $10^{-7}$ to $10^{-10}$ mole fraction as temperature goes from $1455$\,K to $2055$\,K. HCN and C$_2$H$_2$ abundances are almost constant with increase in temperature but decreases after $1900$\,K. We note that the H$_2$O mole fraction is $\sim$ $4\times10^{-4}$ at solar metallicity and solar C/O ratio, when only gas phase chemistry is considered, as adopted by \citet{Barstow2016}. However, if we include condensation with rainout, which is the case for this entire grid, some of the oxygen is taken up by condensate species reducing elemental oxygen available to form H$_2$O \citep{Moses2011}. This leads to H$_2$O abundances at solar metallicity being lower at a value of $\sim$ $1.4\times10^{-5}$.
  
\begin{figure*}
\begin{center}
 \subfloat[]{\includegraphics[width=\columnwidth]{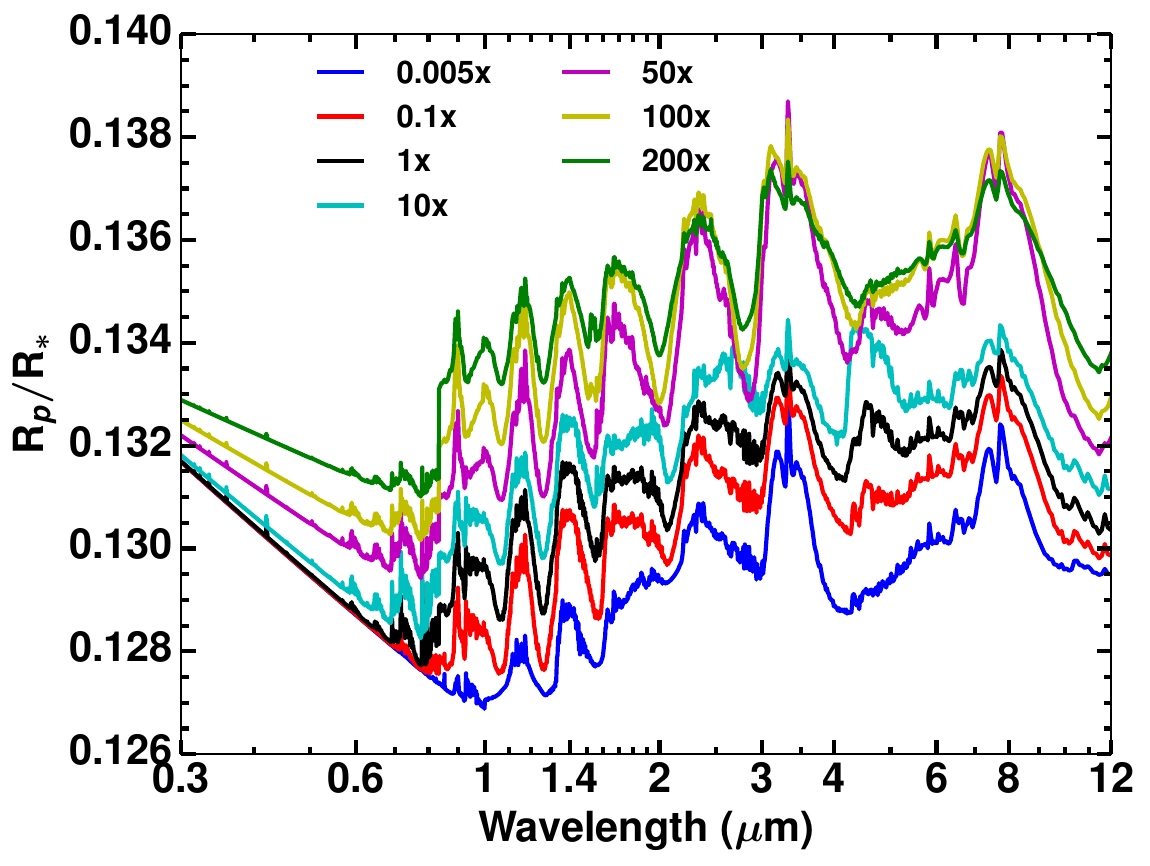}\label{fig:metal_hatp12b}}
 \subfloat[]{\includegraphics[width=\columnwidth]{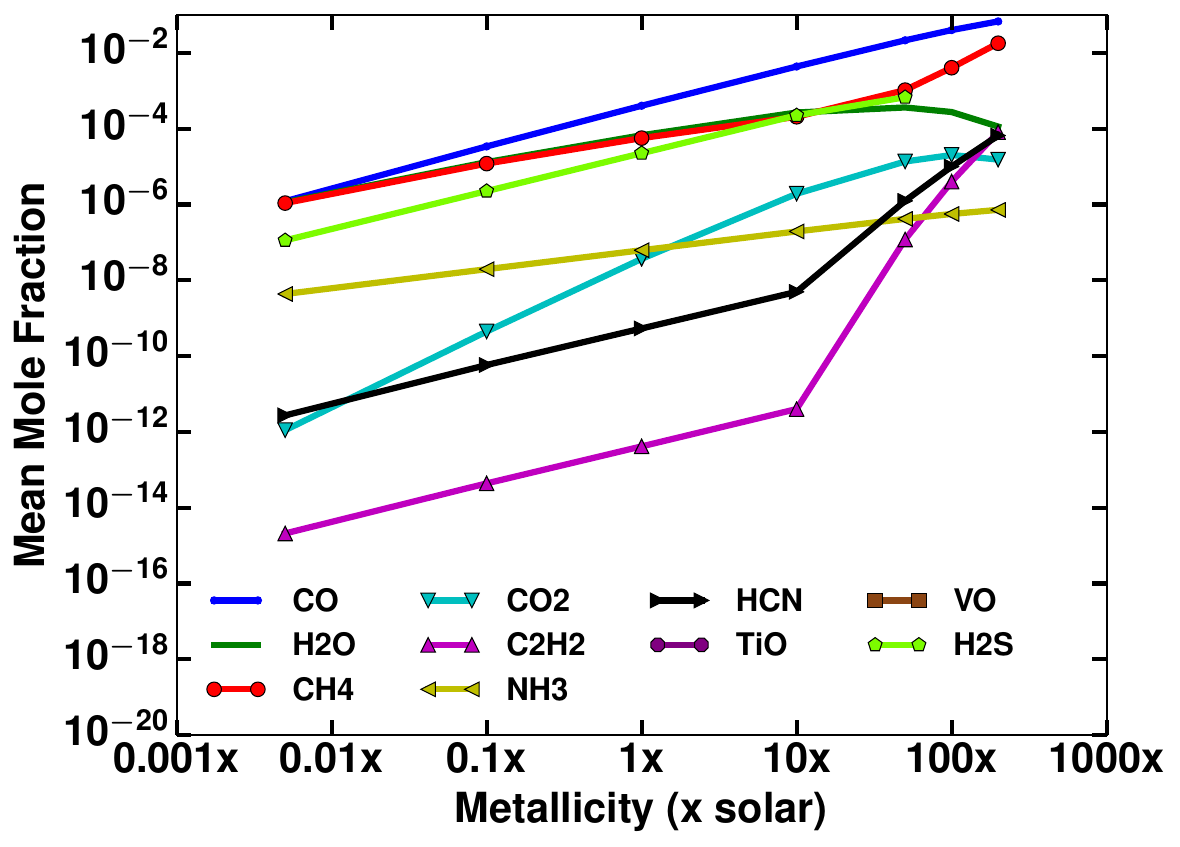}\label{fig:chem_metal_hatp12b}}
\end{center}
 \caption{\textbf{(a)} Figure showing HAT-P-12b transmission spectra for a range of metallicity (times solar) at its equilibrium temperature, solar C/O ratio and clear atmosphere. X-axis is wavelength in \textmu m and Y-axis transit radius ratio ($\RpRs$). \textbf{(b)} Figure showing change in mean chemical abundances between 0.1 and 100 millibar for various molecules, with change in metallicity for HAT-P-12b. X-axis is metallicity ($\times$ solar) while Y-axis shows mean mole fraction.}
\end{figure*}

\begin{figure*}
\begin{center}
 \subfloat[]{\includegraphics[width=\columnwidth]{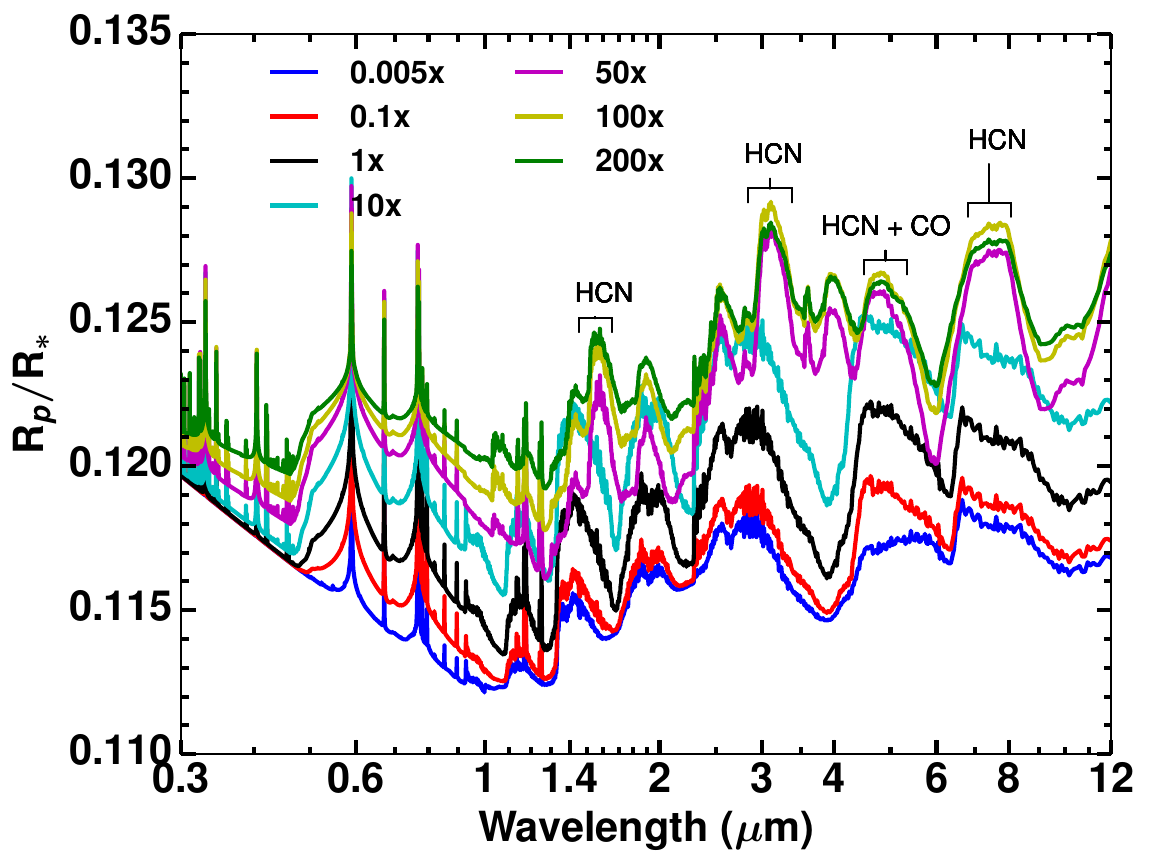}\label{fig:metal_wasp17b}}
 \subfloat[]{\includegraphics[width=\columnwidth]{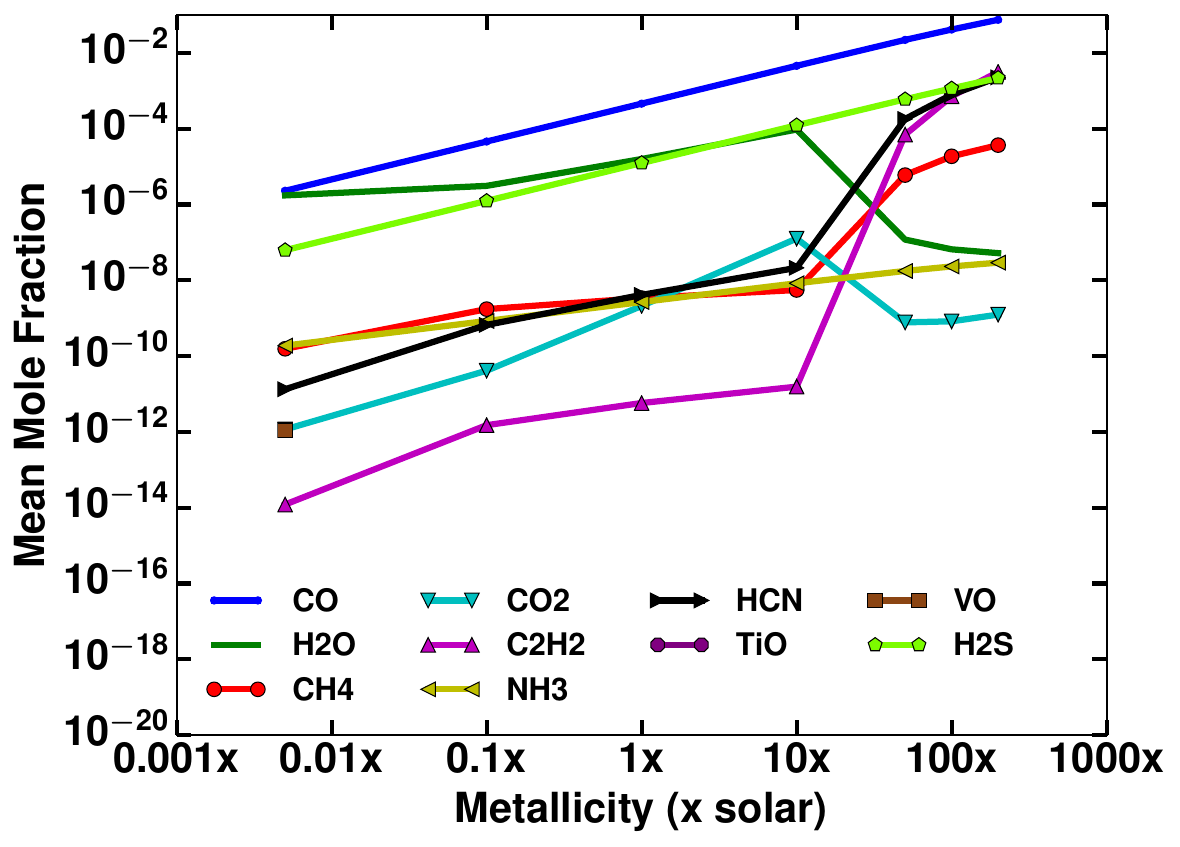}\label{fig:chem_metal_wasp17b}}
\end{center}
 \caption{\textbf{(a)} Figure showing WASP-17b transmission spectra for a range of metallicity (times solar),  similar to Figure \ref{fig:metal_hatp12b}, with major molecular features shown at highest metallicity (200x). \textbf{(b)} Figure showing change in mean chemical abundances between 0.1 and 100 millibar for various molecules, with change in metallicity for WASP-17b, similar to Figure \ref{fig:chem_metal_hatp12b}.}
\end{figure*}

\begin{figure*}
\begin{center}
 \subfloat[]{\includegraphics[width=\columnwidth]{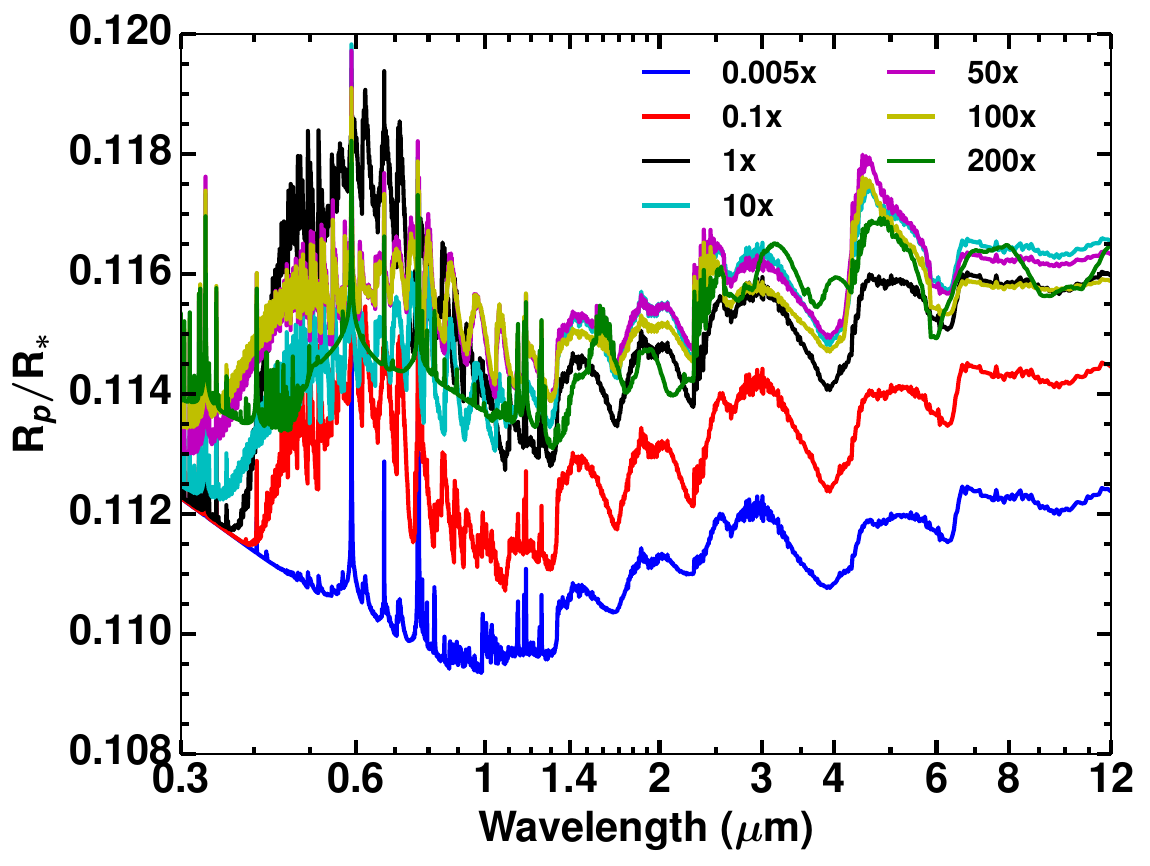}\label{fig:metal_wasp12b}}
 \subfloat[]{\includegraphics[width=\columnwidth]{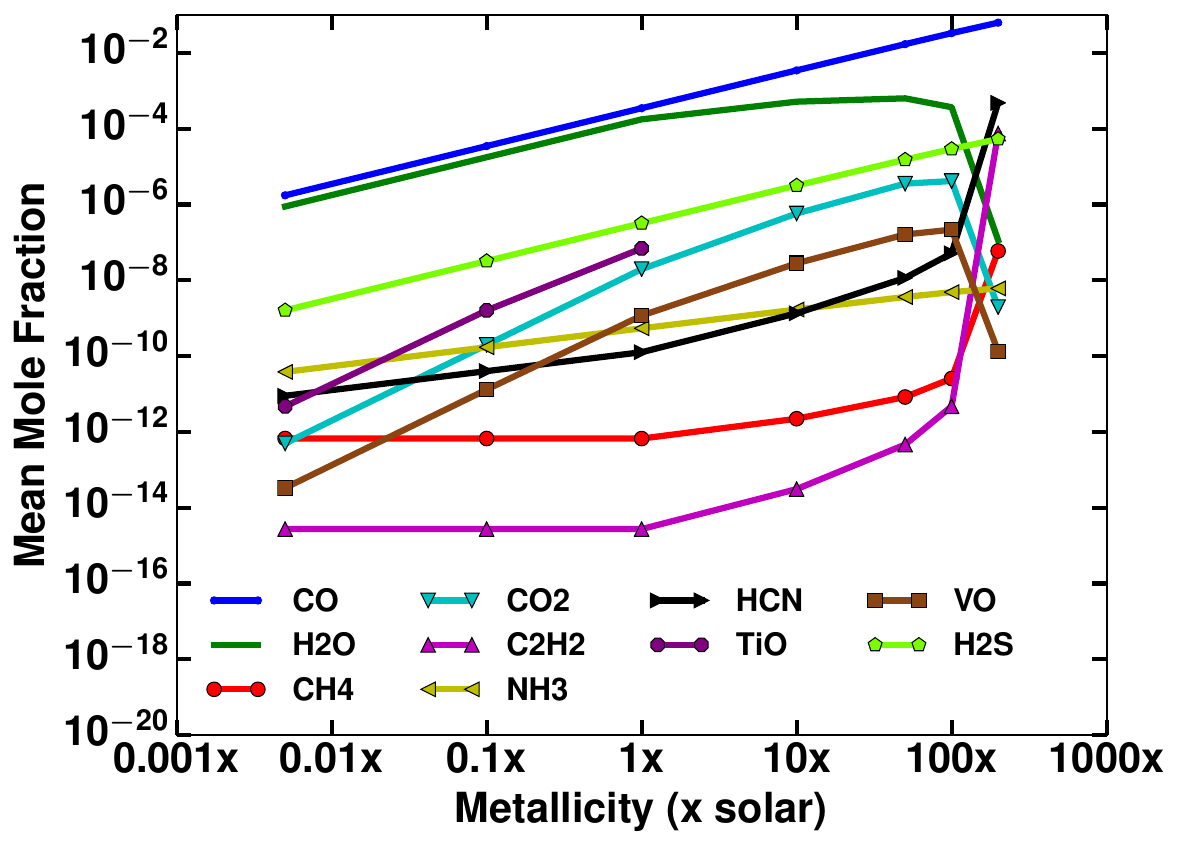}\label{fig:chem_metal_wasp12b}}
\end{center}
 \caption{\textbf{(a)} Figure showing WASP-12b transmission spectra for a range of metallicity (times solar),  similar to Figure \ref{fig:metal_hatp12b}. \textbf{(b)} Figure showing change in mean chemical abundances between 0.1 and 100 millibar for various molecules, with change in metallicity for WASP-12b, similar to Figure \ref{fig:chem_metal_hatp12b}.}
\end{figure*}

\subsection{Effect of Metallicity}
\label {subsection:Effect of Metallicity}
Figure  \ref{fig:metal_hatp12b},  \ref{fig:metal_wasp17b} and  \ref{fig:metal_wasp12b} show changes in the transmission spectra with changes in metallicity for HAT-P-12b, WASP-17b and WASP-12b, respectively. Additionally, their corresponding mean chemical abundances are shown in  \ref{fig:chem_metal_hatp12b}, \ref{fig:chem_metal_wasp17b} and \ref{fig:chem_metal_wasp12b}, respectively. These simulations are at planetary equilibrium temperature, solar C/O ratio and do not include any haze or clouds.  At optical wavelengths, for HAT-P-12b as shown in Figure \ref{fig:metal_hatp12b}, we see the change in the Rayleigh scattering strength as we go from sub-solar metallicities to super-solar metallicities. This is a direct result of inclusion of multi-gas Rayleigh scattering, explained in Section \ref{subsubsection:Multi-gas Rayleigh Scattering}. The larger spectral features (higher $\RpRs$) at higher metallicities for all wavelengths can be attributed to an increase in opacity \citep{Fortney2010}. In the infrared and near infrared we see a trend where increasing metallicity leads to an increase in the strength of spectral features. However, there is a substantial change in the features at 50 times solar metallicity. This can be attributed to a decrease in H$_2$O and CO$_2$ abundances and a corresponding increase in CH$_4$, HCN and C$_2$H$_2$ abundances shown in Figure \ref{fig:chem_metal_hatp12b} for HAT-P-12b. There is a also a dramatic rise in C$_2$H$_2$ abundances. 

For WASP-17b, one of the most important effects is the broadening of Na features with the increase in metallicity as shown in Figure \ref{fig:metal_wasp17b}. We note that this broadening is not due to any broadening mechanism explained in Section \ref{subsection:Line Lists}, but due to an increase in the opacity.  With the increase in metallicity, Na abundance increases while the scale height of the atmosphere decreases. Since the transmission spectra represents the planetary radius at a reference pressure of $\sim$1 millibar, the opacity at this pressure level will therefore increase with increase in metallicity leading to larger features. It enables probing weak absorption wings as seen in Figure \ref{fig:metal_wasp17b} for Na with broadened features.  This could be used as one of the signatures to constrain the metallicities of exoplanet atmospheres. There also appears to be a transition metallicity between 10 and 50 times solar, between which we see a substantial change in the spectral features. To test this, we removed HCN opacities in the model simulation at 50 times solar metallicity, which allowed us to conclude that the major changes in the spectral features were due to HCN, especially between 2 to 4 \textmu m shown in Figure \ref{fig:chem_metal_wasp17b}. Therefore, HCN may well be detectable using the NIRSPEC G395 grism onboard JWST,  which could also aid constraining planetary atmospheric metallicity. Some of the changes in the spectra due to the change in the metallicity are also due to a decrease in H$_2$O and CO$_2$ abundances along with the increase in CH$_4$ abundances as shown in Figure  \ref{fig:chem_metal_wasp17b}.  HCN has more effect in the transmission spectra for hotter planets like WASP-17b as compared to HAT-P-12b which has been tested by switching off HCN opacities in both the cases.

Figure \ref{fig:metal_wasp12b} shows the change in transmission spectra with changes in metallicity for WASP-12b, with equilibrium temperature of  $2580$\,K. At extremely sub-solar metallicity, TiO/VO features are absent, due to their low abundances as seen in Figure \ref{fig:chem_metal_wasp12b}. However, the strength of TiO/VO spectral features increase with increasing metallicity reaching its peak at solar metallicity, before decreasing again. This decrease is due to depleted TiO in the atmosphere, as seen in Figure \ref{fig:chem_metal_wasp12b}.  The drop in TiO mean mole fraction to 0 for metallicities greater than solar can be attributed to formation of Ti$_3$O$_5$ condensate which takes up all the elemental Titanium. However, the presence of VO leads to comparatively weaker features in the optical wavelengths up-to 100 times solar metallicity. These TiO and VO features are completely absent at extremely high metallicity of 200 times solar.

\begin{figure*}
\begin{center}
 \subfloat[]{\includegraphics[width=\columnwidth]{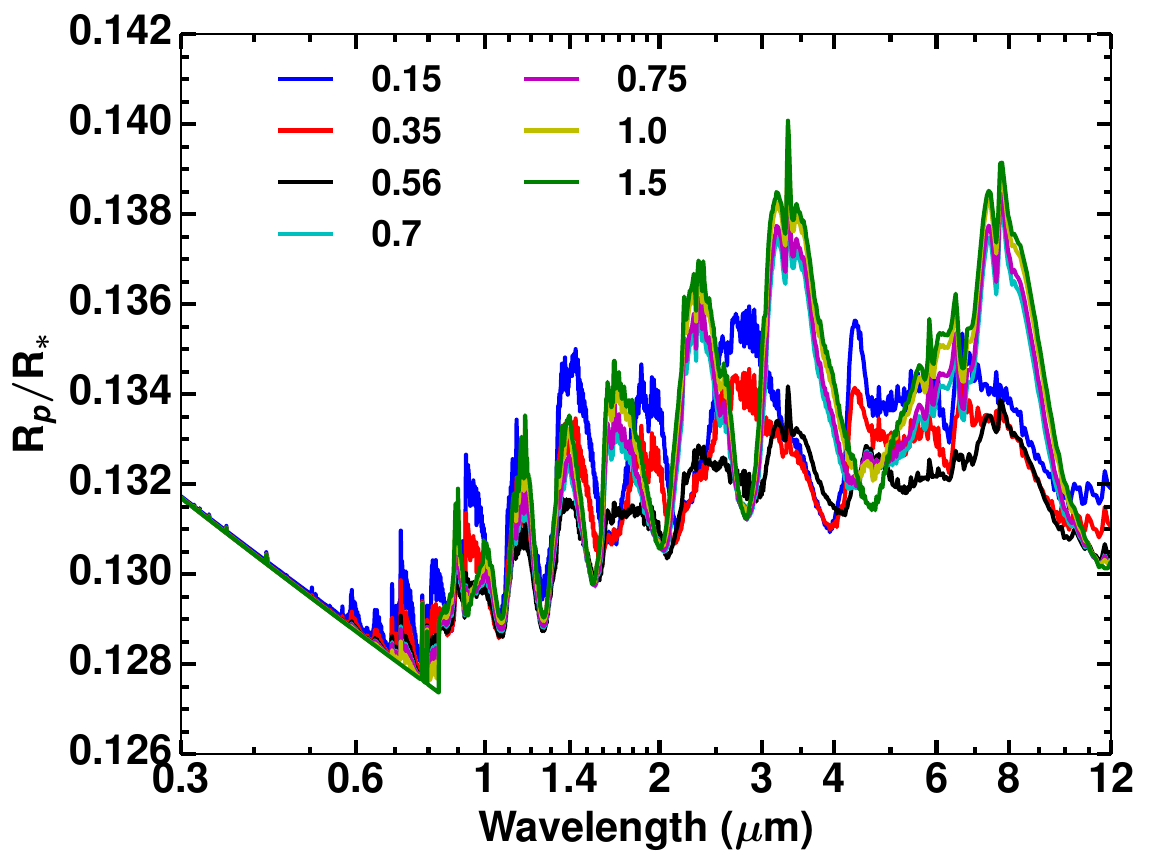}\label{fig:coratio_hatp12b}}
 \subfloat[]{\includegraphics[width=\columnwidth]{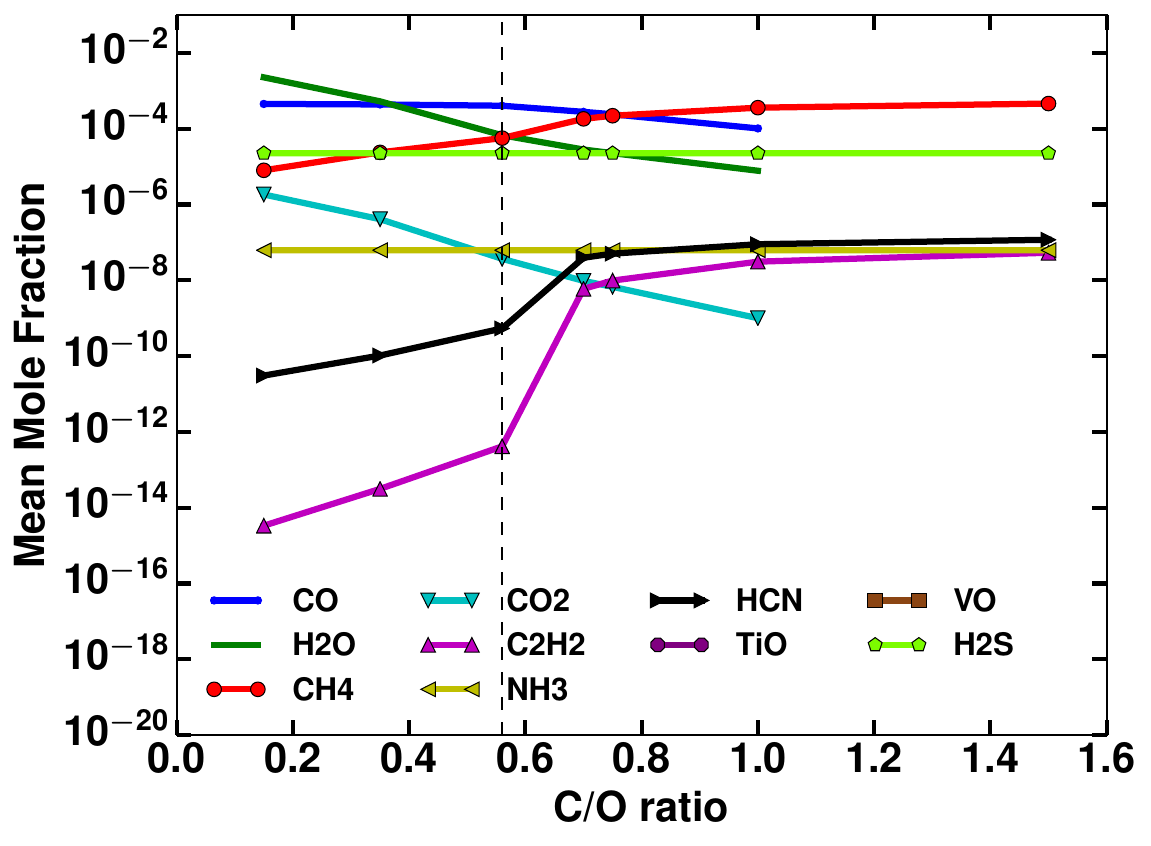}\label{fig:chem_co_hatp12b}}
\end{center}
 \caption{\textbf{(a)} Figure showing HAT-P-12b transmission spectra for a range of C/O ratio at its equilibrium temperature, solar metallicity and clear atmosphere. X-axis is wavelength in \textmu m and Y-axis transit radius ratio ($\RpRs$). \textbf{(b)} Figure showing change in mean chemical abundances between 0.1 and 100 millibar for various molecules, with change in C/O ratio for HAT-P-12b, X-axis is C/O ratio and Y-axis is mean abundances in units of mole fraction. Dashed line indicates solar C/O ratio.}
\end{figure*}

\begin{figure*}
\begin{center}
 \subfloat[]{\includegraphics[width=\columnwidth]{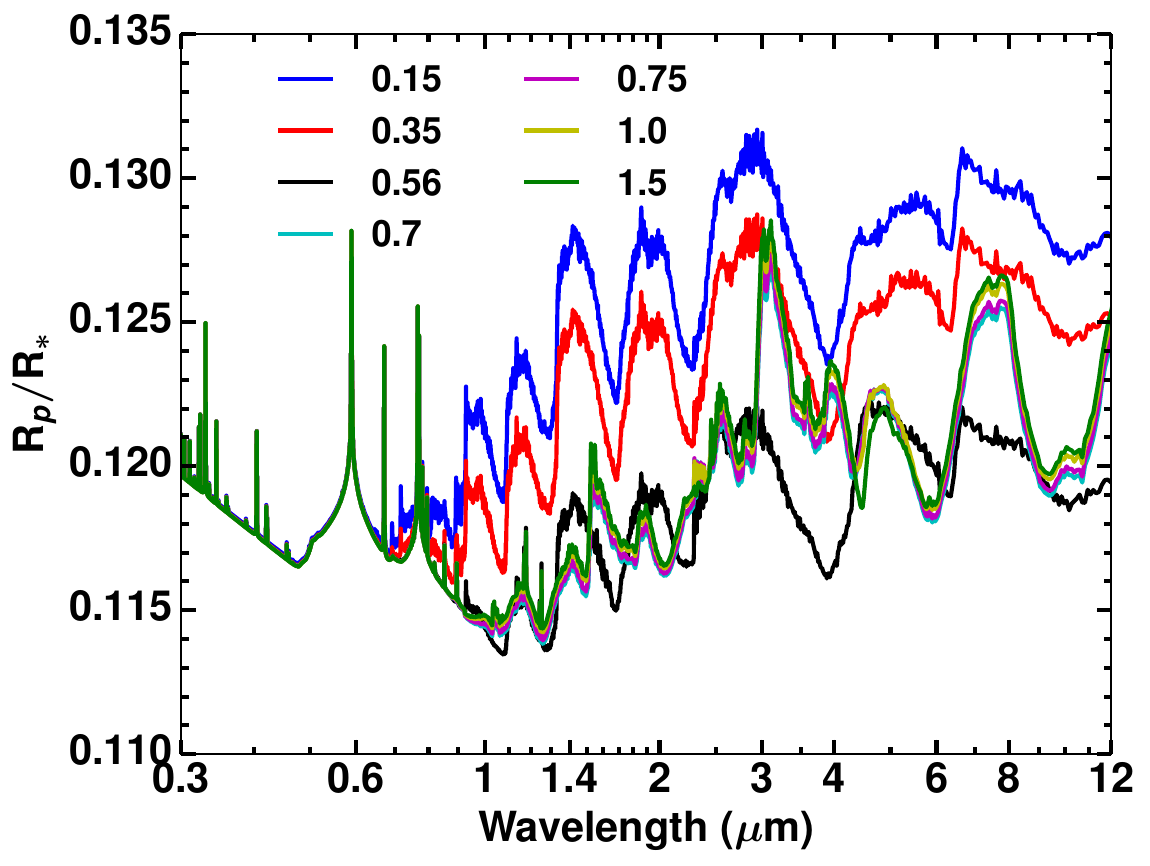}\label{fig:coratio_wasp17b}}
 \subfloat[]{\includegraphics[width=\columnwidth]{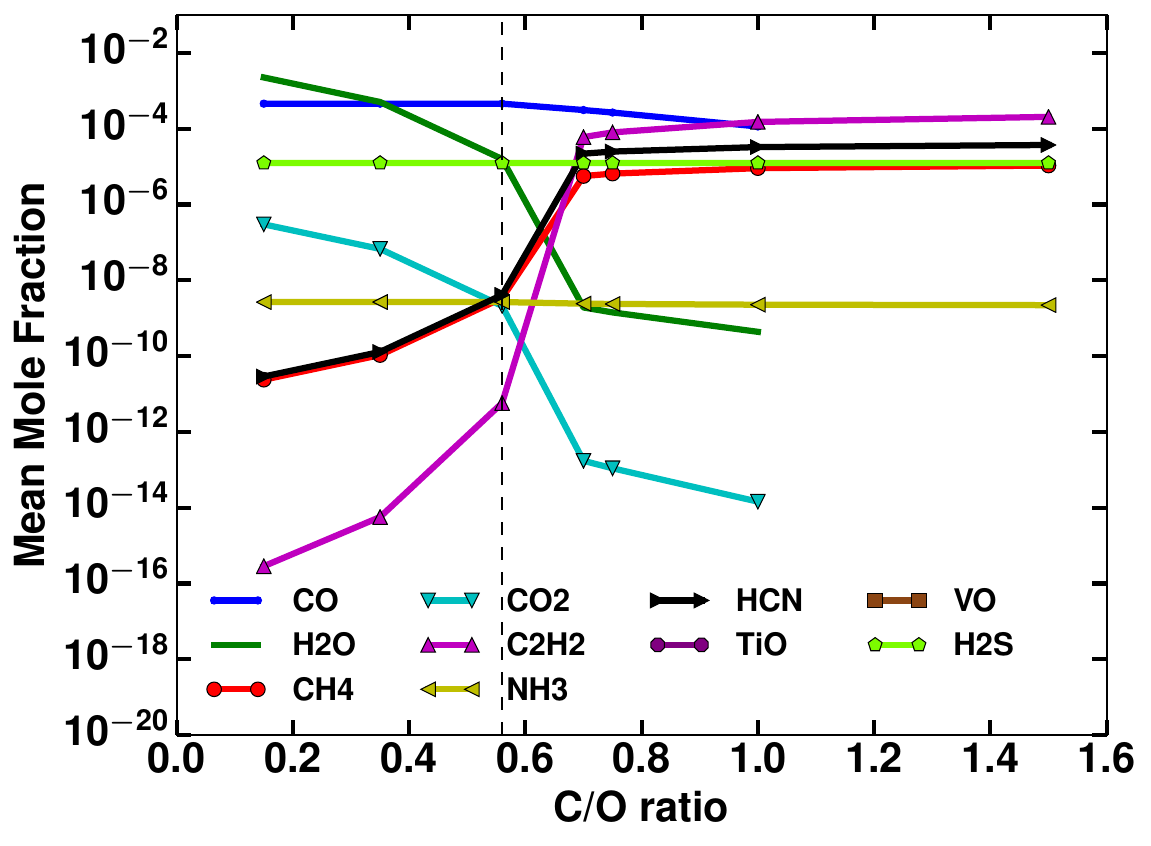}\label{fig:chem_co_wasp17b}}
\end{center}
 \caption{\textbf{(a)} Figure showing WASP-17b transmission spectra for a range of C/O ratio, similar to Figure \ref{fig:coratio_hatp12b}. \textbf{(b)} Figure showing change in mean chemical abundances between 0.1 and 100 millibar for various molecules, with change in C/O ratio for WASP-17b, similar to Figure \ref{fig:chem_co_hatp12b}.}
\end{figure*}

\begin{figure*}
\begin{center}
 \subfloat[]{\includegraphics[width=\columnwidth]{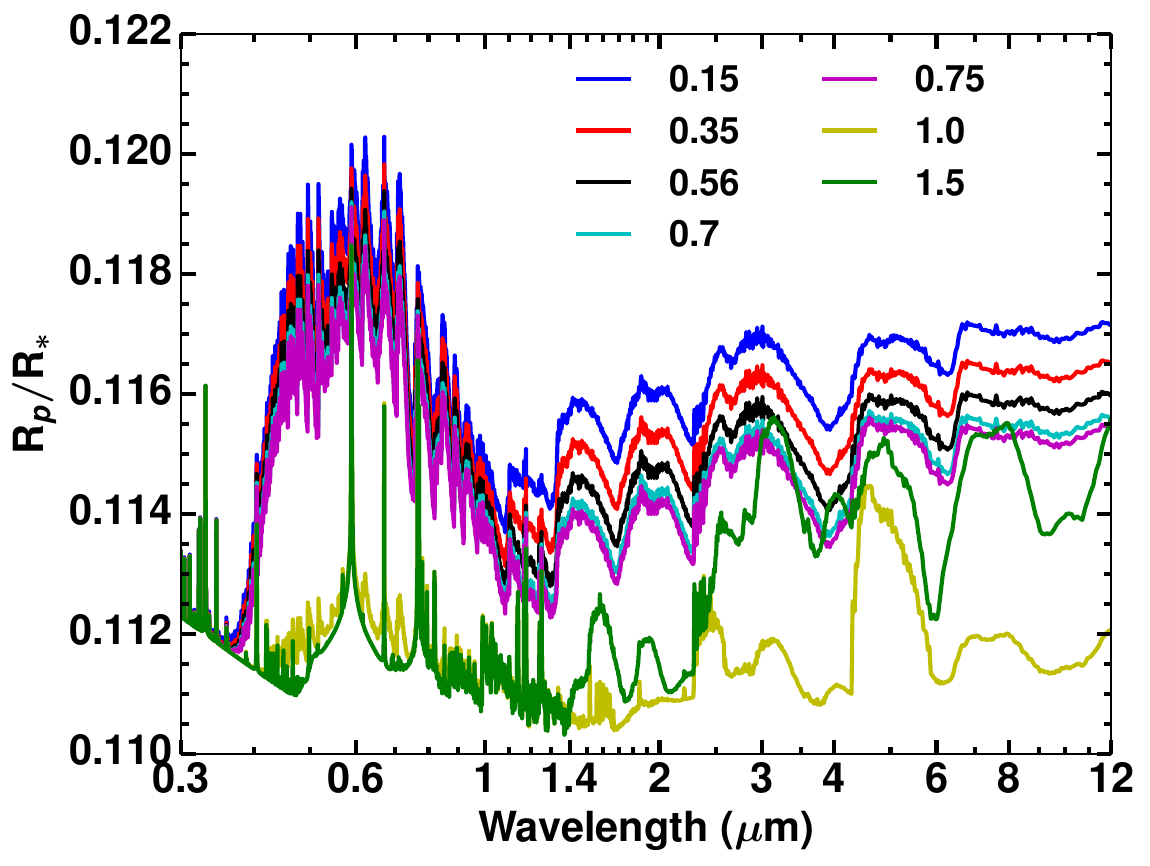}\label{fig:coratio_wasp12b}}
 \subfloat[]{\includegraphics[width=\columnwidth]{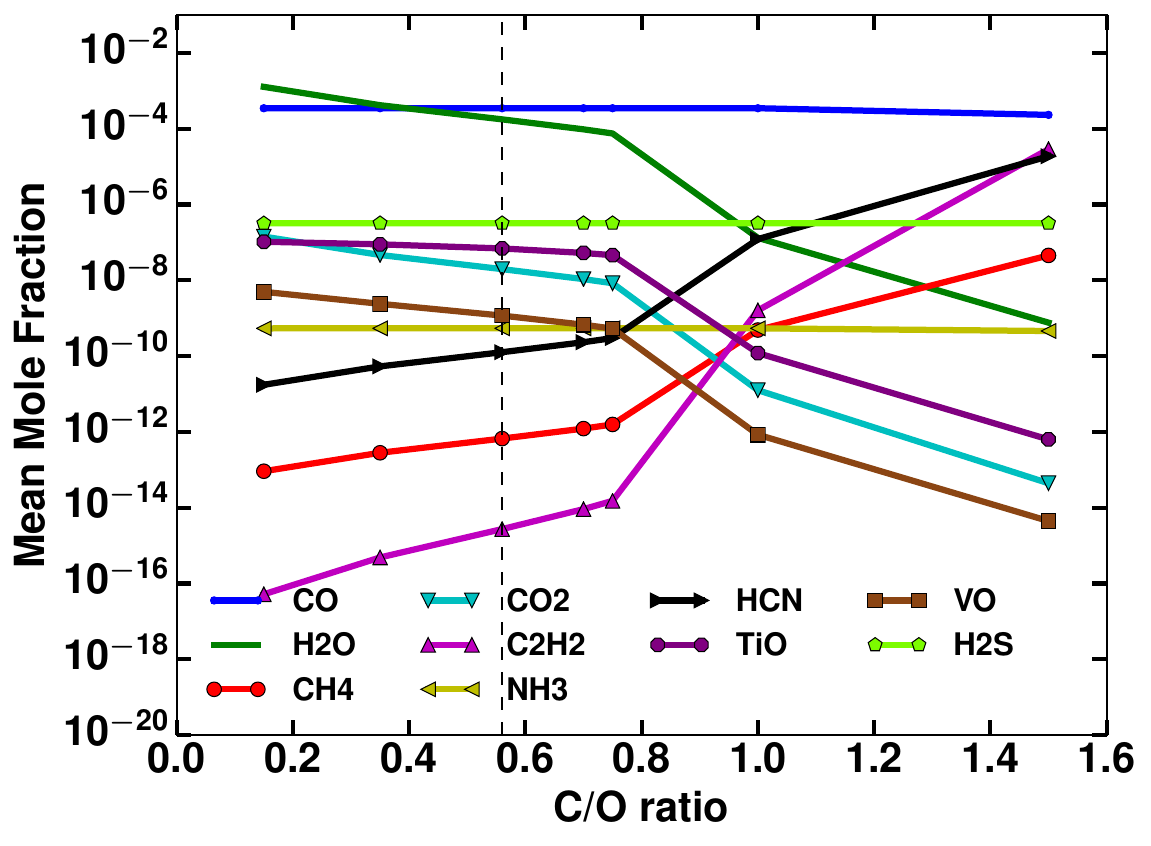}\label{fig:chem_co_wasp12b}}
\end{center}
 \caption{\textbf{(a)} Figure showing WASP-12b transmission spectra for a range of C/O ratio, similar to \ref{fig:coratio_hatp12b}. \textbf{(b)} Figure showing change in mean chemical abundances between 0.1 and 100 millibar for various molecules, with change in C/O ratio for WASP-12b, similar to Figure \ref{fig:chem_co_hatp12b}.}
\end{figure*}

\subsection{Effect of C/O ratio}
\label {subsection:Effect of C/O ratio}
Figures  \ref{fig:coratio_hatp12b},  \ref{fig:coratio_wasp17b} and  \ref{fig:coratio_wasp12b} show changes in the transmission spectra resulting from the changes in the C/O ratio for HAT-P-12b, WASP-17b and WASP-12b, respectively. Additionally, the mean chemical abundances are shown in  \ref{fig:chem_co_hatp12b}, \ref{fig:chem_co_wasp17b} and \ref{fig:chem_co_wasp12b}, respectively. These simulations are at planetary equilibrium temperature, solar metallicity, and do not include any haze or clouds. 

For HAT-P-12b, as the C/O ratio increases CH$_4$ features start dominating over the H$_2$O features, with C/O ratio of $\sim$0.56 being the transition value, as seen in Figure \ref{fig:coratio_hatp12b}. Interestingly, Figure \ref{fig:chem_co_hatp12b} shows that H$_2$O and CH$_4$ abundances are almost equal at the C/O ratio of $\sim$0.56 (solar). However, H$_2$O dominates below $\sim$0.56 and CH$_4$ above it. Note that for lower C/O ratios, oxygen-bearing species such as CO and CO$_2$ dominate, but are replaced by other carbon-bearing species such as HCN and C$_2$H$_2$ as the C/O ratio increases, thereby changing the spectra drastically. The mean mole fraction of CO, CO$_2$ and H$_2$O drop to 0 at C/O ratio of 1.5. This is a combined result of the decrease in elemental oxygen at high C/O ratio and the remaining elemental oxygen taken by more stable condensates such as SiO$_2$, Al$_2$O$_3$, NaAlSi$_3$O$_8$, KAlSi$_2$O$_6$ etc. It must be noted that this result might change with consistent $P$-$T$ profiles, as the temperature will be higher in the higher pressure levels of the atmosphere, affecting the formation of condensates. 

In the case of WASP-17b, as shown in Figure \ref{fig:coratio_wasp17b}, a transition can again be seen from a H$_2$O to a CH$_4$ dominated infrared spectrum as the C/O ratio increases. However, in this case the transition occurs at a higher C/O ratio of $\sim$0.7 (compared to $\sim$0.56 for HAT-P-12b), implying that planets with higher equilibrium temperature have higher transition C/O ratios, in agreement with previous studies \citep{Kopparapu2012, Madhusudhan2012, Moses2013, Venot2015, Molliere2015}. Figure \ref{fig:chem_co_wasp17b} shows the change in mean abundances with C/O ratio for WASP-17b. Here, the transition from H$_2$O to CH$_4$ dominated chemistry occurs at higher C/O ratio compared to HAT-P-12b. It can also be seen that C$_2$H$_2$ and HCN abundances slightly increase even more than CH$_4$ for WASP-17b, for a C/O ratio of 0.7 and higher. This results in a drastic change in transmission spectra, at a C/O ratio of 0.7. Similar to HAT-P-12b, mean mole fraction of CO, CO$_2$ and H$_2$O drop to 0 for WASP-17b at C/O ratio of 1.5.

Figure \ref{fig:coratio_wasp12b} shows spectra for WASP-12b at solar metallicity at various C/O ratios with a clear atmosphere. Figure \ref{fig:coratio_wasp12b} demonstrates an evolution in the TiO/VO features with C/O ratio. For a C/O ratio up to $\sim$0.75, TiO/VO features are dominant but decline thereafter, becoming almost absent by a C/O ratio of 1 and completely absent by 1.5. This is caused by the depletion of oxygen, and subsequent depletion of TiO/VO, as shown in Figure \ref{fig:chem_co_wasp12b}. As found for cooler planets there is a clear transition in the spectra with C/O ratio, as shown in Figure \ref{fig:chem_co_wasp12b}. However, this transition occurs at a higher C/O ratio, $\sim$1-1.3, compared to that found in lower temperature planets (e.g., HAT-P-12b at a C/O ratio of $\sim$0.56). Furthermore, the transition in cooler planets is simply between a H$_2$O and CH$_4$ dominated infrared spectrum, whereas in this hotter case HCN and C$_2$H$_2$ also become more abundant, and therefore spectrally important alongside CH$_4$ at higher C/O ratios. We also performed additional tests adopting radiative-convective equilibrium $P$-$T$ profiles for some planets, to explore whether our conclusions relating to the C/O transition values are robust, and find they remain unchanged, in agreement with previous works \citep{Kopparapu2012, Madhusudhan2012, Moses2013, Venot2015, Molliere2015}.

\begin{figure*}
\begin{center}
 \subfloat[]{\includegraphics[width=\columnwidth]{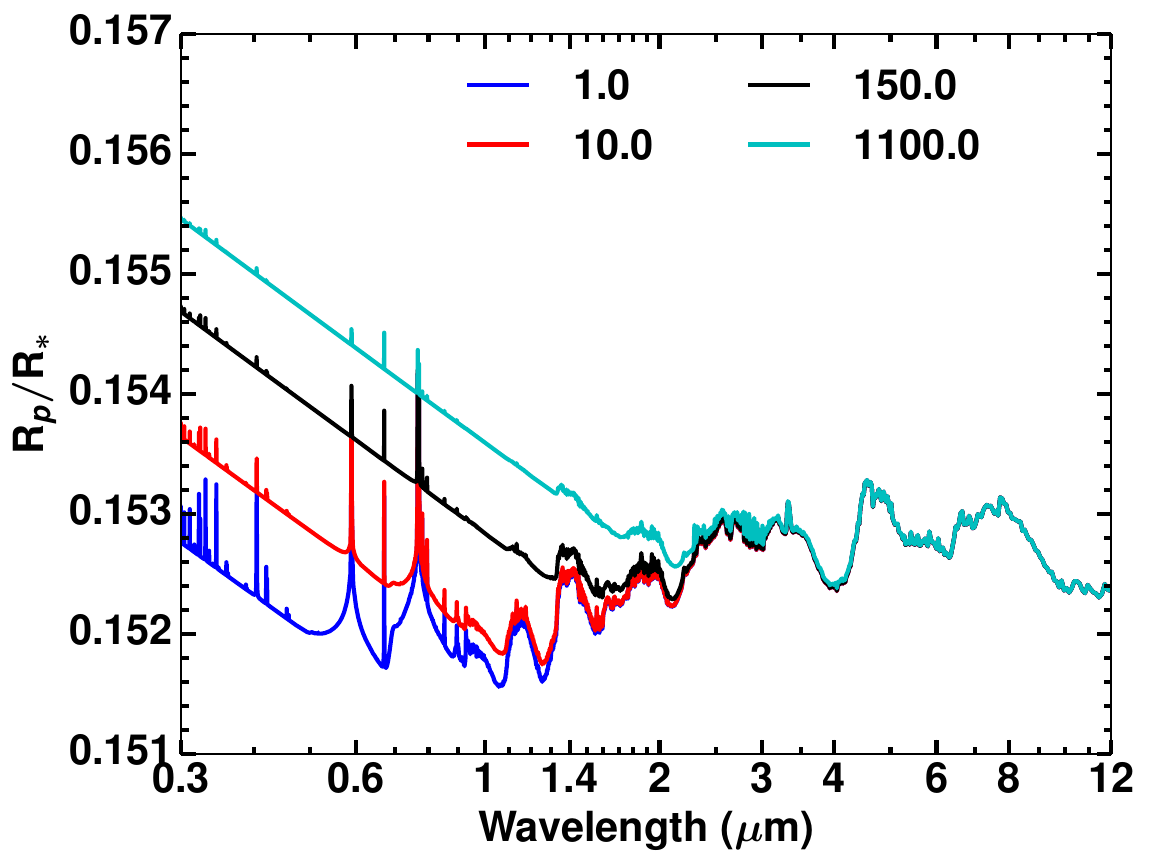}\label{fig:haze_hd189733b}}
 \subfloat[]{\includegraphics[width=\columnwidth]{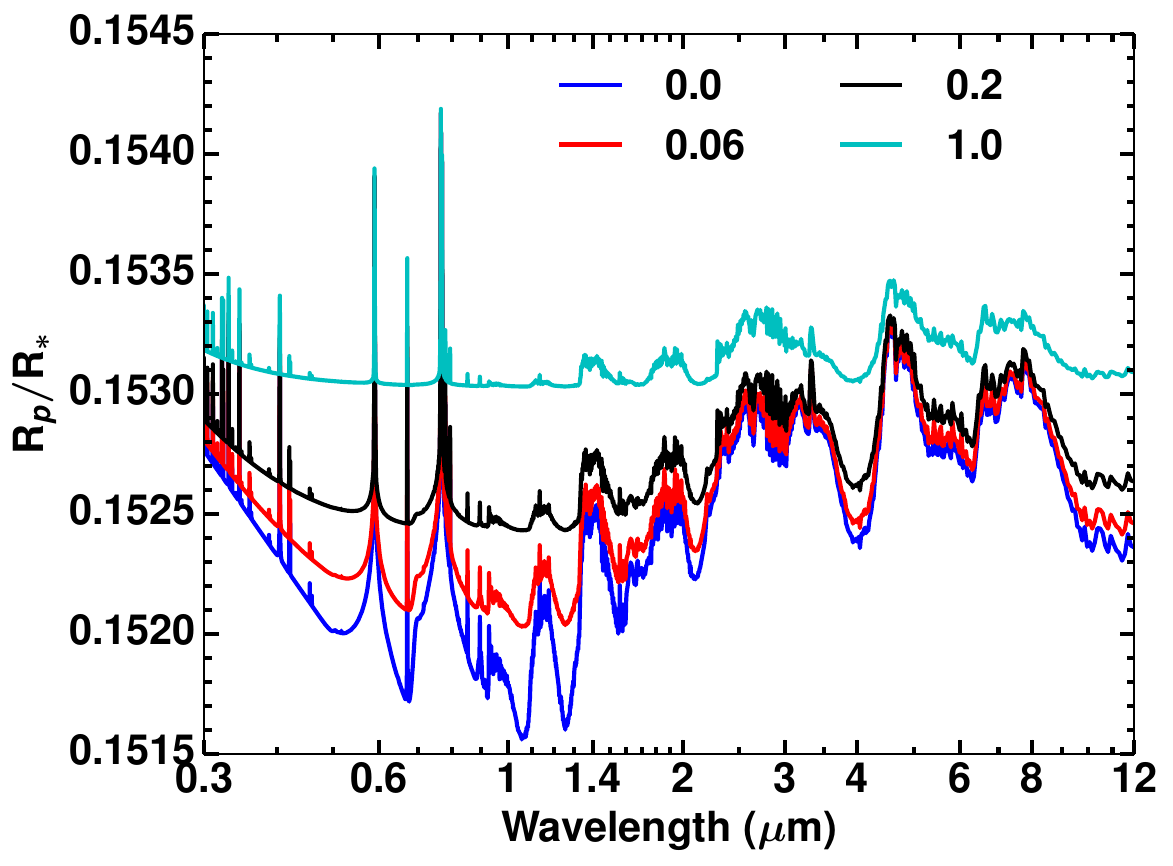}\label{fig:cloud_hd189733b}}
\end{center}
 \caption{\textbf{(a)} Figure showing HD~189733b transmission spectra for a range of haze enhancement factor at its equilibrium temperature, solar C/O ratio, solar metallicity and no clouds. X-axis is wavelength in \textmu m and Y-axis transit radius ratio ($\RpRs$). \textbf{(b)} Figure showing HD~189733b transmission spectra for a range of grey cloud enhancement factor at its equilibrium temperature, solar C/O ratio, solar metallicity and no haze. X-axis is wavelength in \textmu m and Y-axis transit radius ratio ($\RpRs$).}
\end{figure*}

\subsection {Effect of Haze and Clouds}
\label {subsection:Effect of Haze and Clouds}

Haze can be an important source of scattering in planetary atmospheres. Figure \ref{fig:haze_hd189733b} shows the effect of our haze treatment on the transmission spectra of HD~189733b. It shows that as the amount of haze is increased in the atmosphere there is an increase in the amplitude of Rayleigh scattering slope and tendency to mute features, especially at very high values of haze enhancement factor. Haze pre-dominantly effects the optical part of the spectrum due to its scattering nature. 

Figure \ref{fig:cloud_hd189733b} shows the effect of our cloud treatment on the transmission spectra of HD~189733b. An increase in cloud strength ($\alpha_\textup{haze}$) from 0 which indicates no clouds, to 1 which corresponds to grey scattering opacity of $2.5 \times 10^{-3} \textup{cm}^{2}/\textup{g}$ (explained in detail in Section \ref{section:Parameter Space Selection}), increasingly mutes the absorption features at all wavelengths in the transmission spectra. Essentially, increasing cloud cover tends to flatten the spectra. However, interestingly for very hot planets like WASP-12b, the TiO/VO features are so large that even the maximum cloud strength in our parameter space is not able to mute them completely. We note that our model simulations can be used to produce a spectrum that represents patchy clouds using a linear combination of clear and cloudy models \citep[e.g][]{Lineparmentier2016}. 

\section{Simulating JWST observations with ATMO}
\label{Simulating JWST observations with ATMO}
The James Webb Space Telescope (JWST) is an infrared space telescope with a $6.5$-metre primary mirror scheduled for launch in October 2018. The high sensitivity of JWST and its suite of instruments (NIRCam, NIRSpec, NIRISS and MIRI) spanning 0.6-28.3 \textmu m provide the potential to revolutionise our understanding of the atmospheres of extrasolar transiting planets. In preparation of its launch \citet{Batalha2017} have developed a noise simulator, called \texttt{PandExo}\footnote{\url{http://pandexo.science.psu.edu:1111/\#}}, which creates observation simulations of all observatory-supported time-series spectroscopy modes. 

\begin{figure}
\includegraphics[width=\columnwidth]{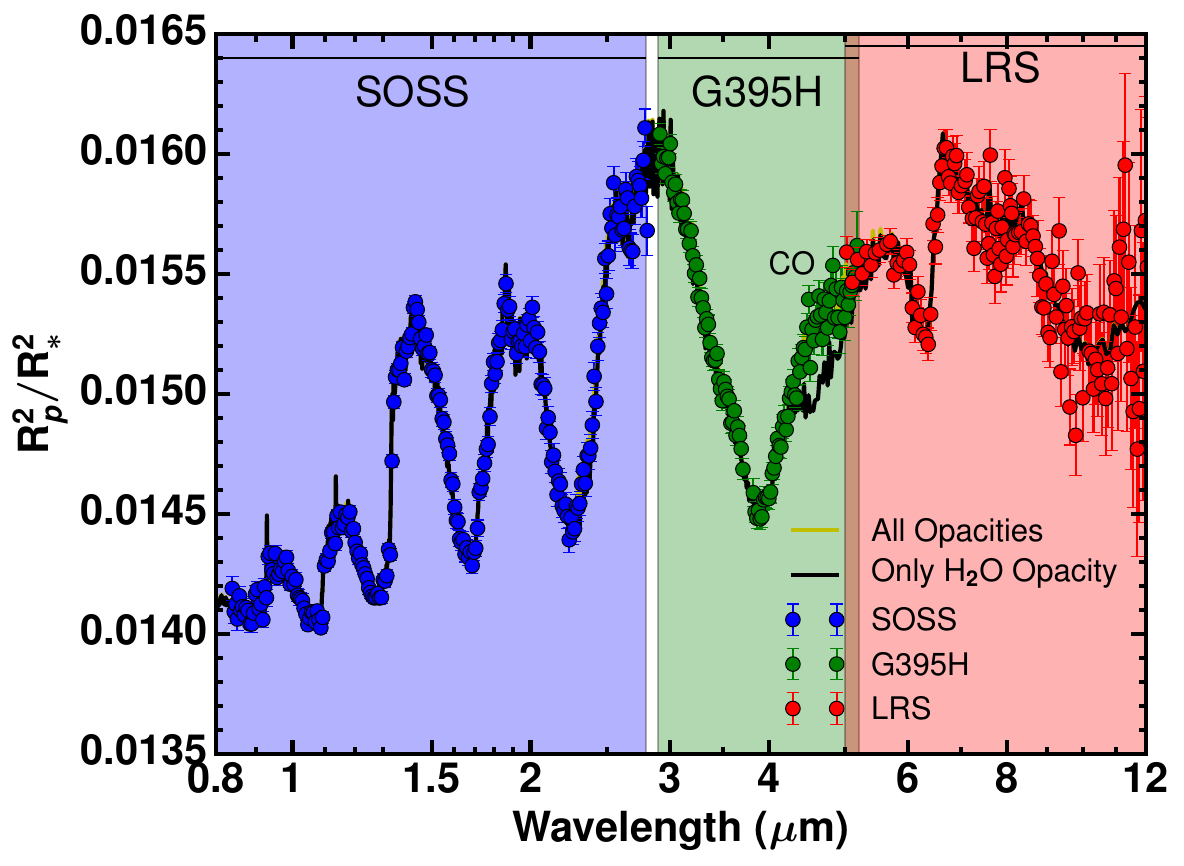}
 \caption{Figure showing ATMO best fit model transmission spectrum (transit depth) for WASP-17b  simulated with \texttt{PandExo} for JWST observations. Model spectrum with all opacities is shown in yellow, which for most of the spectrum is hidden behind only H$_2$O opacity spectrum shown in black. CO (carbon monoxide) feature is marked. Shaded regions and corresponding coloured markers indicate different JWST instrument modes, red  indicates NIRISS SOSS mode, blue indicates NIRSpec G395H mode and green indicates MIRI LRS mode. X-axis is wavelength in \textmu m and Y-axis transit depth ($\Rp^2/\Rs^2$).}
    \label{fig:pandexo_wasp17b}
\end{figure}

We present \texttt{PandExo} simulations of the transmission spectra of WASP-17b for the NIRISS SOSS, NIRSpec G395H and MIRI LRS modes shown in Figure \ref{fig:pandexo_wasp17b} and the \texttt{ATMO} model simulation which best-fits the current HST data shown in Table \ref{tab:grid best fit}. We also over-plotted this with only H$_2$O opacity model spectrum, which shows H$_2$O features explain the spectrum almost completely. Interestingly, between 4 and 6 \textmu m,  just H$_2$O opacity spectrum deviates from all opacity spectrum, which we find is due to CO features, even though this spectrum is for very low C/O ratio of 0.15. This highlights the capability of JWST to detect CO in exoplanet atmospheres and also possibly constrain their C/O ratio. 

The simulation was performed for a single occultation with an equal fraction of in to out of transit observation time, a noise floor of 20 ppm was set for all observation modes and detector saturation was set at 80\% full well. The stellar and planetary parameters necessary for the simulation were retrieved from the TEPCAT database and the stellar spectrum used was identical to the one used for the WASP-17b model grid from the BT-SETTL stellar models. All instrument related parameters, such as subarrays and readout patterns, were kept at the \texttt{PandExo} defaults. The maximum resolution of the \texttt{ATMO} model grid spectrum currently provided is not strictly as high as the achievable resolution of the NIRSpec G395H, however binning of the data will be typically necessary to improve the signal to noise and make resolving certain spectral features possible. As such we do not expect the current model resolution to negatively affect either the current \texttt{PandExo} simulations or any future data analysis. It is evident from these simulations that JWST is likely to provide a dramatic improvement in data quality and wavelength coverage, and the model atmospheres presented, in conjunction with \texttt{PandExo}, are an excellent predictive tool for the planning of future observations. 

 \section{Conclusions}
 \label{section:Conclusions}

We have created an extensive grid\footnote{\url{https://bd-server.astro.ex.ac.uk/exoplanets/}} ($\sim$460,000 simulations) of forward model transmission spectra and the corresponding chemical abundances for 117 observationally significant exoplanets (3920 simulations per planet). The simulated spectra and abundances were produced using a 1D radiative--convective--chemical equilibrium model termed \texttt{ATMO} \citep[described in][]{Tremblin2015, Tremblin2016, Drummond2016}, under the assumption of an isothermal $P$-$T$ profile and including condensation with rainout, varying temperature, metallicity, C/O ratio, haziness and cloudiness. The opacity database used for the simulation \citep{Amundsen2014} is one of the most up-to-date for high temperature planets, including H$_2$ and He broadening wherever possible. The selection of the planets to be modelled was based on their observational transit signal and signal to noise ratio (SNR) in $V_{mag}$ and $K_{mag}$.

In this paper, we explored the validity of the assumption of an isothermal atmosphere, by comparing our simulations with versions including a $P$-$T$ profile in radiative-convective and chemical equilibrium. For a test planet (HD~209458b), we demonstrated that the difference in the transmission spectra between the isothermal and consistent $P$-$T$ profile were small in most cases, except in the temperature regime where spectrally important species condense and potentially rain out of the atmosphere (for example, Na).

We used our set of model simulations to interpret observations of ten exoplanet atmospheres from \citet{Sing2016}. We see a continuum from clear to hazy/cloudy atmospheres as found by \citet{Sing2016}. The data for all the ten planets are consistent with sub-solar to solar C/O ratio, 0.005 to 10 times solar metallicity and a water, rather than a methane dominated atmosphere. The data for  WASP-17b, HAT-P-1b and WASP-6b are consistent with the lowest C/O ratio in out parameter space (0.15), implying the current observations of these planets do not show any clear features, indicative of carbon bearing species. The data for HAT-P-12b and WASP-6b are consistent with extremely high haziness, but without any grey clouds. The data for WASP-12b show extremely muted H$_2$O features leading to the most hazy and cloudy planet of all, while the data for WASP-17b, WASP-39b, WASP-19b and WASP-31b are consistent with a comparatively clear atmosphere. $\chi^{2}$ map for WASP-31b also highlighted the degeneracy existing between the effect of metallicity and clouds/haze on spectral features. $\chi^{2}$ map for HD~209458b revealed a bimodal structure in metallicity, again highlighting the degeneracy between metallicity and all other considered parameters.

We described the variation in transmission spectra with the grid parameters, specifically,  temperature, metallicity, C/O ratio, haziness and cloudiness. We also explored the change in the chemical equilibrium abundances with respect to these parameters in the transmission spectra probed region ($\sim$ 0.1 to 100 millibar). We highlighted spectral features of various chemical species across a range of wavelengths, useful for identifying their signatures in JWST or HST transmission spectra. We find the equilibrium chemical abundances do not change as drastically with temperature, as with metallicity and C/O ratio. CO remains the most abundant chemical species between $\sim$0.1 to 100 millibar, apart from H, H$_2$ and He in all the temperature regimes, except below $800$\,K, where H$_2$O and CH$_4$ are more abundant than CO.

CO also remains the most abundant chemical species apart from H, H$_2$ and He in all the metallicity regimes. CO abundances also increase substantially with increasing metallicity. The transmission spectra and the chemical abundances of many species, change drastically from 10 to 50 times solar metallicity, implying a transition metallicity between these values. The change in spectra is primarily due to HCN and C$_2$H$_2$. Therefore, spectral features of these species could aid constraining planetary atmospheric metallicity. We find the transition C/O ratio, from H$_2$O to CH$_4$ (carbon species)  dominated spectra increases with increasing temperature in agreement with previous studies \citep{Kopparapu2012, Madhusudhan2012, Moses2013, Venot2015,Molliere2015}, but spanning a larger range, with values as low as $\sim$0.56 for low equilibrium temperature (960\,K) planets like HAT-P-12b and $\sim$1-1.3 for very high equilibrium temperature (2580\,K) planets like WASP-12b, where HCN and C$_2$H$_2$ can become more abundant than CH$_4$. 

We also demonstrated the application of our set of model simulations in conjunction with JWST simulator \texttt{PandExo}, as a predictive tool to plan future observations.

We note some of the other major limitations of the current grid. Only the terminator region of the planetary atmosphere is probed using transmission spectra. Therefore, it may not be the representative of the entire planetary atmosphere.  Assumption of equilibrium chemistry becomes less accurate with the decrease in the equilibrium temperature and non-equilibrium effects such as vertical mixing in 1D might become important \citep{Drummond2016}. Current treatment of clouds and haze in our model is very simple without considering any type, shape or distribution of particles which might effect transmission spectra \citep{Wakeford2015, Morley2015}. 1D model is also limited by the absence of various 3D effects like spatial variability, 3D cloud structure, dynamics including horizontal and vertical advection with quenching etc., which can have dramatic effects on observable signatures \citep{Agundez2014, Zellem2014, Kataria2016}.

JWST is expected to constrain the atmospheric $P$-$T$ structure motivating our upcoming work to publish an extended set of model simulations, comprising of transmission spectra, emission spectra and contribution functions with consistent radiative-convective equilibrium $P$-$T$ profiles and equilibrium chemistry. A next step is also to include non-equilibrium chemistry and more realistic clouds, however the computational feasibility is still to be established. The current grid is publicly available online\footnote{\url{https://bd-server.astro.ex.ac.uk/exoplanets/}} and will continuously evolve with the discovery of new observationally significant exoplanets. We encourage the community to use it as a tool to assist them in planning future observations, such as with JWST, HST and various ground based telescopes, along-with interpreting existing datasets. It can provide a useful complement for interpretation, alongside atmospheric retrieval analysis.

\section*{Acknowledgements}
We would like to thank the anonymous reviewer for their constructive comments that improved the paper substantially. We would like to thank Hannah Wakeford for providing very useful comments, John Rowe for making the website and Dave Acreman for helping with computational issues. J.M.G and N.M are part funded by a Leverhulme Trust Research Project Grant, and in part by a University of Exeter College of  Engineering, Mathematics and Physical Sciences PhD studentship. D.K.S, T.E, N.N acknowledges support from the European Research Council under the European Unions Seventh Framework Programme (FP7/2007-2013)/ ERC grant agreement number 336792.  B.D. thanks the University of Exeter for support through a Ph.D. studentship. D.S.A. acknowledges support from the NASA Astrobiology Program through the Nexus for Exoplanet System Science.This  work  used  the  DiRAC  Complexity  system, operated by the University of Leicester IT Services, which forms part of the  STFC  DiRAC  HPC  Facility. This work also used the University of Exeter Supercomputer, a DiRAC Facility jointly funded by STFC, the Large Facilities Capital Fund of BIS and the University of Exeter. 



\bibliographystyle{mnras}
\bibliography{goyal_grid_tosubmit}

\begin{thebibliography}{}
\makeatletter
\relax
\def\mn@urlcharsother{\let\do\@makeother \do\$\do\&\do\#\do\^\do\_\do\%\do\~}
\def\mn@doi{\begingroup\mn@urlcharsother \@ifnextchar [ {\mn@doi@}
  {\mn@doi@[]}}
\def\mn@doi@[#1]#2{\def\@tempa{#1}\ifx\@tempa\@empty \href
  {http://dx.doi.org/#2} {doi:#2}\else \href {http://dx.doi.org/#2} {#1}\fi
  \endgroup}
\def\mn@eprint#1#2{\mn@eprint@#1:#2::\@nil}
\def\mn@eprint@arXiv#1{\href {http://arxiv.org/abs/#1} {{\tt arXiv:#1}}}
\def\mn@eprint@dblp#1{\href {http://dblp.uni-trier.de/rec/bibtex/#1.xml}
  {dblp:#1}}
\def\mn@eprint@#1:#2:#3:#4\@nil{\def\@tempa {#1}\def\@tempb {#2}\def\@tempc
  {#3}\ifx \@tempc \@empty \let \@tempc \@tempb \let \@tempb \@tempa \fi \ifx
  \@tempb \@empty \def\@tempb {arXiv}\fi \@ifundefined
  {mn@eprint@\@tempb}{\@tempb:\@tempc}{\expandafter \expandafter \csname
  mn@eprint@\@tempb\endcsname \expandafter{\@tempc}}}

\bibitem[\protect\citeauthoryear{{Ag{\'u}ndez}, {Parmentier}, {Venot},
  {Hersant}  \& {Selsis}}{{Ag{\'u}ndez} et~al.}{2014}]{Agundez2014}
{Ag{\'u}ndez} M.,  {Parmentier} V.,  {Venot} O.,  {Hersant} F.,   {Selsis} F.,
  2014, \mn@doi [\aap] {10.1051/0004-6361/201322895}, \href
  {http://adsabs.harvard.edu/abs/2014A%26A...564A..73A} {564, A73}

\bibitem[\protect\citeauthoryear{{Allard}, {Royer}, {Kielkopf}  \&
  {Feautrier}}{{Allard} et~al.}{1999}]{Allard1999}
{Allard} N.~F.,  {Royer} A.,  {Kielkopf} J.~F.,   {Feautrier} N.,  1999,
  \mn@doi [\pra] {10.1103/PhysRevA.60.1021}, \href
  {http://adsabs.harvard.edu/abs/1999PhRvA..60.1021A} {60, 1021}

\bibitem[\protect\citeauthoryear{{Allard}, {Allard}, {Hauschildt}, {Kielkopf}
  \& {Machin}}{{Allard} et~al.}{2003}]{Allard2003}
{Allard} N.~F.,  {Allard} F.,  {Hauschildt} P.~H.,  {Kielkopf} J.~F.,
  {Machin} L.,  2003, \mn@doi [\aap] {10.1051/0004-6361:20031299}, \href
  {http://adsabs.harvard.edu/abs/2003A%26A...411L.473A} {411, L473}

\bibitem[\protect\citeauthoryear{{Allard}, {Spiegelman}  \&
  {Kielkopf}}{{Allard} et~al.}{2007}]{Allard2007}
{Allard} N.~F.,  {Spiegelman} F.,   {Kielkopf} J.~F.,  2007, \mn@doi [\aap]
  {10.1051/0004-6361:20066616}, \href
  {http://adsabs.harvard.edu/abs/2007A%26A...465.1085A} {465, 1085}

\bibitem[\protect\citeauthoryear{{Allard}, {Homeier}  \& {Freytag}}{{Allard}
  et~al.}{2012}]{Allard2012}
{Allard} F.,  {Homeier} D.,   {Freytag} B.,  2012, \mn@doi [Philosophical
  Transactions of the Royal Society of London Series A]
  {10.1098/rsta.2011.0269}, \href
  {http://adsabs.harvard.edu/abs/2012RSPTA.370.2765A} {370, 2765}

\bibitem[\protect\citeauthoryear{{Alonso} et~al.,}{{Alonso}
  et~al.}{2004}]{Alonso:2004aa}
{Alonso} R.,  et~al., 2004, \mn@doi [\apjl] {10.1086/425256}, \href
  {http://adsabs.harvard.edu/abs/2004ApJ...613L.153A} {613, L153}

\bibitem[\protect\citeauthoryear{{Amundsen}, {Baraffe}, {Tremblin}, {Manners},
  {Hayek}, {Mayne}  \& {Acreman}}{{Amundsen} et~al.}{2014}]{Amundsen2014}
{Amundsen} D.~S.,  {Baraffe} I.,  {Tremblin} P.,  {Manners} J.,  {Hayek} W.,
  {Mayne} N.~J.,   {Acreman} D.~M.,  2014, \mn@doi [\aap]
  {10.1051/0004-6361/201323169}, \href
  {http://adsabs.harvard.edu/abs/2014A%26A...564A..59A} {564, A59}

\bibitem[\protect\citeauthoryear{{Amundsen}, {Tremblin}, {Manners}, {Baraffe}
  \& {Mayne}}{{Amundsen} et~al.}{2017}]{Amundsen2017}
{Amundsen} D.~S.,  {Tremblin} P.,  {Manners} J.,  {Baraffe} I.,   {Mayne}
  N.~J.,  2017, \mn@doi [\aap] {10.1051/0004-6361/201629322}, \href
  {http://adsabs.harvard.edu/abs/2017A%26A...598A..97A} {598, A97}

\bibitem[\protect\citeauthoryear{{Anderson} et~al.,}{{Anderson}
  et~al.}{2010}]{Anderson:2010aa}
{Anderson} D.~R.,  et~al., 2010, \mn@doi [\apj] {10.1088/0004-637X/709/1/159},
  \href {http://adsabs.harvard.edu/abs/2010ApJ...709..159A} {709, 159}

\bibitem[\protect\citeauthoryear{{Anderson} et~al.,}{{Anderson}
  et~al.}{2011}]{Anderson:2011aa}
{Anderson} D.~R.,  et~al., 2011, \mn@doi [\aap] {10.1051/0004-6361/201016208},
  \href {http://adsabs.harvard.edu/abs/2011A%26A...531A..60A} {531, A60}

\bibitem[\protect\citeauthoryear{{Anderson} et~al.,}{{Anderson}
  et~al.}{2014a}]{Anderson:2014ab}
{Anderson} D.~R.,  et~al., 2014a, preprint, \href
  {http://adsabs.harvard.edu/abs/2014arXiv1410.3449A} {} (\mn@eprint {arXiv}
  {1410.3449})

\bibitem[\protect\citeauthoryear{{Anderson} et~al.,}{{Anderson}
  et~al.}{2014b}]{Anderson:2014ac}
{Anderson} D.~R.,  et~al., 2014b, \mn@doi [\mnras] {10.1093/mnras/stu1737},
  \href {http://adsabs.harvard.edu/abs/2014MNRAS.445.1114A} {445, 1114}

\bibitem[\protect\citeauthoryear{{Anderson} et~al.,}{{Anderson}
  et~al.}{2015a}]{Anderson:2015aa}
{Anderson} D.~R.,  et~al., 2015a, \mn@doi [\aap] {10.1051/0004-6361/201423591},
  \href {http://adsabs.harvard.edu/abs/2015A%26A...575A..61A} {575, A61}

\bibitem[\protect\citeauthoryear{{Anderson} et~al.,}{{Anderson}
  et~al.}{2015b}]{Anderson:2015ab}
{Anderson} D.~R.,  et~al., 2015b, \mn@doi [\apjl] {10.1088/2041-8205/800/1/L9},
  \href {http://adsabs.harvard.edu/abs/2015ApJ...800L...9A} {800, L9}

\bibitem[\protect\citeauthoryear{{Bakos} et~al.,}{{Bakos}
  et~al.}{2007}]{Bakos:2007aa}
{Bakos} G.~{\'A}.,  et~al., 2007, \mn@doi [\apj] {10.1086/509874}, \href
  {http://adsabs.harvard.edu/abs/2007ApJ...656..552B} {656, 552}

\bibitem[\protect\citeauthoryear{{Bakos} et~al.,}{{Bakos}
  et~al.}{2009}]{Bakos:2009aa}
{Bakos} G.~{\'A}.,  et~al., 2009, \mn@doi [\apj] {10.1088/0004-637X/707/1/446},
  \href {http://adsabs.harvard.edu/abs/2009ApJ...707..446B} {707, 446}

\bibitem[\protect\citeauthoryear{{Bakos} et~al.,}{{Bakos}
  et~al.}{2010}]{Bakos:2010aa}
{Bakos} G.~{\'A}.,  et~al., 2010, \mn@doi [\apj]
  {10.1088/0004-637X/710/2/1724}, \href
  {http://adsabs.harvard.edu/abs/2010ApJ...710.1724B} {710, 1724}

\bibitem[\protect\citeauthoryear{{Bakos} et~al.,}{{Bakos}
  et~al.}{2016}]{Bakos:2016aa}
{Bakos} G.~{\'A}.,  et~al., 2016, preprint, \href
  {http://adsabs.harvard.edu/abs/2016arXiv160604556B} {} (\mn@eprint {arXiv}
  {1606.04556})

\bibitem[\protect\citeauthoryear{{Barber}, {Tennyson}, {Harris}  \&
  {Tolchenov}}{{Barber} et~al.}{2006}]{Barber2006}
{Barber} R.~J.,  {Tennyson} J.,  {Harris} G.~J.,   {Tolchenov} R.~N.,  2006,
  \mn@doi [\mnras] {10.1111/j.1365-2966.2006.10184.x}, \href
  {http://adsabs.harvard.edu/abs/2006MNRAS.368.1087B} {368, 1087}

\bibitem[\protect\citeauthoryear{{Barber}, {Strange}, {Hill}, {Polyansky},
  {Mellau}, {Yurchenko}  \& {Tennyson}}{{Barber} et~al.}{2014}]{Barber2014}
{Barber} R.~J.,  {Strange} J.~K.,  {Hill} C.,  {Polyansky} O.~L.,  {Mellau}
  G.~C.,  {Yurchenko} S.~N.,   {Tennyson} J.,  2014, \mn@doi [\mnras]
  {10.1093/mnras/stt2011}, \href
  {http://adsabs.harvard.edu/abs/2014MNRAS.437.1828B} {437, 1828}

\bibitem[\protect\citeauthoryear{{Barros} et~al.,}{{Barros}
  et~al.}{2016}]{Barros:2016aa}
{Barros} S.~C.~C.,  et~al., 2016, \mn@doi [\aap] {10.1051/0004-6361/201526517},
  \href {http://adsabs.harvard.edu/abs/2016A%26A...593A.113B} {593, A113}

\bibitem[\protect\citeauthoryear{{Barshay} \& {Lewis}}{{Barshay} \&
  {Lewis}}{1978}]{Barshay1978}
{Barshay} S.~S.,  {Lewis} J.~S.,  1978, \mn@doi [\icarus]
  {10.1016/0019-1035(78)90192-6}, \href
  {http://adsabs.harvard.edu/abs/1978Icar...33..593B} {33, 593}

\bibitem[\protect\citeauthoryear{{Barstow}, {Aigrain}, {Irwin}  \&
  {Sing}}{{Barstow} et~al.}{2017}]{Barstow2016}
{Barstow} J.~K.,  {Aigrain} S.,  {Irwin} P.~G.~J.,   {Sing} D.~K.,  2017,
  \mn@doi [\apj] {10.3847/1538-4357/834/1/50}, \href
  {http://adsabs.harvard.edu/abs/2017ApJ...834...50B} {834, 50}

\bibitem[\protect\citeauthoryear{{Batalha} et~al.,}{{Batalha}
  et~al.}{2017}]{Batalha2017}
{Batalha} N.~E.,  et~al., 2017, preprint, \href
  {http://adsabs.harvard.edu/abs/2017arXiv170201820B} {} (\mn@eprint {arXiv}
  {1702.01820})

\bibitem[\protect\citeauthoryear{{Beichman} et~al.,}{{Beichman}
  et~al.}{2014}]{Beichman2014}
{Beichman} C.,  et~al., 2014, \mn@doi [\pasp] {10.1086/679566}, \href
  {http://adsabs.harvard.edu/abs/2014PASP..126.1134B} {126, 1134}

\bibitem[\protect\citeauthoryear{{BelBruno}, {Gelfand}, {Radigan}  \&
  {Verges}}{{BelBruno} et~al.}{1982}]{Belbruno1982}
{BelBruno} J.~J.,  {Gelfand} J.,  {Radigan} W.,   {Verges} K.,  1982, \mn@doi
  [Journal of Molecular Spectroscopy] {10.1016/0022-2852(82)90009-1}, \href
  {http://adsabs.harvard.edu/abs/1982JMoSp..94..336B} {94, 336}

\bibitem[\protect\citeauthoryear{{Benneke}}{{Benneke}}{2015}]{Benneke2015}
{Benneke} B.,  2015, preprint, \href
  {http://adsabs.harvard.edu/abs/2015arXiv150407655B} {} (\mn@eprint {arXiv}
  {1504.07655})

\bibitem[\protect\citeauthoryear{{Benneke} \& {Seager}}{{Benneke} \&
  {Seager}}{2012}]{Benneke2012}
{Benneke} B.,  {Seager} S.,  2012, \mn@doi [\apj]
  {10.1088/0004-637X/753/2/100}, \href
  {http://adsabs.harvard.edu/abs/2012ApJ...753..100B} {753, 100}

\bibitem[\protect\citeauthoryear{{Bhatti} et~al.,}{{Bhatti}
  et~al.}{2016}]{Bhatti:2016aa}
{Bhatti} W.,  et~al., 2016, preprint, \href
  {http://adsabs.harvard.edu/abs/2016arXiv160700322B} {} (\mn@eprint {arXiv}
  {1607.00322})

\bibitem[\protect\citeauthoryear{{Biddle} et~al.,}{{Biddle}
  et~al.}{2014}]{Biddle:2014aa}
{Biddle} L.~I.,  et~al., 2014, \mn@doi [\mnras] {10.1093/mnras/stu1199}, \href
  {http://adsabs.harvard.edu/abs/2014MNRAS.443.1810B} {443, 1810}

\bibitem[\protect\citeauthoryear{{Bieryla} et~al.,}{{Bieryla}
  et~al.}{2015}]{Bieryla:2015aa}
{Bieryla} A.,  et~al., 2015, \mn@doi [\aj] {10.1088/0004-6256/150/1/12}, \href
  {http://adsabs.harvard.edu/abs/2015AJ....150...12B} {150, 12}

\bibitem[\protect\citeauthoryear{{Bonfils} et~al.,}{{Bonfils}
  et~al.}{2012}]{Bonfils:2012aa}
{Bonfils} X.,  et~al., 2012, \mn@doi [\aap] {10.1051/0004-6361/201219623},
  \href {http://adsabs.harvard.edu/abs/2012A%26A...546A..27B} {546, A27}

\bibitem[\protect\citeauthoryear{{Bouanich}, {Salem}, {Aroui}, {Walrand}  \&
  {Blanquet}}{{Bouanich} et~al.}{2004}]{Bouanich2004}
{Bouanich} J.-P.,  {Salem} J.,  {Aroui} H.,  {Walrand} J.,   {Blanquet} G.,
  2004, \mn@doi [\jqsrt] {10.1016/S0022-4073(03)00143-2}, \href
  {http://adsabs.harvard.edu/abs/2004JQSRT..84..195B} {84, 195}

\bibitem[\protect\citeauthoryear{{Bouchy} et~al.,}{{Bouchy}
  et~al.}{2005}]{Bouchy:2005aa}
{Bouchy} F.,  et~al., 2005, \mn@doi [\aap] {10.1051/0004-6361:200500201}, \href
  {http://adsabs.harvard.edu/abs/2005A%26A...444L..15B} {444, L15}

\bibitem[\protect\citeauthoryear{{Bouchy} et~al.,}{{Bouchy}
  et~al.}{2010}]{Bouchy:2010aa}
{Bouchy} F.,  et~al., 2010, \mn@doi [\aap] {10.1051/0004-6361/201014817}, \href
  {http://adsabs.harvard.edu/abs/2010A%26A...519A..98B} {519, A98}

\bibitem[\protect\citeauthoryear{{Broeg} et~al.,}{{Broeg}
  et~al.}{2013}]{Broeg2013}
{Broeg} C.,  et~al., 2013, in European Physical Journal Web of Conferences. p.
  03005 (\mn@eprint {arXiv} {1305.2270}), \mn@doi{10.1051/epjconf/20134703005}

\bibitem[\protect\citeauthoryear{{Brown}}{{Brown}}{2001}]{Brown2001}
{Brown} T.~M.,  2001, \mn@doi [\apj] {10.1086/320950}, \href
  {http://adsabs.harvard.edu/abs/2001ApJ...553.1006B} {553, 1006}

\bibitem[\protect\citeauthoryear{{Burke} et~al.,}{{Burke}
  et~al.}{2007}]{Burke:2007aa}
{Burke} C.~J.,  et~al., 2007, \mn@doi [\apj] {10.1086/523087}, \href
  {http://adsabs.harvard.edu/abs/2007ApJ...671.2115B} {671, 2115}

\bibitem[\protect\citeauthoryear{{Burrows} \& {Sharp}}{{Burrows} \&
  {Sharp}}{1999}]{Burrows1999}
{Burrows} A.,  {Sharp} C.~M.,  1999, \mn@doi [\apj] {10.1086/306811}, \href
  {http://adsabs.harvard.edu/abs/1999ApJ...512..843B} {512, 843}

\bibitem[\protect\citeauthoryear{{Burrows} et~al.,}{{Burrows}
  et~al.}{1997}]{Burrows1997}
{Burrows} A.,  et~al., 1997, \mn@doi [\apj] {10.1086/305002}, \href
  {http://adsabs.harvard.edu/abs/1997ApJ...491..856B} {491, 856}

\bibitem[\protect\citeauthoryear{{Burrows}, {Ram}, {Bernath}, {Sharp}  \&
  {Milsom}}{{Burrows} et~al.}{2002}]{Burrows2002}
{Burrows} A.,  {Ram} R.~S.,  {Bernath} P.,  {Sharp} C.~M.,   {Milsom} J.~A.,
  2002, \mn@doi [\apj] {10.1086/342242}, \href
  {http://adsabs.harvard.edu/abs/2002ApJ...577..986B} {577, 986}

\bibitem[\protect\citeauthoryear{{Caffau}, {Ludwig}, {Steffen}, {Freytag}  \&
  {Bonifacio}}{{Caffau} et~al.}{2011}]{Caffau2011}
{Caffau} E.,  {Ludwig} H.-G.,  {Steffen} M.,  {Freytag} B.,   {Bonifacio} P.,
  2011, \mn@doi [\solphys] {10.1007/s11207-010-9541-4}, \href
  {http://adsabs.harvard.edu/abs/2011SoPh..268..255C} {268, 255}

\bibitem[\protect\citeauthoryear{{Carter}, {Winn}, {Gilliland}  \&
  {Holman}}{{Carter} et~al.}{2009}]{Carter:2009aa}
{Carter} J.~A.,  {Winn} J.~N.,  {Gilliland} R.,   {Holman} M.~J.,  2009,
  \mn@doi [\apj] {10.1088/0004-637X/696/1/241}, \href
  {http://adsabs.harvard.edu/abs/2009ApJ...696..241C} {696, 241}

\bibitem[\protect\citeauthoryear{{Charbonneau}, {Brown}, {Noyes}  \&
  {Gilliland}}{{Charbonneau} et~al.}{2002}]{Charbonneau2002}
{Charbonneau} D.,  {Brown} T.~M.,  {Noyes} R.~W.,   {Gilliland} R.~L.,  2002,
  \mn@doi [\apj] {10.1086/338770}, \href
  {http://adsabs.harvard.edu/abs/2002ApJ...568..377C} {568, 377}

\bibitem[\protect\citeauthoryear{{Charbonneau} et~al.,}{{Charbonneau}
  et~al.}{2009}]{Charbonneau:2009aa}
{Charbonneau} D.,  et~al., 2009, \mn@doi [\nat] {10.1038/nature08679}, \href
  {http://adsabs.harvard.edu/abs/2009Natur.462..891C} {462, 891}

\bibitem[\protect\citeauthoryear{{Ciceri} et~al.,}{{Ciceri}
  et~al.}{2013}]{Ciceri:2013aa}
{Ciceri} S.,  et~al., 2013, \mn@doi [\aap] {10.1051/0004-6361/201321669}, \href
  {http://adsabs.harvard.edu/abs/2013A%26A...557A..30C} {557, A30}

\bibitem[\protect\citeauthoryear{{Collier Cameron} et~al.,}{{Collier Cameron}
  et~al.}{2007}]{Collier-Cameron:2007aa}
{Collier Cameron} A.,  et~al., 2007, \mn@doi [\mnras]
  {10.1111/j.1365-2966.2006.11350.x}, \href
  {http://adsabs.harvard.edu/abs/2007MNRAS.375..951C} {375, 951}

\bibitem[\protect\citeauthoryear{{Collier Cameron} et~al.,}{{Collier Cameron}
  et~al.}{2010}]{Collier-Cameron:2010aa}
{Collier Cameron} A.,  et~al., 2010, \mn@doi [\mnras]
  {10.1111/j.1365-2966.2010.16922.x}, \href
  {http://adsabs.harvard.edu/abs/2010MNRAS.407..507C} {407, 507}

\bibitem[\protect\citeauthoryear{{Collins}, {Kielkopf}  \& {Stassun}}{{Collins}
  et~al.}{2015}]{Collins:2015aa}
{Collins} K.~A.,  {Kielkopf} J.~F.,   {Stassun} K.~G.,  2015, preprint, \href
  {http://adsabs.harvard.edu/abs/2015arXiv151200464C} {} (\mn@eprint {arXiv}
  {1512.00464})

\bibitem[\protect\citeauthoryear{{Cox}}{{Cox}}{2000}]{Allen2000}
{Cox} A.~N.,  2000, {Allen's astrophysical quantities}

\bibitem[\protect\citeauthoryear{{Damasso} et~al.,}{{Damasso}
  et~al.}{2015}]{Damasso:2015aa}
{Damasso} M.,  et~al., 2015, \mn@doi [\aap] {10.1051/0004-6361/201425332},
  \href {http://adsabs.harvard.edu/abs/2015A%26A...575A.111D} {575, A111}

\bibitem[\protect\citeauthoryear{{Delrez} et~al.,}{{Delrez}
  et~al.}{2014}]{Delrez:2014aa}
{Delrez} L.,  et~al., 2014, \mn@doi [\aap] {10.1051/0004-6361/201323204}, \href
  {http://adsabs.harvard.edu/abs/2014A%26A...563A.143D} {563, A143}

\bibitem[\protect\citeauthoryear{{Delrez} et~al.,}{{Delrez}
  et~al.}{2016}]{Delrez:2016aa}
{Delrez} L.,  et~al., 2016, \mn@doi [\mnras] {10.1093/mnras/stw522}, \href
  {http://adsabs.harvard.edu/abs/2016MNRAS.458.4025D} {458, 4025}

\bibitem[\protect\citeauthoryear{{Deming} et~al.,}{{Deming}
  et~al.}{2013}]{Deming2013}
{Deming} D.,  et~al., 2013, \mn@doi [\apj] {10.1088/0004-637X/774/2/95}, \href
  {http://adsabs.harvard.edu/abs/2013ApJ...774...95D} {774, 95}

\bibitem[\protect\citeauthoryear{{Demory} et~al.,}{{Demory}
  et~al.}{2016}]{Demory:2016aa}
{Demory} B.-O.,  et~al., 2016, \mn@doi [\nat] {10.1038/nature17169}, \href
  {http://adsabs.harvard.edu/abs/2016Natur.532..207D} {532, 207}

\bibitem[\protect\citeauthoryear{{Dragomir} et~al.,}{{Dragomir}
  et~al.}{2013}]{Dragomir:2013aa}
{Dragomir} D.,  et~al., 2013, \mn@doi [\apjl] {10.1088/2041-8205/772/1/L2},
  \href {http://adsabs.harvard.edu/abs/2013ApJ...772L...2D} {772, L2}

\bibitem[\protect\citeauthoryear{{Drummond}, {Tremblin}, {Baraffe}, {Amundsen},
  {Mayne}, {Venot}  \& {Goyal}}{{Drummond} et~al.}{2016}]{Drummond2016}
{Drummond} B.,  {Tremblin} P.,  {Baraffe} I.,  {Amundsen} D.~S.,  {Mayne}
  N.~J.,  {Venot} O.,   {Goyal} J.,  2016, \mn@doi [\aap]
  {10.1051/0004-6361/201628799}, \href
  {http://adsabs.harvard.edu/abs/2016A%26A...594A..69D} {594, A69}

\bibitem[\protect\citeauthoryear{{Eastman} et~al.,}{{Eastman}
  et~al.}{2016}]{Eastman:2016aa}
{Eastman} J.~D.,  et~al., 2016, \mn@doi [\aj] {10.3847/0004-6256/151/2/45},
  \href {http://adsabs.harvard.edu/abs/2016AJ....151...45E} {151, 45}

\bibitem[\protect\citeauthoryear{{Edwards} \& {Slingo}}{{Edwards} \&
  {Slingo}}{1996}]{Socrates1996}
{Edwards} J.~M.,  {Slingo} A.,  1996, \mn@doi [Quarterly Journal of the Royal
  Meteorological Society] {10.1002/qj.49712253107}, \href
  {http://adsabs.harvard.edu/abs/1996QJRMS.122..689E} {122, 689}

\bibitem[\protect\citeauthoryear{{Enoch} et~al.,}{{Enoch}
  et~al.}{2011a}]{Enoch:2011ab}
{Enoch} B.,  et~al., 2011a, \mn@doi [\aj] {10.1088/0004-6256/142/3/86}, \href
  {http://adsabs.harvard.edu/abs/2011AJ....142...86E} {142, 86}

\bibitem[\protect\citeauthoryear{{Enoch} et~al.,}{{Enoch}
  et~al.}{2011b}]{Enoch:2011aa}
{Enoch} B.,  et~al., 2011b, \mn@doi [\mnras]
  {10.1111/j.1365-2966.2010.17550.x}, \href
  {http://adsabs.harvard.edu/abs/2011MNRAS.410.1631E} {410, 1631}

\bibitem[\protect\citeauthoryear{{Esposito} et~al.,}{{Esposito}
  et~al.}{2014}]{Esposito:2014aa}
{Esposito} M.,  et~al., 2014, \mn@doi [\aap] {10.1051/0004-6361/201423735},
  \href {http://adsabs.harvard.edu/abs/2014A%26A...564L..13E} {564, L13}

\bibitem[\protect\citeauthoryear{{Evans} et~al.,}{{Evans}
  et~al.}{2016a}]{Evans2016}
{Evans} T.~M.,  et~al., 2016a, \mn@doi [\apjl] {10.3847/2041-8205/822/1/L4},
  \href {http://adsabs.harvard.edu/abs/2016ApJ...822L...4E} {822, L4}

\bibitem[\protect\citeauthoryear{{Evans}, {Southworth}  \& {Smalley}}{{Evans}
  et~al.}{2016b}]{Evans:2016aa}
{Evans} D.~F.,  {Southworth} J.,   {Smalley} B.,  2016b, \mn@doi [\apjl]
  {10.3847/2041-8213/833/2/L19}, \href
  {http://adsabs.harvard.edu/abs/2016ApJ...833L..19E} {833, L19}

\bibitem[\protect\citeauthoryear{{Evans} et~al.,}{{Evans}
  et~al.}{2017}]{Evans2017}
{Evans} T.~M.,  et~al., 2017, \mn@doi [\nat] {10.1038/nature23266}, \href
  {http://adsabs.harvard.edu/abs/2017Natur.548...58E} {548, 58}

\bibitem[\protect\citeauthoryear{{Faedi} et~al.,}{{Faedi}
  et~al.}{2011}]{Faedi:2011aa}
{Faedi} F.,  et~al., 2011, \mn@doi [\aap] {10.1051/0004-6361/201116671}, \href
  {http://adsabs.harvard.edu/abs/2011A%26A...531A..40F} {531, A40}

\bibitem[\protect\citeauthoryear{{Faedi} et~al.,}{{Faedi}
  et~al.}{2013}]{Faedi:2013aa}
{Faedi} F.,  et~al., 2013, \mn@doi [\aap] {10.1051/0004-6361/201220520}, \href
  {http://adsabs.harvard.edu/abs/2013A%26A...551A..73F} {551, A73}

\bibitem[\protect\citeauthoryear{{Fischer} et~al.,}{{Fischer}
  et~al.}{2016}]{Fischer2016}
{Fischer} P.~D.,  et~al., 2016, \mn@doi [\apj] {10.3847/0004-637X/827/1/19},
  \href {http://adsabs.harvard.edu/abs/2016ApJ...827...19F} {827, 19}

\bibitem[\protect\citeauthoryear{{Fortney}}{{Fortney}}{2005}]{Fortney2005}
{Fortney} J.~J.,  2005, \mn@doi [\mnras] {10.1111/j.1365-2966.2005.09587.x},
  \href {http://adsabs.harvard.edu/abs/2005MNRAS.364..649F} {364, 649}

\bibitem[\protect\citeauthoryear{{Fortney} \& {Marley}}{{Fortney} \&
  {Marley}}{2007}]{Fortney2007}
{Fortney} J.~J.,  {Marley} M.~S.,  2007, \mn@doi [\apjl] {10.1086/521603},
  \href {http://adsabs.harvard.edu/abs/2007ApJ...666L..45F} {666, L45}

\bibitem[\protect\citeauthoryear{{Fortney}, {Lodders}, {Marley}  \&
  {Freedman}}{{Fortney} et~al.}{2008}]{Fortney2008}
{Fortney} J.~J.,  {Lodders} K.,  {Marley} M.~S.,   {Freedman} R.~S.,  2008,
  \mn@doi [\apj] {10.1086/528370}, \href
  {http://adsabs.harvard.edu/abs/2008ApJ...678.1419F} {678, 1419}

\bibitem[\protect\citeauthoryear{{Fortney}, {Shabram}, {Showman}, {Lian},
  {Freedman}, {Marley}  \& {Lewis}}{{Fortney} et~al.}{2010}]{Fortney2010}
{Fortney} J.~J.,  {Shabram} M.,  {Showman} A.~P.,  {Lian} Y.,  {Freedman}
  R.~S.,  {Marley} M.~S.,   {Lewis} N.~K.,  2010, \mn@doi [\apj]
  {10.1088/0004-637X/709/2/1396}, \href
  {http://adsabs.harvard.edu/abs/2010ApJ...709.1396F} {709, 1396}

\bibitem[\protect\citeauthoryear{{Fortney} et~al.,}{{Fortney}
  et~al.}{2011}]{Fortney:2011aa}
{Fortney} J.~J.,  et~al., 2011, \mn@doi [\apjs] {10.1088/0067-0049/197/1/9},
  \href {http://adsabs.harvard.edu/abs/2011ApJS..197....9F} {197, 9}

\bibitem[\protect\citeauthoryear{{Fortney} et~al.,}{{Fortney}
  et~al.}{2016}]{Fortney2016}
{Fortney} J.~J.,  et~al., 2016, preprint, \href
  {http://adsabs.harvard.edu/abs/2016arXiv160206305F} {} (\mn@eprint {arXiv}
  {1602.06305})

\bibitem[\protect\citeauthoryear{{Fulton} et~al.,}{{Fulton}
  et~al.}{2015}]{Fulton:2015aa}
{Fulton} B.~J.,  et~al., 2015, \mn@doi [\apj] {10.1088/0004-637X/810/1/30},
  \href {http://adsabs.harvard.edu/abs/2015ApJ...810...30F} {810, 30}

\bibitem[\protect\citeauthoryear{{Gamache}, {Lynch}  \& {Brown}}{{Gamache}
  et~al.}{1996}]{Gamache1996}
{Gamache} R.~R.,  {Lynch} R.,   {Brown} L.~R.,  1996, \mn@doi [\jqsrt]
  {10.1016/0022-4073(96)00098-2}, \href
  {http://adsabs.harvard.edu/abs/1996JQSRT..56..471G} {56, 471}

\bibitem[\protect\citeauthoryear{{Gibson}, {Aigrain}, {Barstow}, {Evans},
  {Fletcher}  \& {Irwin}}{{Gibson} et~al.}{2013}]{Gibson:2013aa}
{Gibson} N.~P.,  {Aigrain} S.,  {Barstow} J.~K.,  {Evans} T.~M.,  {Fletcher}
  L.~N.,   {Irwin} P.~G.~J.,  2013, \mn@doi [\mnras] {10.1093/mnras/sts307},
  \href {http://adsabs.harvard.edu/abs/2013MNRAS.428.3680G} {428, 3680}

\bibitem[\protect\citeauthoryear{{Gillon} et~al.,}{{Gillon}
  et~al.}{2007}]{Gillon:2007aa}
{Gillon} M.,  et~al., 2007, \mn@doi [\aap] {10.1051/0004-6361:20077799}, \href
  {http://adsabs.harvard.edu/abs/2007A%26A...472L..13G} {472, L13}

\bibitem[\protect\citeauthoryear{{Gillon} et~al.,}{{Gillon}
  et~al.}{2009}]{Gillon:2009aa}
{Gillon} M.,  et~al., 2009, \mn@doi [\aap] {10.1051/0004-6361/200911749}, \href
  {http://adsabs.harvard.edu/abs/2009A%26A...501..785G} {501, 785}

\bibitem[\protect\citeauthoryear{{Gillon} et~al.,}{{Gillon}
  et~al.}{2012}]{Gillon:2012aa}
{Gillon} M.,  et~al., 2012, \mn@doi [\aap] {10.1051/0004-6361/201218817}, \href
  {http://adsabs.harvard.edu/abs/2012A%26A...542A...4G} {542, A4}

\bibitem[\protect\citeauthoryear{{Gillon} et~al.,}{{Gillon}
  et~al.}{2014}]{Gillon:2014aa}
{Gillon} M.,  et~al., 2014, \mn@doi [\aap] {10.1051/0004-6361/201323014}, \href
  {http://adsabs.harvard.edu/abs/2014A%26A...562L...3G} {562, L3}

\bibitem[\protect\citeauthoryear{{Goody}, {West}, {Chen}  \& {Crisp}}{{Goody}
  et~al.}{1989}]{Goody1989}
{Goody} R.,  {West} R.,  {Chen} L.,   {Crisp} D.,  1989, \mn@doi [\jqsrt]
  {10.1016/0022-4073(89)90044-7}, \href
  {http://adsabs.harvard.edu/abs/1989JQSRT..42..539G} {42, 539}

\bibitem[\protect\citeauthoryear{{Gordon} \& {McBride}}{{Gordon} \&
  {McBride}}{1994}]{Gordon1994}
{Gordon} S.,  {McBride} B.~J.,  1994, NASA Reference Publication, 1311

\bibitem[\protect\citeauthoryear{{Greene}, {Line}, {Montero}, {Fortney},
  {Lustig-Yaeger}  \& {Luther}}{{Greene} et~al.}{2016}]{Greene2016}
{Greene} T.~P.,  {Line} M.~R.,  {Montero} C.,  {Fortney} J.~J.,
  {Lustig-Yaeger} J.,   {Luther} K.,  2016, \mn@doi [\apj]
  {10.3847/0004-637X/817/1/17}, \href
  {http://adsabs.harvard.edu/abs/2016ApJ...817...17G} {817, 17}

\bibitem[\protect\citeauthoryear{{Hadded}, {Aroui}, {Orphal}, {Bouanich}  \&
  {Hartmann}}{{Hadded} et~al.}{2001}]{Hadded2001}
{Hadded} S.,  {Aroui} H.,  {Orphal} J.,  {Bouanich} J.-P.,   {Hartmann} J.-M.,
  2001, \mn@doi [Journal of Molecular Spectroscopy] {10.1006/jmsp.2001.8452},
  \href {http://adsabs.harvard.edu/abs/2001JMoSp.210..275H} {210, 275}

\bibitem[\protect\citeauthoryear{{Harps{\o}e} et~al.,}{{Harps{\o}e}
  et~al.}{2013}]{Harpse:2013aa}
{Harps{\o}e} K.~B.~W.,  et~al., 2013, \mn@doi [\aap]
  {10.1051/0004-6361/201219996}, \href
  {http://adsabs.harvard.edu/abs/2013A%26A...549A..10H} {549, A10}

\bibitem[\protect\citeauthoryear{{Harris}, {Tennyson}, {Kaminsky}, {Pavlenko}
  \& {Jones}}{{Harris} et~al.}{2006}]{Harris2006}
{Harris} G.~J.,  {Tennyson} J.,  {Kaminsky} B.~M.,  {Pavlenko} Y.~V.,   {Jones}
  H.~R.~A.,  2006, \mn@doi [\mnras] {10.1111/j.1365-2966.2005.09960.x}, \href
  {http://adsabs.harvard.edu/abs/2006MNRAS.367..400H} {367, 400}

\bibitem[\protect\citeauthoryear{{Hartman} et~al.,}{{Hartman}
  et~al.}{2009}]{Hartman:2009aa}
{Hartman} J.~D.,  et~al., 2009, \mn@doi [\apj] {10.1088/0004-637X/706/1/785},
  \href {http://adsabs.harvard.edu/abs/2009ApJ...706..785H} {706, 785}

\bibitem[\protect\citeauthoryear{{Hartman} et~al.,}{{Hartman}
  et~al.}{2011a}]{Hartman:2011aa}
{Hartman} J.~D.,  et~al., 2011a, \mn@doi [\apj] {10.1088/0004-637X/726/1/52},
  \href {http://adsabs.harvard.edu/abs/2011ApJ...726...52H} {726, 52}

\bibitem[\protect\citeauthoryear{{Hartman} et~al.,}{{Hartman}
  et~al.}{2011b}]{Hartman:2011ab}
{Hartman} J.~D.,  et~al., 2011b, \mn@doi [\apj] {10.1088/0004-637X/728/2/138},
  \href {http://adsabs.harvard.edu/abs/2011ApJ...728..138H} {728, 138}

\bibitem[\protect\citeauthoryear{{Hartman} et~al.,}{{Hartman}
  et~al.}{2011c}]{Hartman:2011ac}
{Hartman} J.~D.,  et~al., 2011c, \mn@doi [\apj] {10.1088/0004-637X/742/1/59},
  \href {http://adsabs.harvard.edu/abs/2011ApJ...742...59H} {742, 59}

\bibitem[\protect\citeauthoryear{{Hartman} et~al.,}{{Hartman}
  et~al.}{2012}]{Hartman:2012aa}
{Hartman} J.~D.,  et~al., 2012, \mn@doi [\aj] {10.1088/0004-6256/144/5/139},
  \href {http://adsabs.harvard.edu/abs/2012AJ....144..139H} {144, 139}

\bibitem[\protect\citeauthoryear{{Hartman} et~al.,}{{Hartman}
  et~al.}{2014}]{Hartman:2014aa}
{Hartman} J.~D.,  et~al., 2014, \mn@doi [\aj] {10.1088/0004-6256/147/6/128},
  \href {http://adsabs.harvard.edu/abs/2014AJ....147..128H} {147, 128}

\bibitem[\protect\citeauthoryear{{Hartman} et~al.,}{{Hartman}
  et~al.}{2015a}]{Hartman:2015ab}
{Hartman} J.~D.,  et~al., 2015a, \mn@doi [\aj] {10.1088/0004-6256/149/5/166},
  \href {http://adsabs.harvard.edu/abs/2015AJ....149..166H} {149, 166}

\bibitem[\protect\citeauthoryear{{Hartman} et~al.,}{{Hartman}
  et~al.}{2015b}]{Hartman:2015aa}
{Hartman} J.~D.,  et~al., 2015b, \mn@doi [\aj] {10.1088/0004-6256/150/6/168},
  \href {http://adsabs.harvard.edu/abs/2015AJ....150..168H} {150, 168}

\bibitem[\protect\citeauthoryear{{Hartman} et~al.,}{{Hartman}
  et~al.}{2016}]{Hartman:2016aa}
{Hartman} J.~D.,  et~al., 2016, \mn@doi [\aj] {10.3847/0004-6256/152/6/182},
  \href {http://adsabs.harvard.edu/abs/2016AJ....152..182H} {152, 182}

\bibitem[\protect\citeauthoryear{{Hay} et~al.,}{{Hay}
  et~al.}{2016}]{Hay:2016aa}
{Hay} K.~L.,  et~al., 2016, \mn@doi [\mnras] {10.1093/mnras/stw2090}, \href
  {http://adsabs.harvard.edu/abs/2016MNRAS.463.3276H} {463, 3276}

\bibitem[\protect\citeauthoryear{{Hebb} et~al.,}{{Hebb}
  et~al.}{2009}]{Hebb:2009aa}
{Hebb} L.,  et~al., 2009, \mn@doi [\apj] {10.1088/0004-637X/693/2/1920}, \href
  {http://adsabs.harvard.edu/abs/2009ApJ...693.1920H} {693, 1920}

\bibitem[\protect\citeauthoryear{{Hebb} et~al.,}{{Hebb}
  et~al.}{2010}]{Hebb:2010aa}
{Hebb} L.,  et~al., 2010, \mn@doi [\apj] {10.1088/0004-637X/708/1/224}, \href
  {http://adsabs.harvard.edu/abs/2010ApJ...708..224H} {708, 224}

\bibitem[\protect\citeauthoryear{{H{\'e}brard} et~al.,}{{H{\'e}brard}
  et~al.}{2013}]{Hebrard:2013aa}
{H{\'e}brard} G.,  et~al., 2013, \mn@doi [\aap] {10.1051/0004-6361/201220363},
  \href {http://adsabs.harvard.edu/abs/2013A%26A...549A.134H} {549, A134}

\bibitem[\protect\citeauthoryear{{Hedges} \& {Madhusudhan}}{{Hedges} \&
  {Madhusudhan}}{2016}]{Hedges2016}
{Hedges} C.,  {Madhusudhan} N.,  2016, \mn@doi [\mnras] {10.1093/mnras/stw278},
  \href {http://adsabs.harvard.edu/abs/2016MNRAS.458.1427H} {458, 1427}

\bibitem[\protect\citeauthoryear{{Heiter} et~al.,}{{Heiter}
  et~al.}{2008}]{Heiter2008}
{Heiter} U.,  et~al., 2008, in Journal of Physics Conference Series. p. 012011,
  \mn@doi{10.1088/1742-6596/130/1/012011}

\bibitem[\protect\citeauthoryear{{Hellier} et~al.,}{{Hellier}
  et~al.}{2009}]{Hellier:2009aa}
{Hellier} C.,  et~al., 2009, \mn@doi [\apjl] {10.1088/0004-637X/690/1/L89},
  \href {http://adsabs.harvard.edu/abs/2009ApJ...690L..89H} {690, L89}

\bibitem[\protect\citeauthoryear{{Hellier} et~al.,}{{Hellier}
  et~al.}{2010}]{Hellier:2010aa}
{Hellier} C.,  et~al., 2010, \mn@doi [\apjl] {10.1088/2041-8205/723/1/L60},
  \href {http://adsabs.harvard.edu/abs/2010ApJ...723L..60H} {723, L60}

\bibitem[\protect\citeauthoryear{{Hellier} et~al.,}{{Hellier}
  et~al.}{2011}]{Hellier:2011aa}
{Hellier} C.,  et~al., 2011, \mn@doi [\aap] {10.1051/0004-6361/201117081},
  \href {http://adsabs.harvard.edu/abs/2011A%26A...535L...7H} {535, L7}

\bibitem[\protect\citeauthoryear{{Hellier} et~al.,}{{Hellier}
  et~al.}{2012}]{Hellier:2012aa}
{Hellier} C.,  et~al., 2012, \mn@doi [\mnras]
  {10.1111/j.1365-2966.2012.21780.x}, \href
  {http://adsabs.harvard.edu/abs/2012MNRAS.426..739H} {426, 739}

\bibitem[\protect\citeauthoryear{{Hellier} et~al.,}{{Hellier}
  et~al.}{2014}]{Hellier:2014aa}
{Hellier} C.,  et~al., 2014, \mn@doi [\mnras] {10.1093/mnras/stu410}, \href
  {http://adsabs.harvard.edu/abs/2014MNRAS.440.1982H} {440, 1982}

\bibitem[\protect\citeauthoryear{{Hellier} et~al.,}{{Hellier}
  et~al.}{2015}]{Hellier:2015aa}
{Hellier} C.,  et~al., 2015, \mn@doi [\aj] {10.1088/0004-6256/150/1/18}, \href
  {http://adsabs.harvard.edu/abs/2015AJ....150...18H} {150, 18}

\bibitem[\protect\citeauthoryear{{Hellier} et~al.,}{{Hellier}
  et~al.}{2017}]{Hellier:2017aa}
{Hellier} C.,  et~al., 2017, \mn@doi [\mnras] {10.1093/mnras/stw3005}, \href
  {http://adsabs.harvard.edu/abs/2017MNRAS.465.3693H} {465, 3693}

\bibitem[\protect\citeauthoryear{{Heng} \& {Kitzmann}}{{Heng} \&
  {Kitzmann}}{2017}]{Heng2017}
{Heng} K.,  {Kitzmann} D.,  2017, preprint, \href
  {http://adsabs.harvard.edu/abs/2017arXiv170202051H} {} (\mn@eprint {arXiv}
  {1702.02051})

\bibitem[\protect\citeauthoryear{{Heng} \& {Tsai}}{{Heng} \&
  {Tsai}}{2016}]{Heng2016}
{Heng} K.,  {Tsai} S.-M.,  2016, \mn@doi [\apj] {10.3847/0004-637X/829/2/104},
  \href {http://adsabs.harvard.edu/abs/2016ApJ...829..104H} {829, 104}

\bibitem[\protect\citeauthoryear{{Henry}, {Marcy}, {Butler}  \& {Vogt}}{{Henry}
  et~al.}{2000}]{Henry:2000aa}
{Henry} G.~W.,  {Marcy} G.~W.,  {Butler} R.~P.,   {Vogt} S.~S.,  2000, \mn@doi
  [\apjl] {10.1086/312458}, \href
  {http://adsabs.harvard.edu/abs/2000ApJ...529L..41H} {529, L41}

\bibitem[\protect\citeauthoryear{{Howard} et~al.,}{{Howard}
  et~al.}{2012}]{Howard:2012aa}
{Howard} A.~W.,  et~al., 2012, \mn@doi [\apj] {10.1088/0004-637X/749/2/134},
  \href {http://adsabs.harvard.edu/abs/2012ApJ...749..134H} {749, 134}

\bibitem[\protect\citeauthoryear{{Hubbard}, {Fortney}, {Lunine}, {Burrows},
  {Sudarsky}  \& {Pinto}}{{Hubbard} et~al.}{2001}]{Hubbard2001}
{Hubbard} W.~B.,  {Fortney} J.~J.,  {Lunine} J.~I.,  {Burrows} A.,  {Sudarsky}
  D.,   {Pinto} P.,  2001, \mn@doi [\apj] {10.1086/322490}, \href
  {http://adsabs.harvard.edu/abs/2001ApJ...560..413H} {560, 413}

\bibitem[\protect\citeauthoryear{{Huitson}, {Sing}, {Vidal-Madjar},
  {Ballester}, {Lecavelier des Etangs}, {D{\'e}sert}  \& {Pont}}{{Huitson}
  et~al.}{2012}]{Huitson2012}
{Huitson} C.~M.,  {Sing} D.~K.,  {Vidal-Madjar} A.,  {Ballester} G.~E.,
  {Lecavelier des Etangs} A.,  {D{\'e}sert} J.-M.,   {Pont} F.,  2012, \mn@doi
  [\mnras] {10.1111/j.1365-2966.2012.20805.x}, \href
  {http://adsabs.harvard.edu/abs/2012MNRAS.422.2477H} {422, 2477}

\bibitem[\protect\citeauthoryear{{Huitson} et~al.,}{{Huitson}
  et~al.}{2013}]{Huitson2013}
{Huitson} C.~M.,  et~al., 2013, \mn@doi [\mnras] {10.1093/mnras/stt1243}, \href
  {http://adsabs.harvard.edu/abs/2013MNRAS.434.3252H} {434, 3252}

\bibitem[\protect\citeauthoryear{{Johnson} et~al.,}{{Johnson}
  et~al.}{2011}]{Johnson:2011aa}
{Johnson} J.~A.,  et~al., 2011, \mn@doi [\apj] {10.1088/0004-637X/735/1/24},
  \href {http://adsabs.harvard.edu/abs/2011ApJ...735...24J} {735, 24}

\bibitem[\protect\citeauthoryear{{Kataria}, {Sing}, {Lewis}, {Visscher},
  {Showman}, {Fortney}  \& {Marley}}{{Kataria} et~al.}{2016}]{Kataria2016}
{Kataria} T.,  {Sing} D.~K.,  {Lewis} N.~K.,  {Visscher} C.,  {Showman} A.~P.,
  {Fortney} J.~J.,   {Marley} M.~S.,  2016, \mn@doi [\apj]
  {10.3847/0004-637X/821/1/9}, \href
  {http://adsabs.harvard.edu/abs/2016ApJ...821....9K} {821, 9}

\bibitem[\protect\citeauthoryear{{Kopparapu}, {Kasting}  \&
  {Zahnle}}{{Kopparapu} et~al.}{2012}]{Kopparapu2012}
{Kopparapu} R.~k.,  {Kasting} J.~F.,   {Zahnle} K.~J.,  2012, \mn@doi [\apj]
  {10.1088/0004-637X/745/1/77}, \href
  {http://adsabs.harvard.edu/abs/2012ApJ...745...77K} {745, 77}

\bibitem[\protect\citeauthoryear{{Kov{\'a}cs} et~al.,}{{Kov{\'a}cs}
  et~al.}{2007}]{Kovacs:2007aa}
{Kov{\'a}cs} G.,  et~al., 2007, \mn@doi [\apjl] {10.1086/524058}, \href
  {http://adsabs.harvard.edu/abs/2007ApJ...670L..41K} {670, L41}

\bibitem[\protect\citeauthoryear{{Kreidberg} et~al.,}{{Kreidberg}
  et~al.}{2014}]{Kreidberg2014}
{Kreidberg} L.,  et~al., 2014, \mn@doi [\nat] {10.1038/nature12888}, \href
  {http://adsabs.harvard.edu/abs/2014Natur.505...69K} {505, 69}

\bibitem[\protect\citeauthoryear{{Kreidberg} et~al.,}{{Kreidberg}
  et~al.}{2015}]{Kreidberg2015}
{Kreidberg} L.,  et~al., 2015, \mn@doi [\apj] {10.1088/0004-637X/814/1/66},
  \href {http://adsabs.harvard.edu/abs/2015ApJ...814...66K} {814, 66}

\bibitem[\protect\citeauthoryear{{Kuhn} et~al.,}{{Kuhn}
  et~al.}{2016}]{Kuhn:2016aa}
{Kuhn} R.~B.,  et~al., 2016, \mn@doi [\mnras] {10.1093/mnras/stw880}, \href
  {http://adsabs.harvard.edu/abs/2016MNRAS.459.4281K} {459, 4281}

\bibitem[\protect\citeauthoryear{{Lacis} \& {Oinas}}{{Lacis} \&
  {Oinas}}{1991}]{Lacis1991}
{Lacis} A.~A.,  {Oinas} V.,  1991, \mn@doi [\jgr] {10.1029/90JD01945}, \href
  {http://adsabs.harvard.edu/abs/1991JGR....96.9027L} {96, 9027}

\bibitem[\protect\citeauthoryear{{Lam} et~al.,}{{Lam}
  et~al.}{2017}]{Lam:2017aa}
{Lam} K.~W.~F.,  et~al., 2017, \mn@doi [\aap] {10.1051/0004-6361/201629403},
  \href {http://adsabs.harvard.edu/abs/2017A%26A...599A...3L} {599, A3}

\bibitem[\protect\citeauthoryear{{Landrain}, {Blanquet}, {Lep{\`e}re},
  {Walrand}  \& {Bouanich}}{{Landrain} et~al.}{1997}]{Landrain1997}
{Landrain} V.,  {Blanquet} G.,  {Lep{\`e}re} M.,  {Walrand} J.,   {Bouanich}
  J.-P.,  1997, \mn@doi [Journal of Molecular Spectroscopy]
  {10.1006/jmsp.1996.7223}, \href
  {http://adsabs.harvard.edu/abs/1997JMoSp.182..184L} {182, 184}

\bibitem[\protect\citeauthoryear{{Lanotte} et~al.,}{{Lanotte}
  et~al.}{2014}]{Lanotte:2014aa}
{Lanotte} A.~A.,  et~al., 2014, \mn@doi [\aap] {10.1051/0004-6361/201424373},
  \href {http://adsabs.harvard.edu/abs/2014A%26A...572A..73L} {572, A73}

\bibitem[\protect\citeauthoryear{{Le Moal} \& {Severin}}{{Le Moal} \&
  {Severin}}{1986}]{Lemoal1986}
{Le Moal} M.~F.,  {Severin} F.,  1986, \jqsrt, \href
  {http://adsabs.harvard.edu/abs/1986JQSRT..35..145L} {35, 145}

\bibitem[\protect\citeauthoryear{{Lecavelier Des Etangs}, {Pont},
  {Vidal-Madjar}  \& {Sing}}{{Lecavelier Des Etangs}
  et~al.}{2008}]{Lecavelier2008}
{Lecavelier Des Etangs} A.,  {Pont} F.,  {Vidal-Madjar} A.,   {Sing} D.,  2008,
  \mn@doi [\aap] {10.1051/0004-6361:200809388}, \href
  {http://adsabs.harvard.edu/abs/2008A%26A...481L..83L} {481, L83}

\bibitem[\protect\citeauthoryear{{Lee}, {Youn}, {Kim}, {Lee}  \& {Hinse}}{{Lee}
  et~al.}{2012}]{Lee:2012aa}
{Lee} J.~W.,  {Youn} J.-H.,  {Kim} S.-L.,  {Lee} C.-U.,   {Hinse} T.~C.,  2012,
  \mn@doi [\aj] {10.1088/0004-6256/143/4/95}, \href
  {http://adsabs.harvard.edu/abs/2012AJ....143...95L} {143, 95}

\bibitem[\protect\citeauthoryear{{Lehmann}, {Guenther}, {Sebastian},
  {D{\"o}llinger}, {Hartmann}  \& {Mkrtichian}}{{Lehmann}
  et~al.}{2015}]{Lehmann:2015aa}
{Lehmann} H.,  {Guenther} E.,  {Sebastian} D.,  {D{\"o}llinger} M.,  {Hartmann}
  M.,   {Mkrtichian} D.~E.,  2015, \mn@doi [\aap]
  {10.1051/0004-6361/201526176}, \href
  {http://adsabs.harvard.edu/abs/2015A%26A...578L...4L} {578, L4}

\bibitem[\protect\citeauthoryear{{Lendl} et~al.,}{{Lendl}
  et~al.}{2012}]{Lendl:2012aa}
{Lendl} M.,  et~al., 2012, \mn@doi [\aap] {10.1051/0004-6361/201219585}, \href
  {http://adsabs.harvard.edu/abs/2012A%26A...544A..72L} {544, A72}

\bibitem[\protect\citeauthoryear{{Lendl} et~al.,}{{Lendl}
  et~al.}{2014}]{Lendl:2014aa}
{Lendl} M.,  et~al., 2014, \mn@doi [\aap] {10.1051/0004-6361/201424481}, \href
  {http://adsabs.harvard.edu/abs/2014A%26A...568A..81L} {568, A81}

\bibitem[\protect\citeauthoryear{{Lendl} et~al.,}{{Lendl}
  et~al.}{2016}]{Lendl:2016aa}
{Lendl} M.,  et~al., 2016, \mn@doi [\aap] {10.1051/0004-6361/201527594}, \href
  {http://adsabs.harvard.edu/abs/2016A%26A...587A..67L} {587, A67}

\bibitem[\protect\citeauthoryear{{Leonard}}{{Leonard}}{1974}]{Leonard1974}
{Leonard} P.~J.,  1974, \mn@doi [Atomic Data and Nuclear Data Tables]
  {10.1016/S0092-640X(74)80028-8}, \href
  {http://adsabs.harvard.edu/abs/1974ADNDT..14...21L} {14, 21}

\bibitem[\protect\citeauthoryear{{Levy}, {Lacome}  \& {Tarrago}}{{Levy}
  et~al.}{1994}]{Levy1994}
{Levy} A.,  {Lacome} N.,   {Tarrago} G.,  1994, \mn@doi [Journal of Molecular
  Spectroscopy] {10.1006/jmsp.1994.1168}, \href
  {http://adsabs.harvard.edu/abs/1994JMoSp.166...20L} {166, 20}

\bibitem[\protect\citeauthoryear{{Line} \& {Parmentier}}{{Line} \&
  {Parmentier}}{2016}]{Lineparmentier2016}
{Line} M.~R.,  {Parmentier} V.,  2016, \mn@doi [\apj]
  {10.3847/0004-637X/820/1/78}, \href
  {http://adsabs.harvard.edu/abs/2016ApJ...820...78L} {820, 78}

\bibitem[\protect\citeauthoryear{{Liou}}{{Liou}}{1980}]{Liou1980}
{Liou} K.~N.,  1980, {An introduction to atmospheric radiation.}

\bibitem[\protect\citeauthoryear{{Lister} et~al.,}{{Lister}
  et~al.}{2009}]{Lister:2009aa}
{Lister} T.~A.,  et~al., 2009, \mn@doi [\apj] {10.1088/0004-637X/703/1/752},
  \href {http://adsabs.harvard.edu/abs/2009ApJ...703..752L} {703, 752}

\bibitem[\protect\citeauthoryear{{Lodders} \& {Fegley}}{{Lodders} \&
  {Fegley}}{2006}]{Lodders2006}
{Lodders} K.,  {Fegley} Jr. B.,  2006, {Chemistry of Low Mass Substellar
  Objects}.
p.~1, \mn@doi{10.1007/3-540-30313-8_1}

\bibitem[\protect\citeauthoryear{{Maciejewski} et~al.,}{{Maciejewski}
  et~al.}{2014}]{Maciejewski:2014aa}
{Maciejewski} G.,  et~al., 2014, \actaa, \href
  {http://adsabs.harvard.edu/abs/2014AcA....64...27M} {64, 11}

\bibitem[\protect\citeauthoryear{{Madhusudhan}}{{Madhusudhan}}{2012}]{Madhusudhan2012}
{Madhusudhan} N.,  2012, \mn@doi [\apj] {10.1088/0004-637X/758/1/36}, \href
  {http://adsabs.harvard.edu/abs/2012ApJ...758...36M} {758, 36}

\bibitem[\protect\citeauthoryear{{Madhusudhan} \& {Seager}}{{Madhusudhan} \&
  {Seager}}{2009}]{Madhusudhan2009}
{Madhusudhan} N.,  {Seager} S.,  2009, \mn@doi [\apj]
  {10.1088/0004-637X/707/1/24}, \href
  {http://adsabs.harvard.edu/abs/2009ApJ...707...24M} {707, 24}

\bibitem[\protect\citeauthoryear{{Madhusudhan} et~al.,}{{Madhusudhan}
  et~al.}{2011}]{Madhusudhan2011a}
{Madhusudhan} N.,  et~al., 2011, \mn@doi [\nat] {10.1038/nature09602}, \href
  {http://adsabs.harvard.edu/abs/2011Natur.469...64M} {469, 64}

\bibitem[\protect\citeauthoryear{{Madhusudhan}, {Bitsch}, {Johansen}  \&
  {Eriksson}}{{Madhusudhan} et~al.}{2016}]{Madhusudhan2016}
{Madhusudhan} N.,  {Bitsch} B.,  {Johansen} A.,   {Eriksson} L.,  2016,
  preprint, \href {http://adsabs.harvard.edu/abs/2016arXiv161103083M} {}
  (\mn@eprint {arXiv} {1611.03083})

\bibitem[\protect\citeauthoryear{{Malik} et~al.,}{{Malik}
  et~al.}{2017}]{Malik2017}
{Malik} M.,  et~al., 2017, \mn@doi [\aj] {10.3847/1538-3881/153/2/56}, \href
  {http://adsabs.harvard.edu/abs/2017AJ....153...56M} {153, 56}

\bibitem[\protect\citeauthoryear{{Mancini} et~al.,}{{Mancini}
  et~al.}{2013}]{Mancini:2013aa}
{Mancini} L.,  et~al., 2013, \mn@doi [\mnras] {10.1093/mnras/stt1394}, \href
  {http://adsabs.harvard.edu/abs/2013MNRAS.436....2M} {436, 2}

\bibitem[\protect\citeauthoryear{{Mancini} et~al.,}{{Mancini}
  et~al.}{2014a}]{Mancini:2014ab}
{Mancini} L.,  et~al., 2014a, \mn@doi [\aap] {10.1051/0004-6361/201323265},
  \href {http://adsabs.harvard.edu/abs/2014A%26A...562A.126M} {562, A126}

\bibitem[\protect\citeauthoryear{{Mancini} et~al.,}{{Mancini}
  et~al.}{2014b}]{Mancini:2014aa}
{Mancini} L.,  et~al., 2014b, \mn@doi [\aap] {10.1051/0004-6361/201424106},
  \href {http://adsabs.harvard.edu/abs/2014A%26A...568A.127M} {568, A127}

\bibitem[\protect\citeauthoryear{{Mancini} et~al.,}{{Mancini}
  et~al.}{2015}]{Mancini:2015aa}
{Mancini} L.,  et~al., 2015, \mn@doi [\aap] {10.1051/0004-6361/201526030},
  \href {http://adsabs.harvard.edu/abs/2015A%26A...579A.136M} {579, A136}

\bibitem[\protect\citeauthoryear{{Mancini} et~al.,}{{Mancini}
  et~al.}{2017}]{Mancini:2017aa}
{Mancini} L.,  et~al., 2017, \mn@doi [\mnras] {10.1093/mnras/stw1987}, \href
  {http://adsabs.harvard.edu/abs/2017MNRAS.465..843M} {465, 843}

\bibitem[\protect\citeauthoryear{{Mandell}, {Haynes}, {Sinukoff},
  {Madhusudhan}, {Burrows}  \& {Deming}}{{Mandell} et~al.}{2013}]{Mandell2013}
{Mandell} A.~M.,  {Haynes} K.,  {Sinukoff} E.,  {Madhusudhan} N.,  {Burrows}
  A.,   {Deming} D.,  2013, \mn@doi [\apj] {10.1088/0004-637X/779/2/128}, \href
  {http://adsabs.harvard.edu/abs/2013ApJ...779..128M} {779, 128}

\bibitem[\protect\citeauthoryear{{Mandushev} et~al.,}{{Mandushev}
  et~al.}{2007}]{Mandushev:2007aa}
{Mandushev} G.,  et~al., 2007, \mn@doi [\apjl] {10.1086/522115}, \href
  {http://adsabs.harvard.edu/abs/2007ApJ...667L.195M} {667, L195}

\bibitem[\protect\citeauthoryear{{Mansfield} \& {Peck}}{{Mansfield} \&
  {Peck}}{1969}]{Mansfield1969}
{Mansfield} C.~R.,  {Peck} E.~R.,  1969, Journal of the Optical Society of
  America (1917-1983), \href
  {http://adsabs.harvard.edu/abs/1969JOSA...59..199M} {59, 199}

\bibitem[\protect\citeauthoryear{{Mantz}, {Malathy Devi}, {Chris Benner},
  {Smith}, {Predoi-Cross}  \& {Dulick}}{{Mantz} et~al.}{2005}]{Mantz2005}
{Mantz} A.~W.,  {Malathy Devi} V.,  {Chris Benner} D.,  {Smith} M.~A.~H.,
  {Predoi-Cross} A.,   {Dulick} M.,  2005, \mn@doi [Journal of Molecular
  Structure] {10.1016/j.molstruc.2004.11.094}, \href
  {http://adsabs.harvard.edu/abs/2005JMoSt.742...99M} {742, 99}

\bibitem[\protect\citeauthoryear{{Margolis}}{{Margolis}}{1993}]{Margolis1993}
{Margolis} J.~S.,  1993, \mn@doi [\jqsrt] {10.1016/0022-4073(93)90073-Q}, \href
  {http://adsabs.harvard.edu/abs/1993JQSRT..50..431M} {50, 431}

\bibitem[\protect\citeauthoryear{{Maxted} et~al.,}{{Maxted}
  et~al.}{2011}]{Maxted:2011aa}
{Maxted} P.~F.~L.,  et~al., 2011, \mn@doi [\pasp] {10.1086/660007}, \href
  {http://adsabs.harvard.edu/abs/2011PASP..123..547M} {123, 547}

\bibitem[\protect\citeauthoryear{{Maxted} et~al.,}{{Maxted}
  et~al.}{2016}]{Maxted:2016aa}
{Maxted} P.~F.~L.,  et~al., 2016, \mn@doi [\aap] {10.1051/0004-6361/201628250},
  \href {http://adsabs.harvard.edu/abs/2016A%26A...591A..55M} {591, A55}

\bibitem[\protect\citeauthoryear{{Mayor} \& {Queloz}}{{Mayor} \&
  {Queloz}}{1995}]{Mayor1995}
{Mayor} M.,  {Queloz} D.,  1995, \mn@doi [\nat] {10.1038/378355a0}, \href
  {http://adsabs.harvard.edu/abs/1995Natur.378..355M} {378, 355}

\bibitem[\protect\citeauthoryear{{Mbarek} \& {Kempton}}{{Mbarek} \&
  {Kempton}}{2016}]{Mbarek2016}
{Mbarek} R.,  {Kempton} E.~M.-R.,  2016, \mn@doi [\apj]
  {10.3847/0004-637X/827/2/121}, \href
  {http://adsabs.harvard.edu/abs/2016ApJ...827..121M} {827, 121}

\bibitem[\protect\citeauthoryear{{McBride}, {Gordon}  \& {Reno}}{{McBride}
  et~al.}{1993}]{McBride1993}
{McBride} B.~J.,  {Gordon} S.,   {Reno} M.~A.,  1993, NASA Technical
  Memorandum, 4513

\bibitem[\protect\citeauthoryear{{McBride}, {Zehe}  \& {Gordon}}{{McBride}
  et~al.}{2002}]{McBride2002}
{McBride} B.~J.,  {Zehe} M.~J.,   {Gordon} S.,  2002, NASA/TP, 2002-211556

\bibitem[\protect\citeauthoryear{{McCullough} et~al.,}{{McCullough}
  et~al.}{2006}]{McCullough:2006aa}
{McCullough} P.~R.,  et~al., 2006, \mn@doi [\apj] {10.1086/505651}, \href
  {http://adsabs.harvard.edu/abs/2006ApJ...648.1228M} {648, 1228}

\bibitem[\protect\citeauthoryear{{McCullough}, {Crouzet}, {Deming}  \&
  {Madhusudhan}}{{McCullough} et~al.}{2014}]{McCullough2014}
{McCullough} P.~R.,  {Crouzet} N.,  {Deming} D.,   {Madhusudhan} N.,  2014,
  \mn@doi [\apj] {10.1088/0004-637X/791/1/55}, \href
  {http://adsabs.harvard.edu/abs/2014ApJ...791...55M} {791, 55}

\bibitem[\protect\citeauthoryear{{McKemmish}, {Yurchenko}  \&
  {Tennyson}}{{McKemmish} et~al.}{2016}]{Volinelist2016}
{McKemmish} L.~K.,  {Yurchenko} S.~N.,   {Tennyson} J.,  2016, \mn@doi [\mnras]
  {10.1093/mnras/stw1969}, \href
  {http://adsabs.harvard.edu/abs/2016MNRAS.463..771M} {463, 771}

\bibitem[\protect\citeauthoryear{{Molli{\`e}re}, {van Boekel}, {Dullemond},
  {Henning}  \& {Mordasini}}{{Molli{\`e}re} et~al.}{2015}]{Molliere2015}
{Molli{\`e}re} P.,  {van Boekel} R.,  {Dullemond} C.,  {Henning} T.,
  {Mordasini} C.,  2015, \mn@doi [\apj] {10.1088/0004-637X/813/1/47}, \href
  {http://adsabs.harvard.edu/abs/2015ApJ...813...47M} {813, 47}

\bibitem[\protect\citeauthoryear{{Molli{\`e}re}, {van Boekel}, {Bouwman},
  {Henning}, {Lagage}  \& {Min}}{{Molli{\`e}re} et~al.}{2016}]{Molliere2016}
{Molli{\`e}re} P.,  {van Boekel} R.,  {Bouwman} J.,  {Henning} T.,  {Lagage}
  P.-O.,   {Min} M.,  2016, preprint, \href
  {http://adsabs.harvard.edu/abs/2016arXiv161108608M} {} (\mn@eprint {arXiv}
  {1611.08608})

\bibitem[\protect\citeauthoryear{{Morley}, {Fortney}, {Marley}, {Zahnle},
  {Line}, {Kempton}, {Lewis}  \& {Cahoy}}{{Morley} et~al.}{2015}]{Morley2015}
{Morley} C.~V.,  {Fortney} J.~J.,  {Marley} M.~S.,  {Zahnle} K.,  {Line} M.,
  {Kempton} E.,  {Lewis} N.,   {Cahoy} K.,  2015, \mn@doi [\apj]
  {10.1088/0004-637X/815/2/110}, \href
  {http://adsabs.harvard.edu/abs/2015ApJ...815..110M} {815, 110}

\bibitem[\protect\citeauthoryear{{Moses} et~al.,}{{Moses}
  et~al.}{2011}]{Moses2011}
{Moses} J.~I.,  et~al., 2011, \mn@doi [\apj] {10.1088/0004-637X/737/1/15},
  \href {http://adsabs.harvard.edu/abs/2011ApJ...737...15M} {737, 15}

\bibitem[\protect\citeauthoryear{{Moses}, {Madhusudhan}, {Visscher}  \&
  {Freedman}}{{Moses} et~al.}{2013a}]{Moses2013}
{Moses} J.~I.,  {Madhusudhan} N.,  {Visscher} C.,   {Freedman} R.~S.,  2013a,
  \mn@doi [\apj] {10.1088/0004-637X/763/1/25}, \href
  {http://adsabs.harvard.edu/abs/2013ApJ...763...25M} {763, 25}

\bibitem[\protect\citeauthoryear{{Moses} et~al.,}{{Moses}
  et~al.}{2013b}]{Moses2013b}
{Moses} J.~I.,  et~al., 2013b, \mn@doi [\apj] {10.1088/0004-637X/777/1/34},
  \href {http://adsabs.harvard.edu/abs/2013ApJ...777...34M} {777, 34}

\bibitem[\protect\citeauthoryear{{Neveu-VanMalle} et~al.,}{{Neveu-VanMalle}
  et~al.}{2014}]{Neveu-VanMalle:2014aa}
{Neveu-VanMalle} M.,  et~al., 2014, \mn@doi [\aap]
  {10.1051/0004-6361/201424744}, \href
  {http://adsabs.harvard.edu/abs/2014A%26A...572A..49N} {572, A49}

\bibitem[\protect\citeauthoryear{{Niemann} et~al.,}{{Niemann}
  et~al.}{1998}]{Niemann1998}
{Niemann} H.~B.,  et~al., 1998, \mn@doi [\jgr] {10.1029/98JE01050}, \href
  {http://adsabs.harvard.edu/abs/1998JGR...10322831N} {103, 22831}

\bibitem[\protect\citeauthoryear{{Nikolov} et~al.,}{{Nikolov}
  et~al.}{2014}]{Nikolov:2014aa}
{Nikolov} N.,  et~al., 2014, \mn@doi [\mnras] {10.1093/mnras/stt1859}, \href
  {http://adsabs.harvard.edu/abs/2014MNRAS.437...46N} {437, 46}

\bibitem[\protect\citeauthoryear{{Nikolov} et~al.,}{{Nikolov}
  et~al.}{2015}]{Nikolov2015}
{Nikolov} N.,  et~al., 2015, \mn@doi [\mnras] {10.1093/mnras/stu2433}, \href
  {http://adsabs.harvard.edu/abs/2015MNRAS.447..463N} {447, 463}

\bibitem[\protect\citeauthoryear{{Nouri}, {Orphal}, {Aroui}  \&
  {Hartmann}}{{Nouri} et~al.}{2004}]{Nouri2004}
{Nouri} S.,  {Orphal} J.,  {Aroui} H.,   {Hartmann} J.-M.,  2004, \mn@doi
  [Journal of Molecular Spectroscopy] {10.1016/j.jms.2004.05.009}, \href
  {http://adsabs.harvard.edu/abs/2004JMoSp.227...60N} {227, 60}

\bibitem[\protect\citeauthoryear{{Noyes} et~al.,}{{Noyes}
  et~al.}{2008}]{Noyes:2008aa}
{Noyes} R.~W.,  et~al., 2008, \mn@doi [\apjl] {10.1086/527358}, \href
  {http://adsabs.harvard.edu/abs/2008ApJ...673L..79N} {673, L79}

\bibitem[\protect\citeauthoryear{{{\"O}berg}, {Murray-Clay}  \&
  {Bergin}}{{{\"O}berg} et~al.}{2011}]{Oberg2011}
{{\"O}berg} K.~I.,  {Murray-Clay} R.,   {Bergin} E.~A.,  2011, \mn@doi [\apjl]
  {10.1088/2041-8205/743/1/L16}, \href
  {http://adsabs.harvard.edu/abs/2011ApJ...743L..16O} {743, L16}

\bibitem[\protect\citeauthoryear{{Padmanabhan}, {Tzanetakis}, {Chanda}  \&
  {Thomson}}{{Padmanabhan} et~al.}{2014}]{Padmanabhan2014}
{Padmanabhan} A.,  {Tzanetakis} T.,  {Chanda} A.,   {Thomson} M.~J.,  2014,
  \mn@doi [\jqsrt] {10.1016/j.jqsrt.2013.07.016}, \href
  {http://adsabs.harvard.edu/abs/2014JQSRT.133...81P} {133, 81}

\bibitem[\protect\citeauthoryear{{Penndorf}}{{Penndorf}}{1957}]{Penndorf1957}
{Penndorf} R.,  1957, Journal of the Optical Society of America (1917-1983),
  \href {http://adsabs.harvard.edu/abs/1957JOSA...47..176P} {47, 176}

\bibitem[\protect\citeauthoryear{{Pepper} et~al.,}{{Pepper}
  et~al.}{2017}]{Pepper:2017aa}
{Pepper} J.,  et~al., 2017, \mn@doi [\aj] {10.3847/1538-3881/aa6572}, \href
  {http://adsabs.harvard.edu/abs/2017AJ....153..215P} {153, 215}

\bibitem[\protect\citeauthoryear{{Pine}}{{Pine}}{1992}]{Pine1992}
{Pine} A.~S.,  1992, \mn@doi [\jcp] {10.1063/1.463943}, \href
  {http://adsabs.harvard.edu/abs/1992JChPh..97..773P} {97, 773}

\bibitem[\protect\citeauthoryear{{Pine}, {Markov}, {Buffa}  \&
  {Tarrini}}{{Pine} et~al.}{1993}]{Pine1993}
{Pine} A.~S.,  {Markov} V.~N.,  {Buffa} G.,   {Tarrini} O.,  1993, \mn@doi
  [\jqsrt] {10.1016/0022-4073(93)90069-T}, \href
  {http://adsabs.harvard.edu/abs/1993JQSRT..50..337P} {50, 337}

\bibitem[\protect\citeauthoryear{{Plez}}{{Plez}}{1998}]{Plez1998}
{Plez} B.,  1998, \aap, \href
  {http://adsabs.harvard.edu/abs/1998A%26A...337..495P} {337, 495}

\bibitem[\protect\citeauthoryear{{Pont}, {Sing}, {Gibson}, {Aigrain}, {Henry}
  \& {Husnoo}}{{Pont} et~al.}{2013}]{Pont2013}
{Pont} F.,  {Sing} D.~K.,  {Gibson} N.~P.,  {Aigrain} S.,  {Henry} G.,
  {Husnoo} N.,  2013, \mn@doi [\mnras] {10.1093/mnras/stt651}, \href
  {http://adsabs.harvard.edu/abs/2013MNRAS.432.2917P} {432, 2917}

\bibitem[\protect\citeauthoryear{{Quinn} et~al.,}{{Quinn}
  et~al.}{2012}]{Quinn:2012aa}
{Quinn} S.~N.,  et~al., 2012, \mn@doi [\apj] {10.1088/0004-637X/745/1/80},
  \href {http://adsabs.harvard.edu/abs/2012ApJ...745...80Q} {745, 80}

\bibitem[\protect\citeauthoryear{{Rajpurohit}, {Reyl{\'e}}, {Allard},
  {Homeier}, {Schultheis}, {Bessell}  \& {Robin}}{{Rajpurohit}
  et~al.}{2013}]{Rajpurohit2013}
{Rajpurohit} A.~S.,  {Reyl{\'e}} C.,  {Allard} F.,  {Homeier} D.,  {Schultheis}
  M.,  {Bessell} M.~S.,   {Robin} A.~C.,  2013, \mn@doi [\aap]
  {10.1051/0004-6361/201321346}, \href
  {http://adsabs.harvard.edu/abs/2013A%26A...556A..15R} {556, A15}

\bibitem[\protect\citeauthoryear{{Rayleigh}}{{Rayleigh}}{1919}]{Rayleigh1919}
{Rayleigh} 1919, \mn@doi [\nat] {10.1038/104276c0}, \href
  {http://adsabs.harvard.edu/abs/1919Natur.104R.276R} {104, 276}

\bibitem[\protect\citeauthoryear{{Redfield}, {Endl}, {Cochran}  \&
  {Koesterke}}{{Redfield} et~al.}{2008}]{Redfield2008}
{Redfield} S.,  {Endl} M.,  {Cochran} W.~D.,   {Koesterke} L.,  2008, \mn@doi
  [\apjl] {10.1086/527475}, \href
  {http://adsabs.harvard.edu/abs/2008ApJ...673L..87R} {673, L87}

\bibitem[\protect\citeauthoryear{{R{\'e}galia-Jarlot}, {Thomas}, {von der
  Heyden}  \& {Barbe}}{{R{\'e}galia-Jarlot} et~al.}{2005}]{Regalia2005}
{R{\'e}galia-Jarlot} L.,  {Thomas} X.,  {von der Heyden} P.,   {Barbe} A.,
  2005, \mn@doi [\jqsrt] {10.1016/j.jqsrt.2004.05.042}, \href
  {http://adsabs.harvard.edu/abs/2005JQSRT..91..121R} {91, 121}

\bibitem[\protect\citeauthoryear{{Richard} et~al.,}{{Richard}
  et~al.}{2012}]{Ciahitranpaper2012}
{Richard} C.,  et~al., 2012, \mn@doi [\jqsrt] {10.1016/j.jqsrt.2011.11.004},
  \href {http://adsabs.harvard.edu/abs/2012JQSRT.113.1276R} {113, 1276}

\bibitem[\protect\citeauthoryear{{Ricker} et~al.,}{{Ricker}
  et~al.}{2014}]{Ricker2014}
{Ricker} G.~R.,  et~al., 2014, in Space Telescopes and Instrumentation 2014:
  Optical, Infrared, and Millimeter Wave. p. 914320 (\mn@eprint {arXiv}
  {1406.0151}), \mn@doi{10.1117/12.2063489}

\bibitem[\protect\citeauthoryear{{Rodriguez} et~al.,}{{Rodriguez}
  et~al.}{2016}]{Rodriguez:2016aa}
{Rodriguez} J.~E.,  et~al., 2016, \mn@doi [\aj] {10.3847/0004-6256/151/6/138},
  \href {http://adsabs.harvard.edu/abs/2016AJ....151..138R} {151, 138}

\bibitem[\protect\citeauthoryear{{Rothman} et~al.,}{{Rothman}
  et~al.}{2009}]{Rothman2009}
{Rothman} L.~S.,  et~al., 2009, \mn@doi [\jqsrt] {10.1016/j.jqsrt.2009.02.013},
  \href {http://adsabs.harvard.edu/abs/2009JQSRT.110..533R} {110, 533}

\bibitem[\protect\citeauthoryear{{Rothman} et~al.,}{{Rothman}
  et~al.}{2010}]{Rothman2010}
{Rothman} L.~S.,  et~al., 2010, \mn@doi [\jqsrt] {10.1016/j.jqsrt.2010.05.001},
  \href {http://adsabs.harvard.edu/abs/2010JQSRT.111.2139R} {111, 2139}

\bibitem[\protect\citeauthoryear{{Rothman} et~al.,}{{Rothman}
  et~al.}{2013}]{Rothman2013}
{Rothman} L.~S.,  et~al., 2013, \mn@doi [\jqsrt] {10.1016/j.jqsrt.2013.07.002},
  \href {http://adsabs.harvard.edu/abs/2013JQSRT.130....4R} {130, 4}

\bibitem[\protect\citeauthoryear{{Salem}, {Bouanich}, {Walrand}, {Aroui}  \&
  {Blanquet}}{{Salem} et~al.}{2005}]{Salem2005}
{Salem} J.,  {Bouanich} J.-P.,  {Walrand} J.,  {Aroui} H.,   {Blanquet} G.,
  2005, \mn@doi [Journal of Molecular Spectroscopy]
  {10.1016/j.jms.2005.04.014}, \href
  {http://adsabs.harvard.edu/abs/2005JMoSp.232..247S} {232, 247}

\bibitem[\protect\citeauthoryear{{Sato} et~al.,}{{Sato}
  et~al.}{2005}]{Sato:2005aa}
{Sato} B.,  et~al., 2005, \mn@doi [\apj] {10.1086/449306}, \href
  {http://adsabs.harvard.edu/abs/2005ApJ...633..465S} {633, 465}

\bibitem[\protect\citeauthoryear{{Sauval} \& {Tatum}}{{Sauval} \&
  {Tatum}}{1984}]{Sauval1984}
{Sauval} A.~J.,  {Tatum} J.~B.,  1984, \mn@doi [\apjs] {10.1086/190980}, \href
  {http://adsabs.harvard.edu/abs/1984ApJS...56..193S} {56, 193}

\bibitem[\protect\citeauthoryear{{Seager} \& {Sasselov}}{{Seager} \&
  {Sasselov}}{2000}]{Seager2000}
{Seager} S.,  {Sasselov} D.~D.,  2000, \mn@doi [\apj] {10.1086/309088}, \href
  {http://adsabs.harvard.edu/abs/2000ApJ...537..916S} {537, 916}

\bibitem[\protect\citeauthoryear{{Seager}, {Richardson}, {Hansen}, {Menou},
  {Cho}  \& {Deming}}{{Seager} et~al.}{2005}]{Seager2005}
{Seager} S.,  {Richardson} L.~J.,  {Hansen} B.~M.~S.,  {Menou} K.,  {Cho}
  J.~Y.-K.,   {Deming} D.,  2005, \mn@doi [\apj] {10.1086/444411}, \href
  {http://adsabs.harvard.edu/abs/2005ApJ...632.1122S} {632, 1122}

\bibitem[\protect\citeauthoryear{{Sharp} \& {Burrows}}{{Sharp} \&
  {Burrows}}{2007}]{Sharp2007}
{Sharp} C.~M.,  {Burrows} A.,  2007, \mn@doi [\apjs] {10.1086/508708}, \href
  {http://adsabs.harvard.edu/abs/2007ApJS..168..140S} {168, 140}

\bibitem[\protect\citeauthoryear{{Sing}, {Vidal-Madjar}, {Lecavelier des
  Etangs}, {D{\'e}sert}, {Ballester}  \& {Ehrenreich}}{{Sing}
  et~al.}{2008}]{Sing2008}
{Sing} D.~K.,  {Vidal-Madjar} A.,  {Lecavelier des Etangs} A.,  {D{\'e}sert}
  J.-M.,  {Ballester} G.,   {Ehrenreich} D.,  2008, \mn@doi [\apj]
  {10.1086/590076}, \href {http://adsabs.harvard.edu/abs/2008ApJ...686..667S}
  {686, 667}

\bibitem[\protect\citeauthoryear{{Sing} et~al.,}{{Sing}
  et~al.}{2011}]{Sing2011}
{Sing} D.~K.,  et~al., 2011, \mn@doi [\aap] {10.1051/0004-6361/201015579},
  \href {http://adsabs.harvard.edu/abs/2011A%26A...527A..73S} {527, A73}

\bibitem[\protect\citeauthoryear{{Sing} et~al.,}{{Sing}
  et~al.}{2013}]{Sing2013}
{Sing} D.~K.,  et~al., 2013, \mn@doi [\mnras] {10.1093/mnras/stt1782}, \href
  {http://adsabs.harvard.edu/abs/2013MNRAS.436.2956S} {436, 2956}

\bibitem[\protect\citeauthoryear{{Sing} et~al.,}{{Sing}
  et~al.}{2015}]{Sing2015}
{Sing} D.~K.,  et~al., 2015, \mn@doi [\mnras] {10.1093/mnras/stu2279}, \href
  {http://adsabs.harvard.edu/abs/2015MNRAS.446.2428S} {446, 2428}

\bibitem[\protect\citeauthoryear{{Sing} et~al.,}{{Sing}
  et~al.}{2016}]{Sing2016}
{Sing} D.~K.,  et~al., 2016, \mn@doi [\nat] {10.1038/nature16068}, \href
  {http://adsabs.harvard.edu/abs/2016Natur.529...59S} {529, 59}

\bibitem[\protect\citeauthoryear{{Skillen} et~al.,}{{Skillen}
  et~al.}{2009}]{Skillen:2009aa}
{Skillen} I.,  et~al., 2009, \mn@doi [\aap] {10.1051/0004-6361/200912018},
  \href {http://adsabs.harvard.edu/abs/2009A%26A...502..391S} {502, 391}

\bibitem[\protect\citeauthoryear{{Smalley} et~al.,}{{Smalley}
  et~al.}{2011}]{Smalley:2011aa}
{Smalley} B.,  et~al., 2011, \mn@doi [\aap] {10.1051/0004-6361/201015992},
  \href {http://adsabs.harvard.edu/abs/2011A%26A...526A.130S} {526, A130}

\bibitem[\protect\citeauthoryear{{Smalley} et~al.,}{{Smalley}
  et~al.}{2012}]{Smalley:2012aa}
{Smalley} B.,  et~al., 2012, \mn@doi [\aap] {10.1051/0004-6361/201219731},
  \href {http://adsabs.harvard.edu/abs/2012A%26A...547A..61S} {547, A61}

\bibitem[\protect\citeauthoryear{{Smith}}{{Smith}}{2015}]{Smith:2015aa}
{Smith} A.~M.~S.,  2015, \actaa, \href
  {http://adsabs.harvard.edu/abs/2015AcA....65..117S} {65}

\bibitem[\protect\citeauthoryear{{Sneep} \& {Ubachs}}{{Sneep} \&
  {Ubachs}}{2005}]{Sneep2005}
{Sneep} M.,  {Ubachs} W.,  2005, \mn@doi [\jqsrt]
  {10.1016/j.jqsrt.2004.07.025}, \href
  {http://adsabs.harvard.edu/abs/2005JQSRT..92..293S} {92, 293}

\bibitem[\protect\citeauthoryear{{Snellen}, {Albrecht}, {de Mooij}  \& {Le
  Poole}}{{Snellen} et~al.}{2008}]{Snellen2008}
{Snellen} I.~A.~G.,  {Albrecht} S.,  {de Mooij} E.~J.~W.,   {Le Poole} R.~S.,
  2008, \mn@doi [\aap] {10.1051/0004-6361:200809762}, \href
  {http://adsabs.harvard.edu/abs/2008A%26A...487..357S} {487, 357}

\bibitem[\protect\citeauthoryear{{Solodov} \& {Starikov}}{{Solodov} \&
  {Starikov}}{2009}]{Solodov2009}
{Solodov} A.~M.,  {Starikov} V.~I.,  2009, \mn@doi [Molecular Physics]
  {10.1080/00268970802698655}, \href
  {http://adsabs.harvard.edu/abs/2009MolPh.107...43S} {107, 43}

\bibitem[\protect\citeauthoryear{{Sousa-Silva}, {Al-Refaie}, {Tennyson}  \&
  {Yurchenko}}{{Sousa-Silva} et~al.}{2014}]{Ph3linelist2014}
{Sousa-Silva} C.,  {Al-Refaie} A.~F.,  {Tennyson} J.,   {Yurchenko} S.~N.,
  2014, VizieR Online Data Catalog, \href
  {http://adsabs.harvard.edu/abs/2014yCat..74462337S} {744}

\bibitem[\protect\citeauthoryear{{Southworth}}{{Southworth}}{2010}]{Southworth:2010aa}
{Southworth} J.,  2010, \mn@doi [\mnras] {10.1111/j.1365-2966.2010.17231.x},
  \href {http://adsabs.harvard.edu/abs/2010MNRAS.408.1689S} {408, 1689}

\bibitem[\protect\citeauthoryear{{Southworth}}{{Southworth}}{2011a}]{Tepcat2011}
{Southworth} J.,  2011a, \mn@doi [\mnras] {10.1111/j.1365-2966.2011.19399.x},
  \href {http://adsabs.harvard.edu/abs/2011MNRAS.417.2166S} {417, 2166}

\bibitem[\protect\citeauthoryear{{Southworth}}{{Southworth}}{2011b}]{Southworth:2011aa}
{Southworth} J.,  2011b, \mn@doi [\mnras] {10.1111/j.1365-2966.2011.19399.x},
  \href {http://adsabs.harvard.edu/abs/2011MNRAS.417.2166S} {417, 2166}

\bibitem[\protect\citeauthoryear{{Southworth}}{{Southworth}}{2012}]{Southworth:2012aa}
{Southworth} J.,  2012, \mn@doi [\mnras] {10.1111/j.1365-2966.2012.21756.x},
  \href {http://adsabs.harvard.edu/abs/2012MNRAS.426.1291S} {426, 1291}

\bibitem[\protect\citeauthoryear{{Southworth} \& {Evans}}{{Southworth} \&
  {Evans}}{2016}]{Southworth:2016ab}
{Southworth} J.,  {Evans} D.~F.,  2016, \mn@doi [\mnras]
  {10.1093/mnras/stw1943}, \href
  {http://adsabs.harvard.edu/abs/2016MNRAS.463...37S} {463, 37}

\bibitem[\protect\citeauthoryear{{Southworth}, {Bruni}, {Mancini}  \&
  {Gregorio}}{{Southworth} et~al.}{2012a}]{Southworth:2012ab}
{Southworth} J.,  {Bruni} I.,  {Mancini} L.,   {Gregorio} J.,  2012a, \mn@doi
  [\mnras] {10.1111/j.1365-2966.2011.20230.x}, \href
  {http://adsabs.harvard.edu/abs/2012MNRAS.420.2580S} {420, 2580}

\bibitem[\protect\citeauthoryear{{Southworth} et~al.,}{{Southworth}
  et~al.}{2012b}]{Southworth:2012ac}
{Southworth} J.,  et~al., 2012b, \mn@doi [\mnras]
  {10.1111/j.1365-2966.2012.21781.x}, \href
  {http://adsabs.harvard.edu/abs/2012MNRAS.426.1338S} {426, 1338}

\bibitem[\protect\citeauthoryear{{Southworth} et~al.,}{{Southworth}
  et~al.}{2013}]{Southworth:2013aa}
{Southworth} J.,  et~al., 2013, \mn@doi [\mnras] {10.1093/mnras/stt1089}, \href
  {http://adsabs.harvard.edu/abs/2013MNRAS.434.1300S} {434, 1300}

\bibitem[\protect\citeauthoryear{{Southworth} et~al.,}{{Southworth}
  et~al.}{2014}]{Southworth:2014aa}
{Southworth} J.,  et~al., 2014, \mn@doi [\mnras] {10.1093/mnras/stu1492}, \href
  {http://adsabs.harvard.edu/abs/2014MNRAS.444..776S} {444, 776}

\bibitem[\protect\citeauthoryear{{Southworth} et~al.,}{{Southworth}
  et~al.}{2016}]{Southworth:2016aa}
{Southworth} J.,  et~al., 2016, \mn@doi [\mnras] {10.1093/mnras/stw279}, \href
  {http://adsabs.harvard.edu/abs/2016MNRAS.457.4205S} {457, 4205}

\bibitem[\protect\citeauthoryear{{Sozzetti} et~al.,}{{Sozzetti}
  et~al.}{2015}]{Sozzetti:2015aa}
{Sozzetti} A.,  et~al., 2015, \mn@doi [\aap] {10.1051/0004-6361/201425570},
  \href {http://adsabs.harvard.edu/abs/2015A%26A...575L..15S} {575, L15}

\bibitem[\protect\citeauthoryear{{Spiegel}, {Silverio}  \& {Burrows}}{{Spiegel}
  et~al.}{2009}]{Spiegel2009}
{Spiegel} D.~S.,  {Silverio} K.,   {Burrows} A.,  2009, \mn@doi [\apj]
  {10.1088/0004-637X/699/2/1487}, \href
  {http://adsabs.harvard.edu/abs/2009ApJ...699.1487S} {699, 1487}

\bibitem[\protect\citeauthoryear{{Stevens} et~al.,}{{Stevens}
  et~al.}{2017}]{Stevens:2017aa}
{Stevens} D.~J.,  et~al., 2017, \mn@doi [\aj] {10.3847/1538-3881/aa5ffb}, \href
  {http://adsabs.harvard.edu/abs/2017AJ....153..178S} {153, 178}

\bibitem[\protect\citeauthoryear{{Stevenson} et~al.,}{{Stevenson}
  et~al.}{2014}]{Stevenson2014}
{Stevenson} K.~B.,  et~al., 2014, \mn@doi [Science] {10.1126/science.1256758},
  \href {http://adsabs.harvard.edu/abs/2014Sci...346..838S} {346, 838}

\bibitem[\protect\citeauthoryear{{Steyert}, {Wang}, {Sirota}, {Donahue}  \&
  {Reuter}}{{Steyert} et~al.}{2004}]{Steyert2004}
{Steyert} D.~W.,  {Wang} W.~F.,  {Sirota} J.~M.,  {Donahue} N.~M.,   {Reuter}
  D.~C.,  2004, \mn@doi [\jqsrt] {10.1016/S0022-4073(02)00300-X}, \href
  {http://adsabs.harvard.edu/abs/2004JQSRT..83..183S} {83, 183}

\bibitem[\protect\citeauthoryear{{Sudarsky}, {Burrows}  \& {Hubeny}}{{Sudarsky}
  et~al.}{2003}]{Sudarsky2003}
{Sudarsky} D.,  {Burrows} A.,   {Hubeny} I.,  2003, \mn@doi [\apj]
  {10.1086/374331}, \href {http://adsabs.harvard.edu/abs/2003ApJ...588.1121S}
  {588, 1121}

\bibitem[\protect\citeauthoryear{{Tashkun} \& {Perevalov}}{{Tashkun} \&
  {Perevalov}}{2011}]{Tashkun2011}
{Tashkun} S.~A.,  {Perevalov} V.~I.,  2011, \mn@doi [\jqsrt]
  {10.1016/j.jqsrt.2011.03.005}, \href
  {http://adsabs.harvard.edu/abs/2011JQSRT.112.1403T} {112, 1403}

\bibitem[\protect\citeauthoryear{{Tennyson} \& {Yurchenko}}{{Tennyson} \&
  {Yurchenko}}{2012}]{Tennyson2012}
{Tennyson} J.,  {Yurchenko} S.~N.,  2012, \mn@doi [\mnras]
  {10.1111/j.1365-2966.2012.21440.x}, \href
  {http://adsabs.harvard.edu/abs/2012MNRAS.425...21T} {425, 21}

\bibitem[\protect\citeauthoryear{{Tennyson} et~al.,}{{Tennyson}
  et~al.}{2016}]{Tennyson2016}
{Tennyson} J.,  et~al., 2016, \mn@doi [Journal of Molecular Spectroscopy]
  {10.1016/j.jms.2016.05.002}, \href
  {http://adsabs.harvard.edu/abs/2016JMoSp.327...73T} {327, 73}

\bibitem[\protect\citeauthoryear{{Thibault}, {Boissoles}, {Le Doucen},
  {Bouanich}, {Arcas}  \& {Boulet}}{{Thibault} et~al.}{1992}]{Thibault1992}
{Thibault} F.,  {Boissoles} J.,  {Le Doucen} R.,  {Bouanich} J.~P.,  {Arcas}
  P.,   {Boulet} C.,  1992, \mn@doi [\jcp] {10.1063/1.462737}, \href
  {http://adsabs.harvard.edu/abs/1992JChPh..96.4945T} {96, 4945}

\bibitem[\protect\citeauthoryear{{Thibault}, {Calil}, {Boissoles}  \&
  {Launay}}{{Thibault} et~al.}{2000}]{Thibault2000}
{Thibault} F.,  {Calil} B.,  {Boissoles} J.,   {Launay} J.~M.,  2000, \mn@doi
  [Physical Chemistry Chemical Physics (Incorporating Faraday Transactions)]
  {10.1039/B006224N}, \href {http://adsabs.harvard.edu/abs/2000PCCP....2.5404T}
  {2, 5404}

\bibitem[\protect\citeauthoryear{{Torres} et~al.,}{{Torres}
  et~al.}{2007}]{Torres:2007aa}
{Torres} G.,  et~al., 2007, \mn@doi [\apjl] {10.1086/521792}, \href
  {http://adsabs.harvard.edu/abs/2007ApJ...666L.121T} {666, L121}

\bibitem[\protect\citeauthoryear{{Tregloan-Reed} et~al.,}{{Tregloan-Reed}
  et~al.}{2015}]{Tregloan-Reed:2015aa}
{Tregloan-Reed} J.,  et~al., 2015, \mn@doi [\mnras] {10.1093/mnras/stv730},
  \href {http://adsabs.harvard.edu/abs/2015MNRAS.450.1760T} {450, 1760}

\bibitem[\protect\citeauthoryear{{Tremblin}, {Amundsen}, {Mourier}, {Baraffe},
  {Chabrier}, {Drummond}, {Homeier}  \& {Venot}}{{Tremblin}
  et~al.}{2015}]{Tremblin2015}
{Tremblin} P.,  {Amundsen} D.~S.,  {Mourier} P.,  {Baraffe} I.,  {Chabrier} G.,
   {Drummond} B.,  {Homeier} D.,   {Venot} O.,  2015, \mn@doi [\apjl]
  {10.1088/2041-8205/804/1/L17}, \href
  {http://adsabs.harvard.edu/abs/2015ApJ...804L..17T} {804, L17}

\bibitem[\protect\citeauthoryear{{Tremblin}, {Amundsen}, {Chabrier}, {Baraffe},
  {Drummond}, {Hinkley}, {Mourier}  \& {Venot}}{{Tremblin}
  et~al.}{2016}]{Tremblin2016}
{Tremblin} P.,  {Amundsen} D.~S.,  {Chabrier} G.,  {Baraffe} I.,  {Drummond}
  B.,  {Hinkley} S.,  {Mourier} P.,   {Venot} O.,  2016, \mn@doi [\apjl]
  {10.3847/2041-8205/817/2/L19}, \href
  {http://adsabs.harvard.edu/abs/2016ApJ...817L..19T} {817, L19}

\bibitem[\protect\citeauthoryear{{Triaud} et~al.,}{{Triaud}
  et~al.}{2013}]{Triaud:2013}
{Triaud} A.~H.~M.~J.,  et~al., 2013, \mn@doi [\aap]
  {10.1051/0004-6361/201220900}, \href
  {http://adsabs.harvard.edu/abs/2013A%26A...551A..80T} {551, A80}

\bibitem[\protect\citeauthoryear{{Turner} et~al.,}{{Turner}
  et~al.}{2016}]{Turner:2016aa}
{Turner} O.~D.,  et~al., 2016, \mn@doi [\pasp]
  {10.1088/1538-3873/128/964/064401}, \href
  {http://adsabs.harvard.edu/abs/2016PASP..128f4401T} {128, 064401}

\bibitem[\protect\citeauthoryear{{Underwood}, {Tennyson}, {Yurchenko}, {Huang},
  {Schwenke}, {Lee}, {Clausen}  \& {Fateev}}{{Underwood}
  et~al.}{2016}]{Underwood2016}
{Underwood} D.~S.,  {Tennyson} J.,  {Yurchenko} S.~N.,  {Huang} X.,  {Schwenke}
  D.~W.,  {Lee} T.~J.,  {Clausen} S.,   {Fateev} A.,  2016, \mn@doi [\mnras]
  {10.1093/mnras/stw849}, \href
  {http://adsabs.harvard.edu/abs/2016MNRAS.459.3890U} {459, 3890}

\bibitem[\protect\citeauthoryear{{Van Grootel} et~al.,}{{Van Grootel}
  et~al.}{2014}]{VanGrootel:2014aa}
{Van Grootel} V.,  et~al., 2014, \mn@doi [\apj] {10.1088/0004-637X/786/1/2},
  \href {http://adsabs.harvard.edu/abs/2014ApJ...786....2V} {786, 2}

\bibitem[\protect\citeauthoryear{{Varanasi} \& {Chudamani}}{{Varanasi} \&
  {Chudamani}}{1990}]{Varanasi1990}
{Varanasi} P.,  {Chudamani} S.,  1990, \mn@doi [\jqsrt]
  {10.1016/0022-4073(90)90060-J}, \href
  {http://adsabs.harvard.edu/abs/1990JQSRT..43....1V} {43, 1}

\bibitem[\protect\citeauthoryear{{Venot}, {H{\'e}brard}, {Ag{\'u}ndez},
  {Dobrijevic}, {Selsis}, {Hersant}, {Iro}  \& {Bounaceur}}{{Venot}
  et~al.}{2012}]{Venot2012}
{Venot} O.,  {H{\'e}brard} E.,  {Ag{\'u}ndez} M.,  {Dobrijevic} M.,  {Selsis}
  F.,  {Hersant} F.,  {Iro} N.,   {Bounaceur} R.,  2012, \mn@doi [\aap]
  {10.1051/0004-6361/201219310}, \href
  {http://adsabs.harvard.edu/abs/2012A%26A...546A..43V} {546, A43}

\bibitem[\protect\citeauthoryear{{Venot}, {H{\'e}brard}, {Ag{\'u}ndez}, {Decin}
   \& {Bounaceur}}{{Venot} et~al.}{2015}]{Venot2015}
{Venot} O.,  {H{\'e}brard} E.,  {Ag{\'u}ndez} M.,  {Decin} L.,   {Bounaceur}
  R.,  2015, \mn@doi [\aap] {10.1051/0004-6361/201425311}, \href
  {http://adsabs.harvard.edu/abs/2015A%26A...577A..33V} {577, A33}

\bibitem[\protect\citeauthoryear{{Wakeford} \& {Sing}}{{Wakeford} \&
  {Sing}}{2015}]{Wakeford2015}
{Wakeford} H.~R.,  {Sing} D.~K.,  2015, \mn@doi [\aap]
  {10.1051/0004-6361/201424207}, \href
  {http://adsabs.harvard.edu/abs/2015A%26A...573A.122W} {573, A122}

\bibitem[\protect\citeauthoryear{{Wakeford} et~al.,}{{Wakeford}
  et~al.}{2013}]{Wakeford2013}
{Wakeford} H.~R.,  et~al., 2013, \mn@doi [\mnras] {10.1093/mnras/stt1536},
  \href {http://adsabs.harvard.edu/abs/2013MNRAS.435.3481W} {435, 3481}

\bibitem[\protect\citeauthoryear{{Wakeford} et~al.,}{{Wakeford}
  et~al.}{2017}]{Wakeford2017}
{Wakeford} H.~R.,  et~al., 2017, preprint, \href
  {http://adsabs.harvard.edu/abs/2017arXiv170504354W} {} (\mn@eprint {arXiv}
  {1705.04354})

\bibitem[\protect\citeauthoryear{{Wende}, {Reiners}, {Seifahrt}  \&
  {Bernath}}{{Wende} et~al.}{2010}]{Fehlinelist2010}
{Wende} S.,  {Reiners} A.,  {Seifahrt} A.,   {Bernath} P.~F.,  2010, \mn@doi
  [\aap] {10.1051/0004-6361/201015220}, \href
  {http://adsabs.harvard.edu/abs/2010A%26A...523A..58W} {523, A58}

\bibitem[\protect\citeauthoryear{{West} et~al.,}{{West}
  et~al.}{2009a}]{West:2009aa}
{West} R.~G.,  et~al., 2009a, \mn@doi [\aj] {10.1088/0004-6256/137/6/4834},
  \href {http://adsabs.harvard.edu/abs/2009AJ....137.4834W} {137, 4834}

\bibitem[\protect\citeauthoryear{{West} et~al.,}{{West}
  et~al.}{2009b}]{West:2009ab}
{West} R.~G.,  et~al., 2009b, \mn@doi [\aap] {10.1051/0004-6361/200810973},
  \href {http://adsabs.harvard.edu/abs/2009A%26A...502..395W} {502, 395}

\bibitem[\protect\citeauthoryear{{West} et~al.,}{{West}
  et~al.}{2016}]{West:2016aa}
{West} R.~G.,  et~al., 2016, \mn@doi [\aap] {10.1051/0004-6361/201527276},
  \href {http://adsabs.harvard.edu/abs/2016A%26A...585A.126W} {585, A126}

\bibitem[\protect\citeauthoryear{{Wilson} et~al.,}{{Wilson}
  et~al.}{2008}]{Wilson:2008aa}
{Wilson} D.~M.,  et~al., 2008, \mn@doi [\apjl] {10.1086/586735}, \href
  {http://adsabs.harvard.edu/abs/2008ApJ...675L.113W} {675, L113}

\bibitem[\protect\citeauthoryear{{Winn}}{{Winn}}{2010}]{Winn2010}
{Winn} J.~N.,  2010, {Exoplanet Transits and Occultations}.
University of Arizona Press, pp 55--77

\bibitem[\protect\citeauthoryear{{Winn} et~al.,}{{Winn}
  et~al.}{2011}]{Winn:2011aa}
{Winn} J.~N.,  et~al., 2011, \mn@doi [\apjl] {10.1088/2041-8205/737/1/L18},
  \href {http://adsabs.harvard.edu/abs/2011ApJ...737L..18W} {737, L18}

\bibitem[\protect\citeauthoryear{{Yurchenko} \& {Tennyson}}{{Yurchenko} \&
  {Tennyson}}{2014}]{Yurchenko2014}
{Yurchenko} S.~N.,  {Tennyson} J.,  2014, \mn@doi [\mnras]
  {10.1093/mnras/stu326}, \href
  {http://adsabs.harvard.edu/abs/2014MNRAS.440.1649Y} {440, 1649}

\bibitem[\protect\citeauthoryear{{Yurchenko}, {Barber}  \&
  {Tennyson}}{{Yurchenko} et~al.}{2011}]{Yurchenko2011}
{Yurchenko} S.~N.,  {Barber} R.~J.,   {Tennyson} J.,  2011, \mn@doi [\mnras]
  {10.1111/j.1365-2966.2011.18261.x}, \href
  {http://adsabs.harvard.edu/abs/2011MNRAS.413.1828Y} {413, 1828}

\bibitem[\protect\citeauthoryear{{Zellem} et~al.,}{{Zellem}
  et~al.}{2014}]{Zellem2014}
{Zellem} R.~T.,  et~al., 2014, \mn@doi [\apj] {10.1088/0004-637X/790/1/53},
  \href {http://adsabs.harvard.edu/abs/2014ApJ...790...53Z} {790, 53}

\bibitem[\protect\citeauthoryear{{Zhou} et~al.,}{{Zhou}
  et~al.}{2014}]{Zhou:2014aa}
{Zhou} G.,  et~al., 2014, \mn@doi [\aj] {10.1088/0004-6256/147/6/144}, \href
  {http://adsabs.harvard.edu/abs/2014AJ....147..144Z} {147, 144}

\bibitem[\protect\citeauthoryear{{Zhou} et~al.,}{{Zhou}
  et~al.}{2016}]{Zhou:2016aa}
{Zhou} G.,  et~al., 2016, \mn@doi [\aj] {10.3847/0004-6256/152/5/136}, \href
  {http://adsabs.harvard.edu/abs/2016AJ....152..136Z} {152, 136}

\makeatother
\end{thebibliography}


\begin{thebibliography}{}
\makeatletter
\relax
\def\mn@urlcharsother{\let\do\@makeother \do\$\do\&\do\#\do\^\do\_\do\%\do\~}
\def\mn@doi{\begingroup\mn@urlcharsother \@ifnextchar [ {\mn@doi@}
  {\mn@doi@[]}}
\def\mn@doi@[#1]#2{\def\@tempa{#1}\ifx\@tempa\@empty \href
  {http://dx.doi.org/#2} {doi:#2}\else \href {http://dx.doi.org/#2} {#1}\fi
  \endgroup}
\def\mn@eprint#1#2{\mn@eprint@#1:#2::\@nil}
\def\mn@eprint@arXiv#1{\href {http://arxiv.org/abs/#1} {{\tt arXiv:#1}}}
\def\mn@eprint@dblp#1{\href {http://dblp.uni-trier.de/rec/bibtex/#1.xml}
  {dblp:#1}}
\def\mn@eprint@#1:#2:#3:#4\@nil{\def\@tempa {#1}\def\@tempb {#2}\def\@tempc
  {#3}\ifx \@tempc \@empty \let \@tempc \@tempb \let \@tempb \@tempa \fi \ifx
  \@tempb \@empty \def\@tempb {arXiv}\fi \@ifundefined
  {mn@eprint@\@tempb}{\@tempb:\@tempc}{\expandafter \expandafter \csname
  mn@eprint@\@tempb\endcsname \expandafter{\@tempc}}}

\bibitem[\protect\citeauthoryear{{Asplund}, {Grevesse}, {Sauval}  \&
  {Scott}}{{Asplund} et~al.}{2009}]{Asplund2009}
{Asplund} M.,  {Grevesse} N.,  {Sauval} A.~J.,   {Scott} P.,  2009, \mn@doi
  [\araa] {10.1146/annurev.astro.46.060407.145222}, \href
  {http://adsabs.harvard.edu/abs/2009ARA%26A..47..481A} {47, 481}

\bibitem[\protect\citeauthoryear{{Barklem} \& {Collet}}{{Barklem} \&
  {Collet}}{2016}]{Barklem2016}
{Barklem} P.~S.,  {Collet} R.,  2016, \mn@doi [\aap]
  {10.1051/0004-6361/201526961}, \href
  {https://ui.adsabs.harvard.edu/\#abs/2016A&A...588A..96B} {588, A96}

\bibitem[\protect\citeauthoryear{{Baudino}, {B{\'e}zard}, {Boccaletti},
  {Bonnefoy}, {Lagrange}  \& {Galicher}}{{Baudino} et~al.}{2015}]{Baudino2015}
{Baudino} J.-L.,  {B{\'e}zard} B.,  {Boccaletti} A.,  {Bonnefoy} M.,
  {Lagrange} A.-M.,   {Galicher} R.,  2015, \mn@doi [\aap]
  {10.1051/0004-6361/201526332}, \href
  {http://adsabs.harvard.edu/abs/2015A%26A...582A..83B} {582, A83}

\bibitem[\protect\citeauthoryear{{Baudino}, {Molli{\`e}re}, {Venot},
  {Tremblin}, {B{\'e}zard}  \& {Lagage}}{{Baudino} et~al.}{2017}]{Baudino2017}
{Baudino} J.-L.,  {Molli{\`e}re} P.,  {Venot} O.,  {Tremblin} P.,  {B{\'e}zard}
  B.,   {Lagage} P.-O.,  2017, \mn@doi [\apj] {10.3847/1538-4357/aa95be}, \href
  {http://adsabs.harvard.edu/abs/2017ApJ...850..150B} {850, 150}

\bibitem[\protect\citeauthoryear{{Burrows} \& {Sharp}}{{Burrows} \&
  {Sharp}}{1999}]{Burrows1999}
{Burrows} A.,  {Sharp} C.~M.,  1999, \mn@doi [\apj] {10.1086/306811}, \href
  {http://adsabs.harvard.edu/abs/1999ApJ...512..843B} {512, 843}

\bibitem[\protect\citeauthoryear{{Chase}}{{Chase}}{1986}]{Chase1986}
{Chase} M.~W.,  1986, {JANAF thermochemical tables}

\bibitem[\protect\citeauthoryear{{Drummond}, {Tremblin}, {Baraffe}, {Amundsen},
  {Mayne}, {Venot}  \& {Goyal}}{{Drummond} et~al.}{2016}]{Drummond2016}
{Drummond} B.,  {Tremblin} P.,  {Baraffe} I.,  {Amundsen} D.~S.,  {Mayne}
  N.~J.,  {Venot} O.,   {Goyal} J.,  2016, \mn@doi [\aap]
  {10.1051/0004-6361/201628799}, \href
  {http://adsabs.harvard.edu/abs/2016A%26A...594A..69D} {594, A69}

\bibitem[\protect\citeauthoryear{{Fischer} et~al.,}{{Fischer}
  et~al.}{2016}]{Fischer2016}
{Fischer} P.~D.,  et~al., 2016, \mn@doi [\apj] {10.3847/0004-637X/827/1/19},
  \href {http://adsabs.harvard.edu/abs/2016ApJ...827...19F} {827, 19}

\bibitem[\protect\citeauthoryear{{Goyal} et~al.,}{{Goyal}
  et~al.}{2018}]{Goyal2018}
{Goyal} J.~M.,  et~al., 2018, \mn@doi [\mnras] {10.1093/mnras/stx3015}, \href
  {https://ui.adsabs.harvard.edu/#abs/2018MNRAS.474.5158G} {474, 5158}

\bibitem[\protect\citeauthoryear{{Goyal}, {Wakeford}, {Mayne}, {Lewis},
  {Drummond}  \& {Sing}}{{Goyal} et~al.}{2019}]{Goyal2019}
{Goyal} J.~M.,  {Wakeford} H.~R.,  {Mayne} N.~J.,  {Lewis} N.~K.,  {Drummond}
  B.,   {Sing} D.~K.,  2019, \mn@doi [\mnras] {10.1093/mnras/sty3001}, \href
  {https://ui.adsabs.harvard.edu/\#abs/2019MNRAS.482.4503G} {482, 4503}

\bibitem[\protect\citeauthoryear{{Heng} \& {Tsai}}{{Heng} \&
  {Tsai}}{2016}]{Heng2016}
{Heng} K.,  {Tsai} S.-M.,  2016, \mn@doi [\apj] {10.3847/0004-637X/829/2/104},
  \href {http://adsabs.harvard.edu/abs/2016ApJ...829..104H} {829, 104}

\bibitem[\protect\citeauthoryear{{Huitson} et~al.,}{{Huitson}
  et~al.}{2013}]{Huitson2013}
{Huitson} C.~M.,  et~al., 2013, \mn@doi [\mnras] {10.1093/mnras/stt1243}, \href
  {http://adsabs.harvard.edu/abs/2013MNRAS.434.3252H} {434, 3252}

\bibitem[\protect\citeauthoryear{{Kempton}, {Lupu}, {Owusu-Asare}, {Slough}  \&
  {Cale}}{{Kempton} et~al.}{2017}]{Kempton2017}
{Kempton} E.~M.-R.,  {Lupu} R.,  {Owusu-Asare} A.,  {Slough} P.,   {Cale} B.,
  2017, \mn@doi [\pasp] {10.1088/1538-3873/aa61ef}, \href
  {http://adsabs.harvard.edu/abs/2017PASP..129d4402K} {129, 044402}

\bibitem[\protect\citeauthoryear{{Lodders}}{{Lodders}}{2003}]{Lodders2003}
{Lodders} K.,  2003, \mn@doi [\apj] {10.1086/375492}, \href
  {https://ui.adsabs.harvard.edu/#abs/2003ApJ...591.1220L} {591, 1220}

\bibitem[\protect\citeauthoryear{{Lodders} \& {Fegley}}{{Lodders} \&
  {Fegley}}{2002}]{Lodders2002}
{Lodders} K.,  {Fegley} B.,  2002, \mn@doi [\icarus] {10.1006/icar.2001.6740},
  \href {https://ui.adsabs.harvard.edu/\#abs/2002Icar..155..393L} {155, 393}

\bibitem[\protect\citeauthoryear{{Mbarek} \& {Kempton}}{{Mbarek} \&
  {Kempton}}{2016}]{Mbarek2016}
{Mbarek} R.,  {Kempton} E.~M.-R.,  2016, \mn@doi [\apj]
  {10.3847/0004-637X/827/2/121}, \href
  {http://adsabs.harvard.edu/abs/2016ApJ...827..121M} {827, 121}

\bibitem[\protect\citeauthoryear{{McCullough}, {Crouzet}, {Deming}  \&
  {Madhusudhan}}{{McCullough} et~al.}{2014}]{McCullough2014}
{McCullough} P.~R.,  {Crouzet} N.,  {Deming} D.,   {Madhusudhan} N.,  2014,
  \mn@doi [\apj] {10.1088/0004-637X/791/1/55}, \href
  {http://adsabs.harvard.edu/abs/2014ApJ...791...55M} {791, 55}

\bibitem[\protect\citeauthoryear{{Molli{\`e}re}, {van Boekel}, {Dullemond},
  {Henning}  \& {Mordasini}}{{Molli{\`e}re} et~al.}{2015}]{Molliere2015}
{Molli{\`e}re} P.,  {van Boekel} R.,  {Dullemond} C.,  {Henning} T.,
  {Mordasini} C.,  2015, \mn@doi [\apj] {10.1088/0004-637X/813/1/47}, \href
  {http://adsabs.harvard.edu/abs/2015ApJ...813...47M} {813, 47}

\bibitem[\protect\citeauthoryear{{Molli{\`e}re}, {van Boekel}, {Bouwman},
  {Henning}, {Lagage}  \& {Min}}{{Molli{\`e}re} et~al.}{2017}]{Molliere2017}
{Molli{\`e}re} P.,  {van Boekel} R.,  {Bouwman} J.,  {Henning} T.,  {Lagage}
  P.-O.,   {Min} M.,  2017, \mn@doi [\aap] {10.1051/0004-6361/201629800e},
  \href {http://adsabs.harvard.edu/abs/2017A%26A...605C...3M} {605, C3}

\bibitem[\protect\citeauthoryear{{Nikolov} et~al.,}{{Nikolov}
  et~al.}{2014}]{Nikolov:2014aa}
{Nikolov} N.,  et~al., 2014, \mn@doi [\mnras] {10.1093/mnras/stt1859}, \href
  {http://adsabs.harvard.edu/abs/2014MNRAS.437...46N} {437, 46}

\bibitem[\protect\citeauthoryear{{Nikolov} et~al.,}{{Nikolov}
  et~al.}{2015}]{Nikolov2015}
{Nikolov} N.,  et~al., 2015, \mn@doi [\mnras] {10.1093/mnras/stu2433}, \href
  {http://adsabs.harvard.edu/abs/2015MNRAS.447..463N} {447, 463}

\bibitem[\protect\citeauthoryear{{Pont}, {Sing}, {Gibson}, {Aigrain}, {Henry}
  \& {Husnoo}}{{Pont} et~al.}{2013}]{Pont2013}
{Pont} F.,  {Sing} D.~K.,  {Gibson} N.~P.,  {Aigrain} S.,  {Henry} G.,
  {Husnoo} N.,  2013, \mn@doi [\mnras] {10.1093/mnras/stt651}, \href
  {http://adsabs.harvard.edu/abs/2013MNRAS.432.2917P} {432, 2917}

\bibitem[\protect\citeauthoryear{{Sing} et~al.,}{{Sing}
  et~al.}{2013}]{Sing2013}
{Sing} D.~K.,  et~al., 2013, \mn@doi [\mnras] {10.1093/mnras/stt1782}, \href
  {http://adsabs.harvard.edu/abs/2013MNRAS.436.2956S} {436, 2956}

\bibitem[\protect\citeauthoryear{{Sing} et~al.,}{{Sing}
  et~al.}{2015}]{Sing2015}
{Sing} D.~K.,  et~al., 2015, \mn@doi [\mnras] {10.1093/mnras/stu2279}, \href
  {http://adsabs.harvard.edu/abs/2015MNRAS.446.2428S} {446, 2428}

\bibitem[\protect\citeauthoryear{{Sing} et~al.,}{{Sing}
  et~al.}{2016}]{Sing2016}
{Sing} D.~K.,  et~al., 2016, \mn@doi [\nat] {10.1038/nature16068}, \href
  {http://adsabs.harvard.edu/abs/2016Natur.529...59S} {529, 59}

\bibitem[\protect\citeauthoryear{{Tsuji}}{{Tsuji}}{1973}]{Tsuji1973}
{Tsuji} T.,  1973, \aap, \href
  {https://ui.adsabs.harvard.edu/\#abs/1973A&A....23..411T} {23, 411}

\bibitem[\protect\citeauthoryear{{Wakeford} et~al.,}{{Wakeford}
  et~al.}{2013}]{Wakeford2013}
{Wakeford} H.~R.,  et~al., 2013, \mn@doi [\mnras] {10.1093/mnras/stt1536},
  \href {http://adsabs.harvard.edu/abs/2013MNRAS.435.3481W} {435, 3481}

\bibitem[\protect\citeauthoryear{{Woitke}, {Helling}, {Hunter}, {Millard},
  {Turner}, {Worters}, {Blecic}  \& {Stock}}{{Woitke}
  et~al.}{2018}]{Woitke2018}
{Woitke} P.,  {Helling} C.,  {Hunter} G.~H.,  {Millard} J.~D.,  {Turner} G.~E.,
   {Worters} M.,  {Blecic} J.,   {Stock} J.~W.,  2018, \mn@doi [\aap]
  {10.1051/0004-6361/201732193}, \href
  {https://ui.adsabs.harvard.edu/\#abs/2018A&A...614A...1W} {614, A1}

\makeatother
\end{thebibliography}

 % if your bibtex file is called example.bib

\clearpage
\newpage

%\begin{thebibliography}{99}

%%%%%%%%%%%%%%%%%%%%%%%%%%%%%%%%%%%%%%%%%%%%%%%%%%

%%%%%%%%%%%%%%%%% APPENDICES %%%%%%%%%%%%%%%%%%%%%
\appendix

\section{$\chi^{2}$ maps of all planets}

\begin{figure}
\includegraphics[width=\columnwidth]{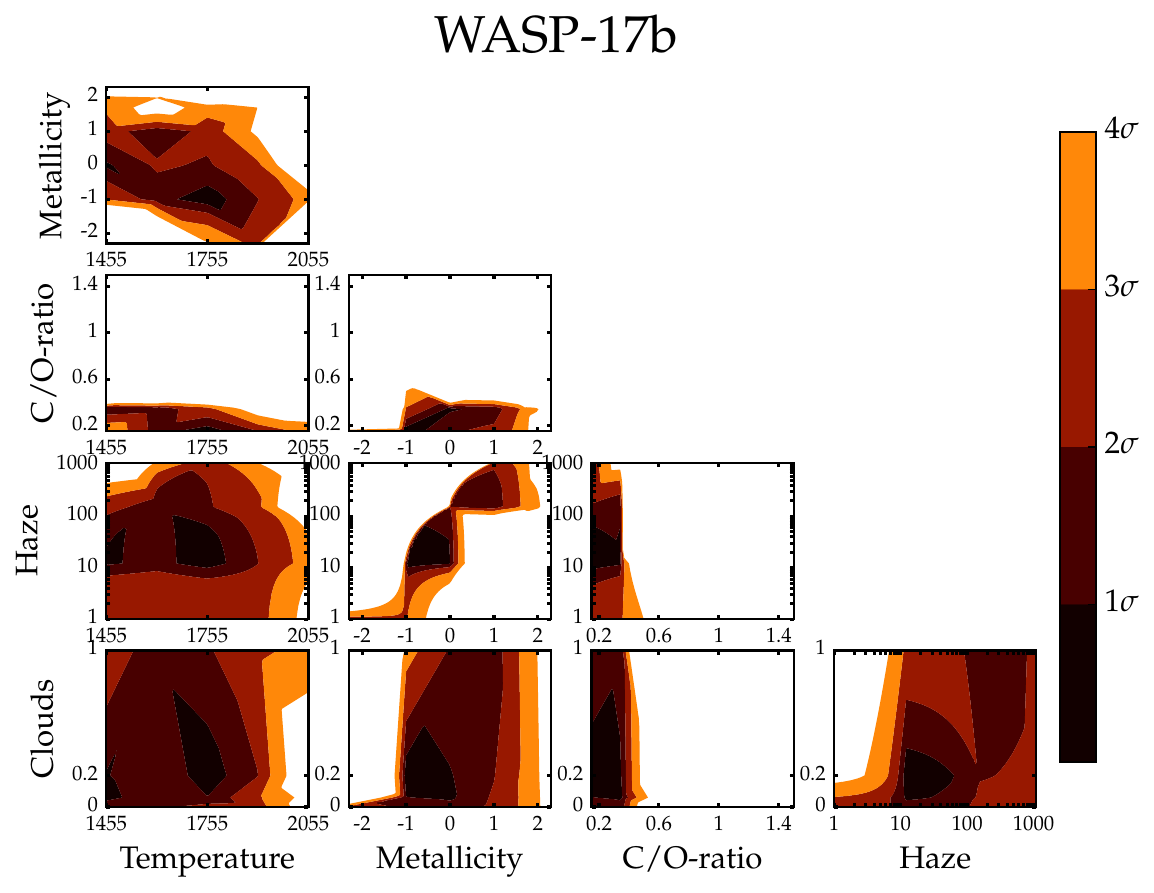}
 \caption{Figure showing WASP-17b $\chi^{2}$ Map, with same format as Figure \ref{fig:chimap_wasp39b}}
    \label{fig:chimap_wasp17b}
\end{figure}

\begin{figure}
\includegraphics[width=\columnwidth]{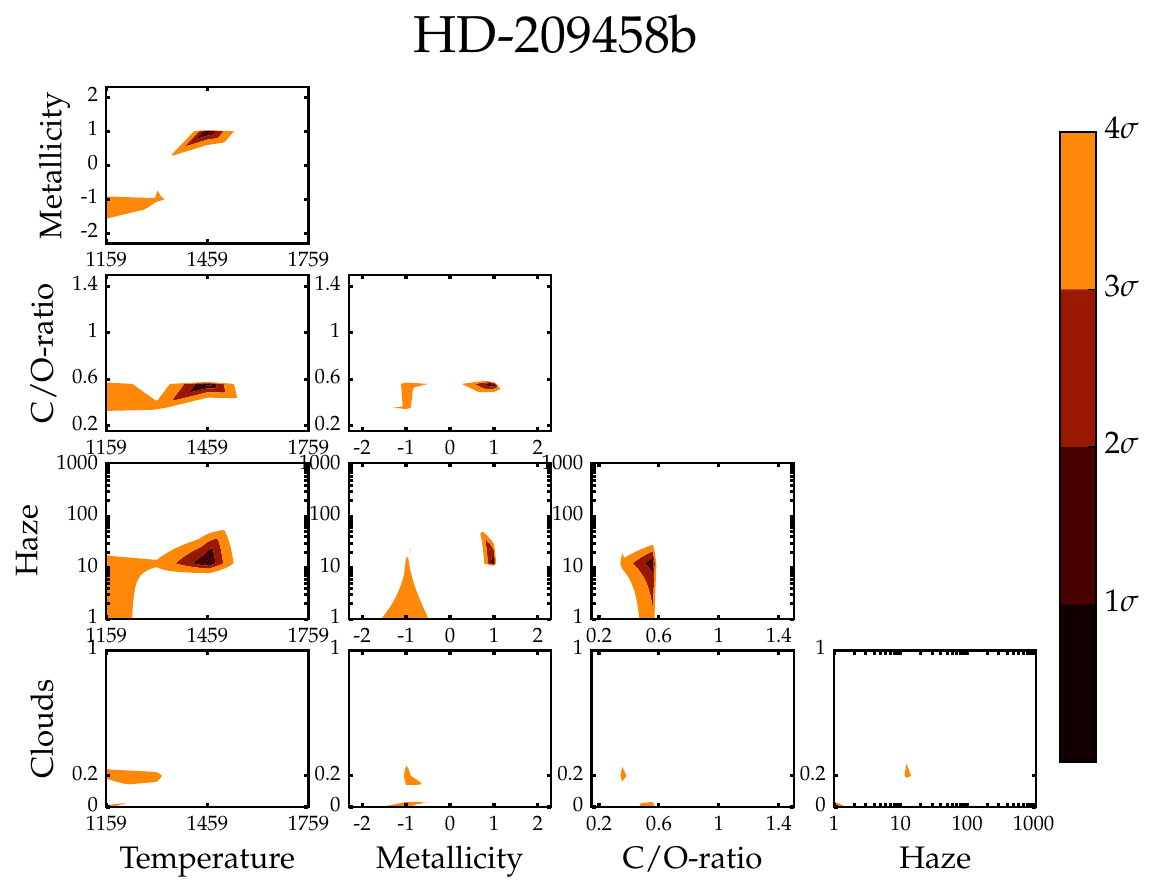}
 \caption{Figure showing HD~209458b $\chi^{2}$ Map, with same format as \ref{fig:chimap_wasp39b}}
    \label{fig:chimap_hd209458b}
\end{figure}

\begin{figure}
\includegraphics[width=\columnwidth]{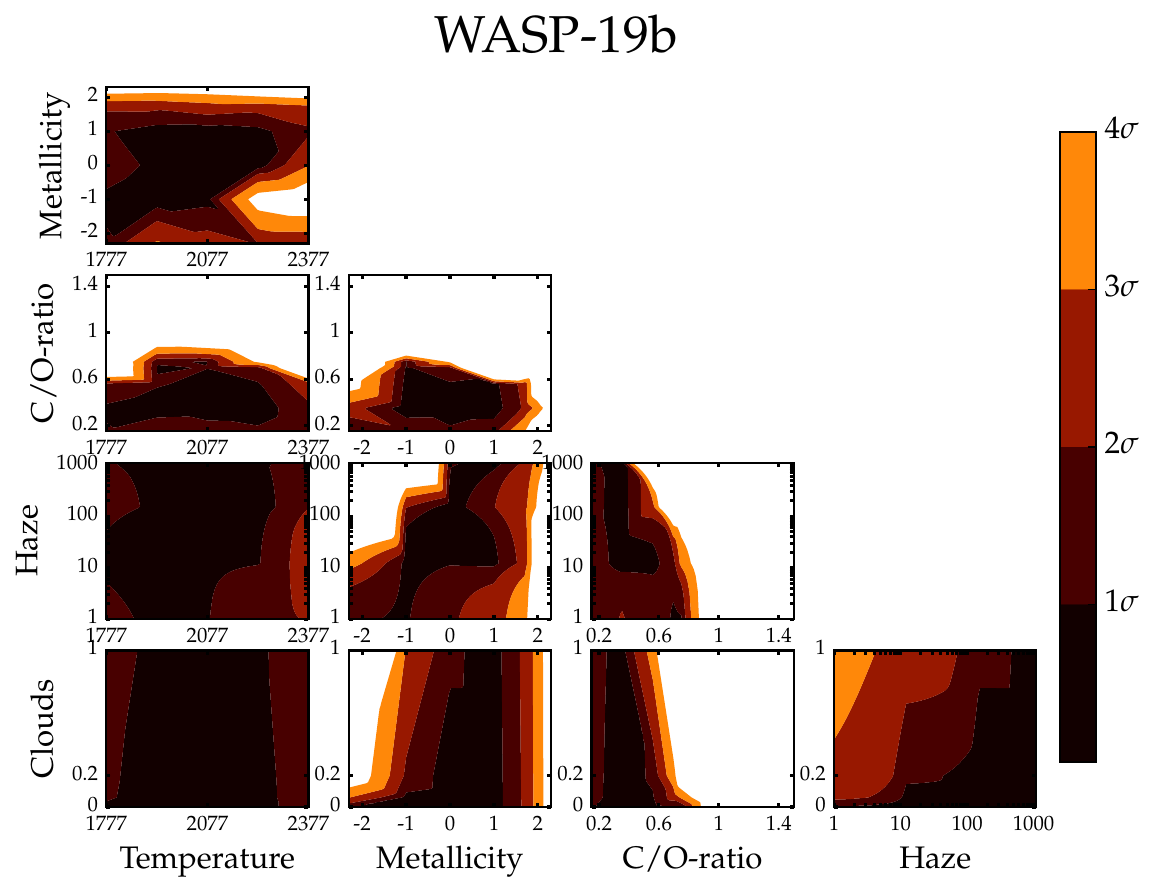}
 \caption{Figure showing WASP-19b $\chi^{2}$ Map, with same format as Figure \ref{fig:chimap_wasp39b}}
    \label{fig:chimap_wasp19b}
\end{figure}

\begin{figure}
\includegraphics[width=\columnwidth]{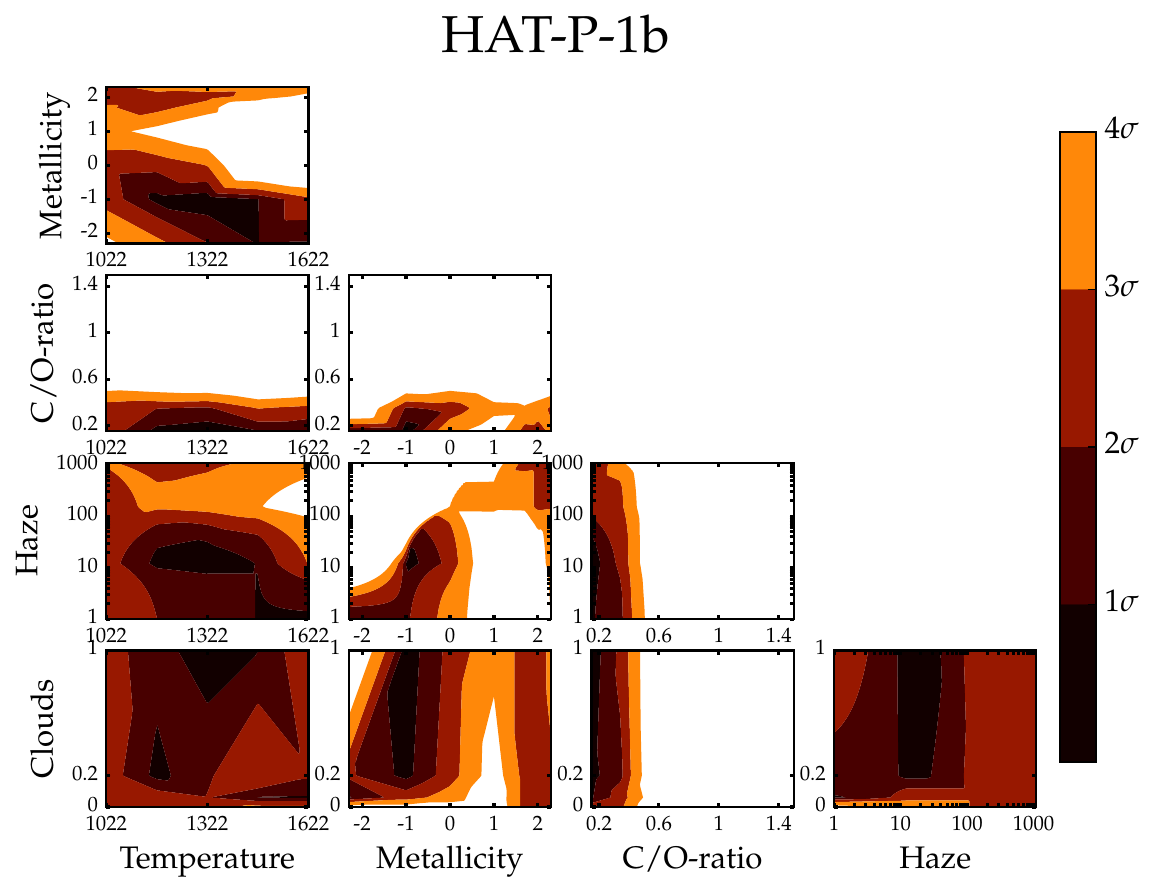}
 \caption{Figure showing HAT-P-1b $\chi^{2}$ Map, with same format as Figure \ref{fig:chimap_wasp39b}}
    \label{fig:chimap_hatp01}
\end{figure}

\begin{figure}
\includegraphics[width=\columnwidth]{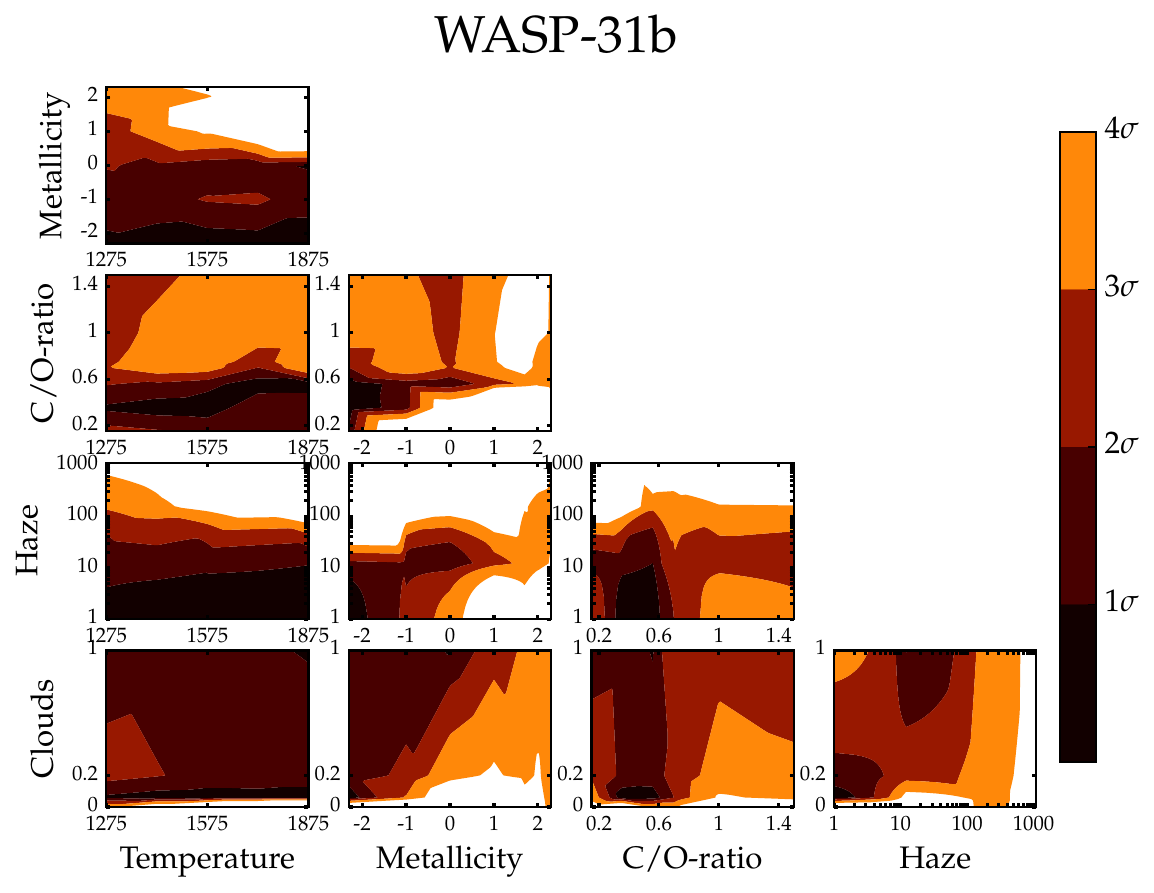}
 \caption{Figure showing WASP-31b $\chi^{2}$ Map, with same format as Figure \ref{fig:chimap_wasp39b}}
    \label{fig:chimap_wasp31b}
\end{figure}

\begin{figure}
\includegraphics[width=\columnwidth]{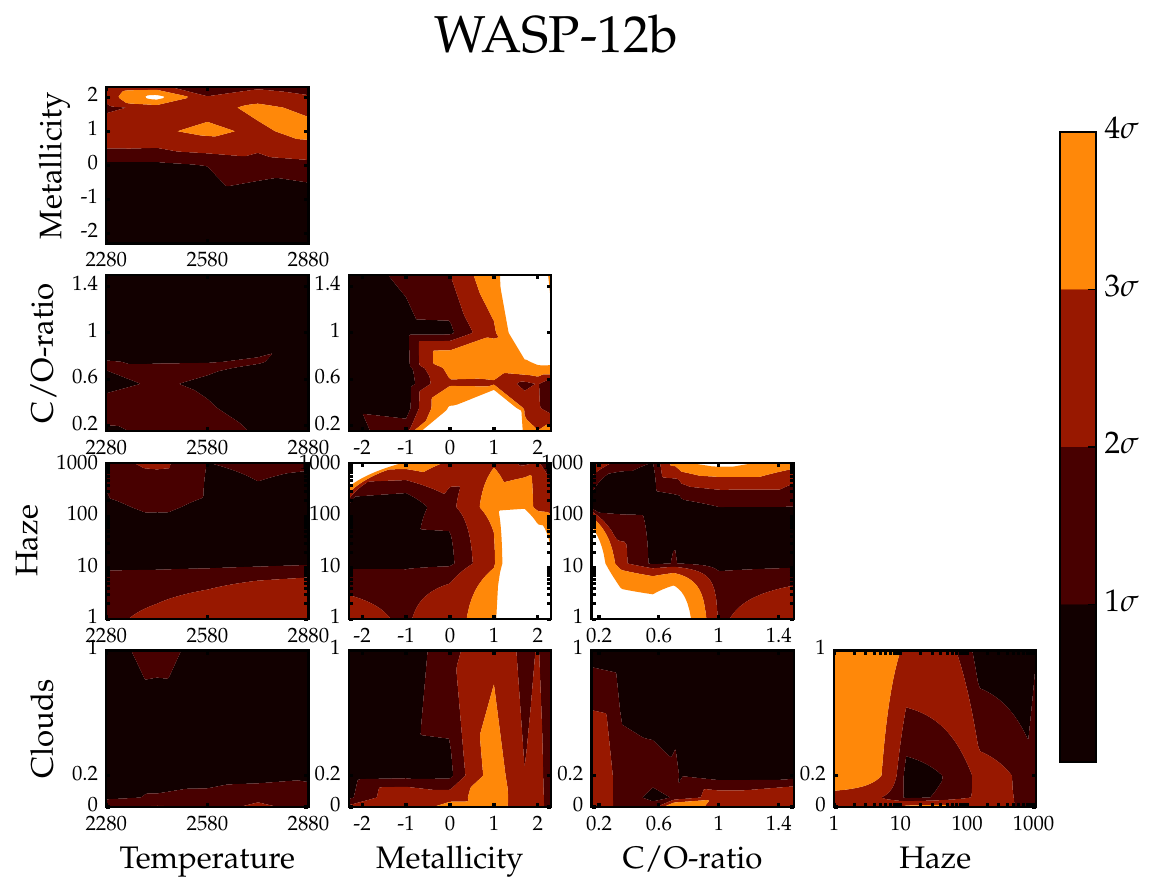}
 \caption{Figure showing WASP-12b $\chi^{2}$ Map, with same format as Figure \ref{fig:chimap_wasp39b}}
    \label{fig:chimap_wasp12b}
\end{figure}

\begin{figure}
\includegraphics[width=\columnwidth]{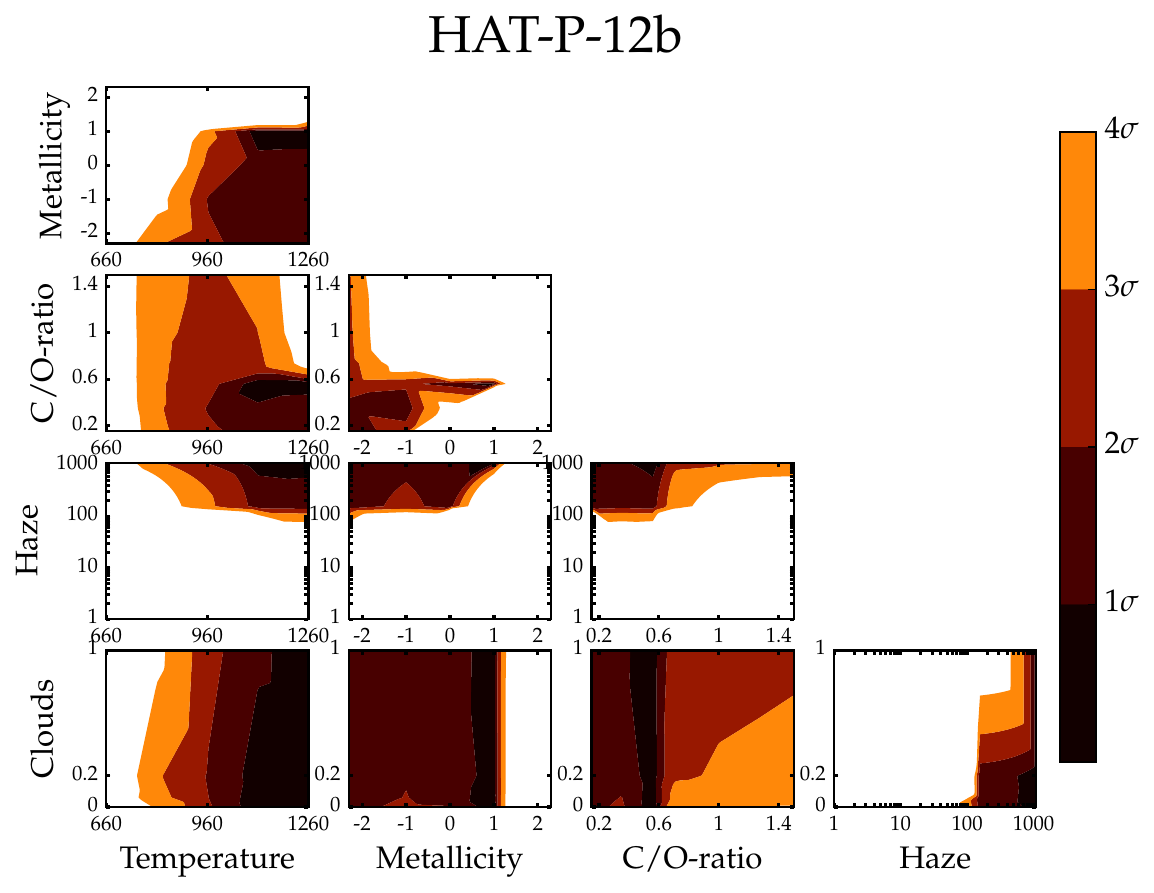}
 \caption{Figure showing HAT-P-12b $\chi^{2}$ Map, with same format as Figure \ref{fig:chimap_wasp39b}}
    \label{fig:chimap_hatp12b}
\end{figure}

\begin{figure}
\includegraphics[width=\columnwidth]{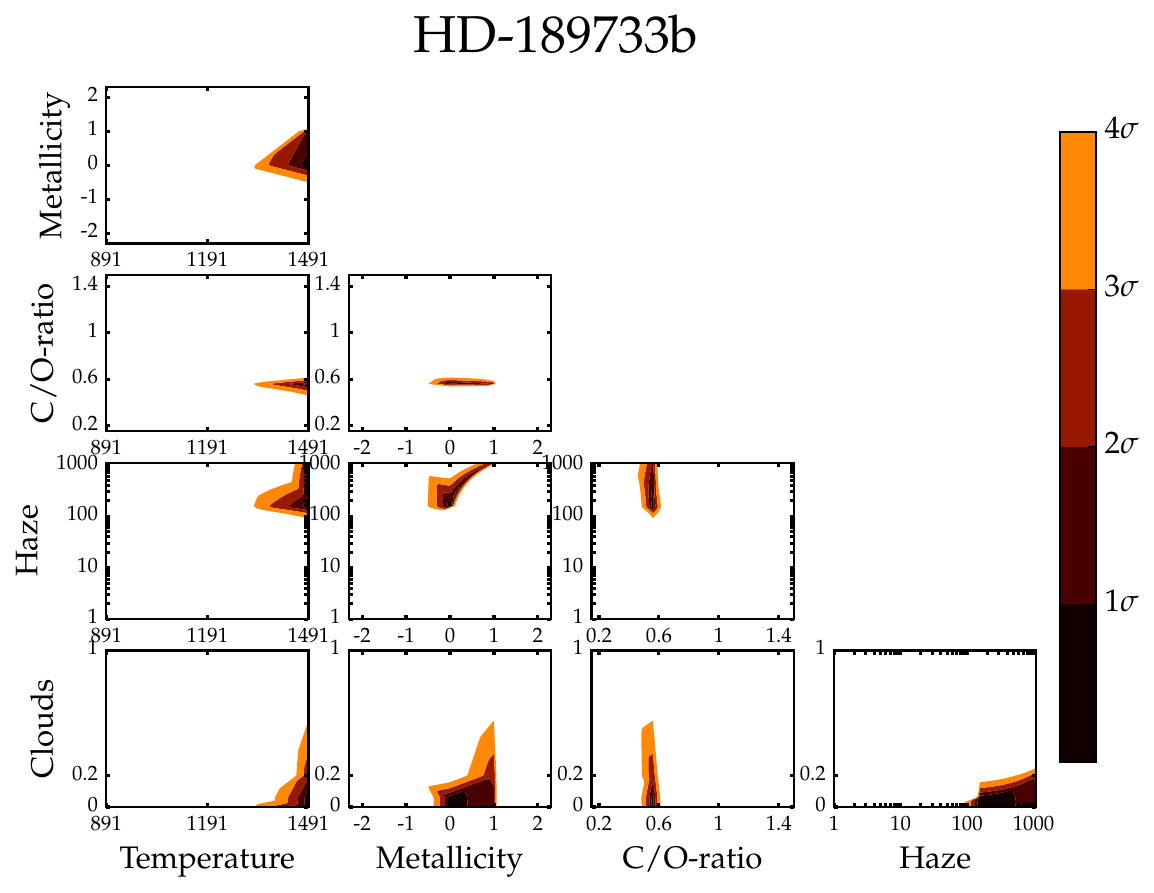}
 \caption{Figure showing HD~189733b $\chi^{2}$ Map, with same format as Figure \ref{fig:chimap_wasp39b}}
    \label{fig:chimap_hd189733b}
\end{figure}

\begin{figure}
\includegraphics[width=\columnwidth]{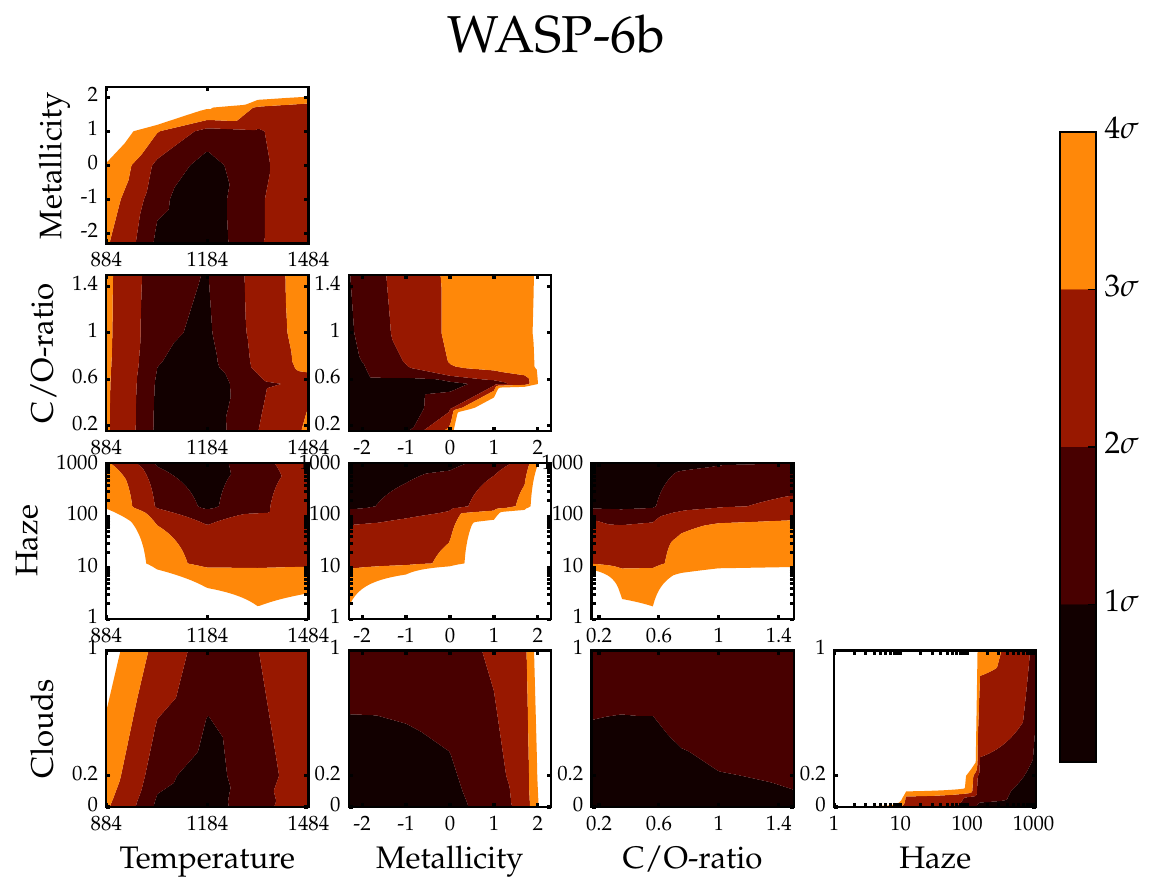}
 \caption{Figure showing WASP-6b $\chi^{2}$ Map, with same format as Figure \ref{fig:chimap_wasp39b}}
    \label{fig:chimap_wasp6b}
\end{figure}

As explained in Section \ref{section:Interpretation of Observations}, $\chi^{2}$ maps of 9 exoplanets are shown here in Figures \ref{fig:chimap_wasp17b} to  \ref{fig:chimap_wasp6b}. 
\label{app:chi-maps of all planets}

\clearpage
\newpage

\begin{figure*}
\begin{center}
 \subfloat[]{\includegraphics[width=\columnwidth]{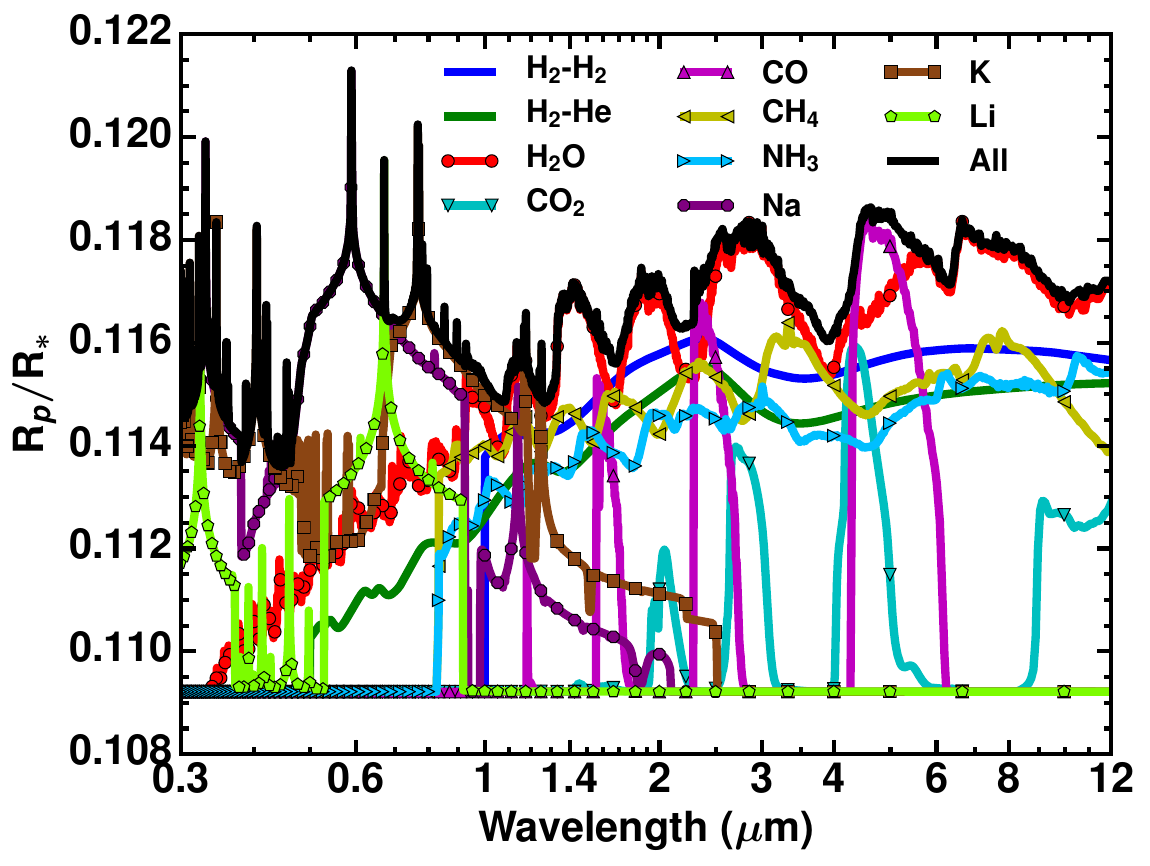}\label{fig:opacity_1to10}}
 \subfloat[]{\includegraphics[width=\columnwidth]{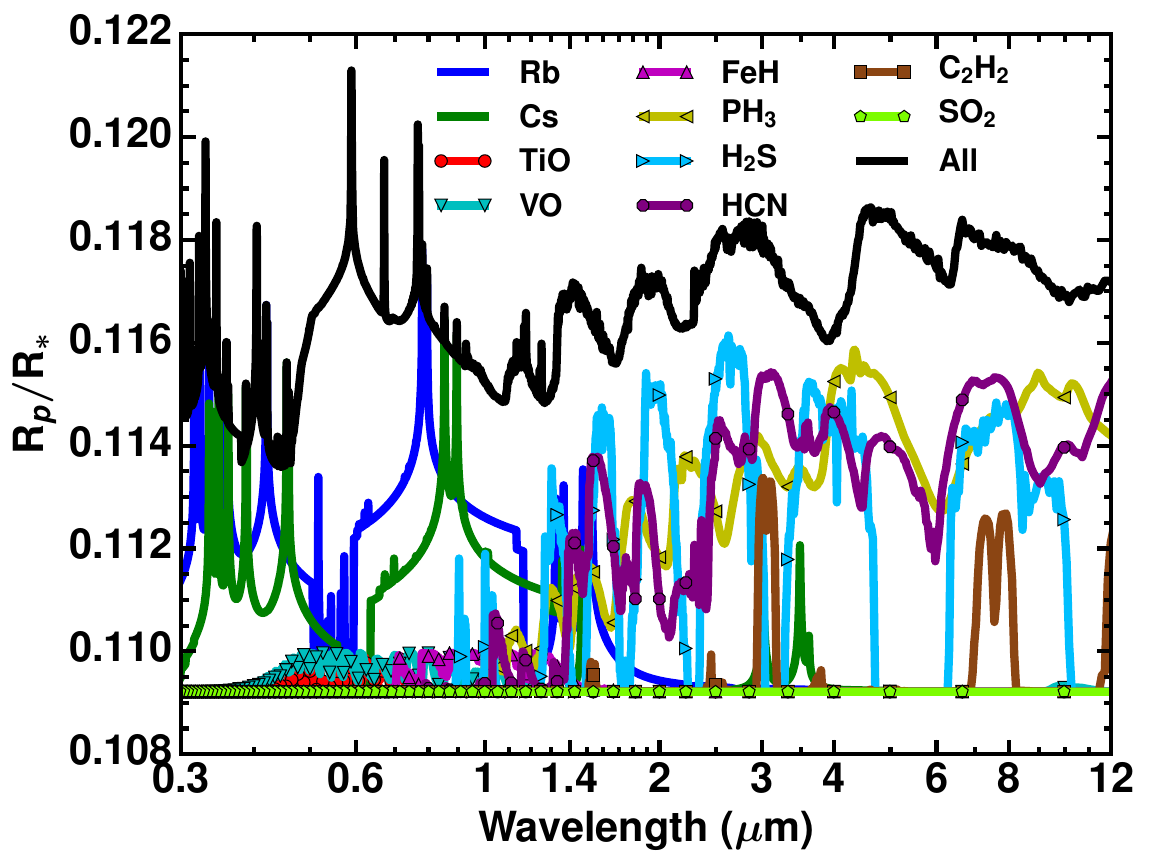}\label{fig:opacity_10to20}}
\end{center}
 \caption{\textbf{(a)} Figure showing transmission spectra features of each individual molecule used in \texttt{ATMO} (1 to 10). H$_2$-H$_2$ (blue), H$_2$-He (green), H$_2$O (red), CO$_2$ (cyan), CO (magenta), CH$_4$ (yellow), NH$_3$ (lightblue), Na (purple), K (brown), Li (lightgreen) and all 20 opacities (black). \textbf{(b)} Figure showing transmission spectra features of each individual molecule used in \texttt{ATMO} (11 to 20). Rb (blue), Cs (green), TiO (red), VO (cyan), FeH (magenta), PH$_3$ (yellow), H$_2$S (lightblue), HCN (purple), C$_2$H$_2$ (brown), SO$_2$ (lightgreen) and all 20 opacities (black). No $\RpRs$ offset was applied while plotting. Individual simulations are divided into blocks of 10 while plotting for clarity.}
\end{figure*}

\section{Transmission Spectral Features} 
\label{subsection:Molecular Features}
This section describes the major spectral features in the transmission spectra of exoplanets. Figures,  \ref{fig:opacity_1to10} and  \ref{fig:opacity_10to20} show the transmission spectra with opacities of just individual species included in the simulation. All the individual opacity model runs are for HD~209458b with $1609$\,K equilibrium temperature, solar metallicity, solar C/O ratio and a clear atmosphere. We omit Rayleigh scattering to avoid absorption features being masked at optical wavelengths. The simulation depicted in black in both the figures include all the 20 opacities in the \texttt{ATMO}, hereafter termed \enquote{all opacity simulation}. This allows identification of major species which contribute to final transmission spectra. . 

With just H$_2$-H$_2$ or H$_2$-He collision induced absorption (CIA) opacities included in the simulation, shown in blue and green, respectively in  Figure \ref{fig:opacity_1to10}, we mainly see broad band collision induced signatures of absorption, primarily in the near infrared regions. The simulation with just  water (H$_{2}$O) in red (Figure \ref{fig:opacity_1to10}), leads to spectral features in almost all parts of the spectrum, one of the major ones being at 1.4 \textmu m, which has been detected in many exoplanet atmospheres. Comparing the all opacity simulation in black, and the simulation in red with just H$_{2}$O opacity we can clearly see that H$_2$O dominates the final transmission spectra at solar metallicity. This changes with change in metallicity and C/O ratio, as explained in detail in Sections \ref{subsection:Effect of Metallicity} and \ref{subsection:Effect of C/O ratio}. Carbon dioxide (CO$_2$) also has many significant features with the strongest one at $\sim$4.2 \textmu m. Carbon monoxide (CO) has extremely large spectral signatures at around 1.6 \textmu m, 2.2-2.8 \textmu m and a wide-band 4 to 6 \textmu m feature. The comparison with the all opacity simulation also shows the substantial contribution of CO to the final planetary transmission spectrum, especially near the 2.5 and 4-5 \textmu m region. Except in the optical, methane (CH$_4$) also has many important features similar to H$_2$O , with major features in the 1.7-1.8, 2-2.8, 3-4 and 7-9 \textmu m bands. Depending on the C/O ratios, infrared spectra can either be H$_2$O dominated or CH$_4$ (carbon species) dominated \citep{Kopparapu2012, Madhusudhan2012, Moses2013, Venot2015, Molliere2015}. Moreover, since the primary absorption features between 1 to 5 \textmu m region, alternate between H$_2$O and CH$_4$ as a function of wavelength, they are in principle readily distinguishable.  Ammonia (NH$_3$) has some wide band spectral features, but smaller than H$_2$O and CH$_4$. Sodium (Na) has one of the strongest signature in hot Jupiter spectra at around 0.58 \textmu m and has been detected conclusively in many of these planets. Potassium (K) is the other alkali metal with very strong spectral features, the strongest being at 0.76 \textmu m along with many narrow features which are extremely difficult to resolve with current instruments. We can also see from Figure \ref{fig:opacity_1to10}, the all opacity simulation matches the individual Na and K model spectra around 0.58 \textmu m and 0.76 \textmu m, respectively, demonstrating their dominance at these wavelengths.

Alkali metal elements like Lithium (Li) (shown in Figure \ref{fig:opacity_1to10}), Rubidium (Rb) and Cesium (Cs) (shown in Figure \ref{fig:opacity_10to20})  have very narrow features in the all opacity simulation, making it challenging to detect them observationally. However, broadband features can be seen in the optical wavelengths in the individual spectra of these species, at high pressure levels (deeper) in the atmosphere, making them important opacity sources to obtain accurate heating rates (depending on their abundances) and thereby $P$-$T$ profiles. 

Titanium oxide (TiO) and vanadium oxide (VO) don't have any major features at  temperatures of $1600$\,K due to their low concentrations.  However, at very high temperatures, above $2200$\,K, TiO and VO dominate the visible region of the spectrum suppressing the Na and K features, as seen in Figure \ref{fig:temp_wasp12b}. Due to their high optical opacity, their presence could lead to a thermal inversion in the planetary atmosphere \citep{Spiegel2009, Evans2016}. Iron Hydride (FeH) features are also visible only at high temperatures similar to TiO/VO. Phosphine (PH$_3$) has its primary features in the infrared with the major one between 4 to 5 \textmu m. Hydrogen sulphide (H$_{2}$S) and hydrogen cyanide (HCN) also have many strong spectral features especially in the infrared. Equilibrium chemistry calculations show H$_{2}$S is a quite abundant species for all temperature and metallicity regimes when rainout condensation is included (see Section \ref{subsection:Effect of temperature}, \ref{subsection:Effect of Metallicity}, \ref {subsection:Effect of C/O ratio}), therefore its primary spectral peaks at 1.5, 2 \textmu m and between 2.5 to 3, 3.3 to 5, 6 to 10 \textmu m will be interesting to observe with JWST. The detection of H$_2$S in Jupiter \citep{Niemann1998}, emphasised the importance of condensation with rainout, since without condensation FeS takes up all sulphur inhibiting H$_2$S formation. At higher metallicities, HCN becomes important due to an increase in its concentration. This leads to many HCN features around 1.1\textmu m, 1.4 to 1.5 \textmu m , 2.5 to 2.7 \textmu m and 3 to 3.2 \textmu m along with a possible broadband feature between 6 and 9 \textmu m. Acetylene (C$_{2}$H$_{2}$) features are negligible at solar metallicity but increase substantially at higher metallicities due to an increase in its concentration. Sulphur dioxide (SO$_2$) does not have any features when chemical equilibrium is considered due to its low concentration, however it tends to be  important in non-equilibrium conditions. We note that this analysis is based on a particular planet, therefore the strength of the features might change with change in planetary and grid parameters. However, the position of peaks in wavelength will remain unchanged since they are inherent characteristics of each species/molecule.

\clearpage
\newpage

\onecolumn

\section{Pressure Broadening Sources}
As explained in Section \ref{subsection:Line Lists}, sources of pressure broadening parameters for all the opacity species used in \texttt{ATMO} are shown in Table \ref{table:Broadening Source}.
\label{app:Pressure Broadening Sources}
\begin{table*}
	\centering	
	\begin{tabular}{|p{2cm}|p{2cm}|p{8cm}lp{4cm}l}        
		\hline
		Molecule & Broadener & Line Width Source & Exponent Source \\
	        \hline
		& H$_2$ & \citet{Gamache1996} & \citet{Gamache1996}  \\ [-1ex]\raisebox{1.5ex}{H$_2$O} & He & \citet{Solodov2009,Steyert2004} & \citet{Gamache1996}  \\
		\hline
	       & H$_2$ & \citet{Padmanabhan2014} & \citet{Sharp2007}  \\ [-1ex]\raisebox{1.5ex}{CO$_2$} & He & \citet{Thibault1992} & \citet{Thibault2000}  \\       
	       \hline
	       & H$_2$ & \citet{Regalia2005} & \citet{Lemoal1986}  \\ [-1ex]\raisebox{1.5ex}{CO} & He & \citet{Belbruno1982, Mantz2005} & \citet{Mantz2005}  \\
	       \hline
	       & H$_2$ & \citet{Pine1992, Margolis1993} & \citet{Margolis1993}  \\ [-1ex]\raisebox{1.5ex}{CH$_4$} & He & \citet{Pine1992} & \citet{Varanasi1990}  \\
	       \hline
	       & H$_2$ & \citet{Hadded2001, Pine1993} & \citet{Nouri2004}  \\ [-1ex]\raisebox{1.5ex}{NH$_3$} & He & \citet{Hadded2001, Pine1993} & \citet{Sharp2007}  \\
	       \hline
	       & H$_2$ & \citet{Allard1999, Allard2003, Allard2007} & \citet{Sharp2007}  \\ [-1ex]\raisebox{1.5ex}{Na} & He & \citet{Allard1999, Allard2003, Allard2007} & \citet{Sharp2007}  \\
	       \hline
	       & H$_2$ & \citet{Allard1999, Allard2003, Allard2007} & \citet{Sharp2007}  \\ [-1ex]\raisebox{1.5ex}{K} & He & \citet{Allard1999, Allard2003, Allard2007} & \citet{Sharp2007}  \\
	       \hline
	       & H$_2$ & \citet{Allard1999} & \citet{Sharp2007}  \\ [-1ex]\raisebox{1.5ex}{Li, Rb, Cs} & He & \citet{Allard1999} & \citet{Sharp2007}  \\
	       \hline
	       & H$_2$ & \citet{Sharp2007} & \citet{Sharp2007}  \\ [-1ex]\raisebox{1.5ex}{TiO, VO} & He & \citet{Sharp2007} & \citet{Sharp2007}  \\
	       \hline
	       & H$_2$ & \citet{Sharp2007} & \citet{Sharp2007}  \\ [-1ex]\raisebox{1.5ex}{FeH, CrH} & He & \citet{Sharp2007} & \citet{Sharp2007}  \\       
	       \hline
	       & H$_2$ & \citet{Bouanich2004} & \citet{Levy1994}  \\ [-1ex]\raisebox{1.5ex}{PH$_3$} & He & \citet{Salem2005} & \citet{Levy1994}  \\
	       \hline
	       & H$_2$ & \citet{Landrain1997} & \citet{Sharp2007}  \\ [-1ex]\raisebox{1.5ex}{HCN} & He & \citet{Landrain1997} & \citet{Sharp2007}  \\
	       \hline
	       {C$_2$H$_2$,H$_2$S, SO$_2$}  & Air & \citet{Rothman2009} & \citet{Rothman2009}\\
		\hline
	\end{tabular}
	\centering
	\caption{Type and source of pressure broadening for all opacities used in \texttt{ATMO}.}
	\label{table:Broadening Source}
\end{table*}

\section{Planets and their parameters in Grid}
\label{app:Planets and their parameters in Grid}
All the stellar and planetary parameters adopted from TEPCAT \citep{Tepcat2011} database, for the model simulations of 117 exoplanets in the grid are listed here. First column shows planet names with 'b' omitted indicating first planet of the stellar system as in TEPCAT database. Subsequent columns show, stellar temperature (T$_{\textup{star}}$) in Kelvin, stellar metallicity ([Fe/H]$_{\textup{star}}$ ), stellar mass (M$_{\textup{star}}$ ) in units of solar mass,  stellar radius (R$_{\textup{star}}$ ) in units of solar radius, logarithmic (base 10) stellar gravity (logg$_{\textup{star}}$ ) in $m/s^2$, semi-major axis (a) in AU, planetary mass (M$_\textup{p}$) in units of Jupiter mass,  planetary radius (R$_\textup{p}$) in units of Jupiter radius, planetary surface gravity (g$_\textup{p}$) in $m/s^2$, planetary equilibrium temperature (Teq$_\textup{p}$) in Kelvin assuming 0 albedo and efficient redistribution, V magnitude (V$_{\textup{mag}}$) of the host star, discovery paper reference (Discovery Paper) and finally the most updated reference.

\begin{landscape}
\begin{table}
\begin{tabular}{|p{1.5cm}|p{1cm}|p{1.25cm}|p{0.75cm}|p{0.75cm}|p{0.75cm}|p{0.75cm}|p{0.75cm}|p{0.75cm}|p{0.75cm}|p{0.75cm}|p{0.75cm}|p{4cm}|p{4cm}|}
\hline
System&T$_{\textup{star}}$ &[Fe/H]$_{\textup{star}}$ &M$_{\textup{star}}$ &R$_{\textup{star}}$ &logg$_{\textup{star}}$ & a &M$_\textup{p}$&R$_\textup{p}$&g$_\textup{p}$&Teq$_\textup{p}$&V$_{\textup{mag}}$&Discovery Paper&Updated Reference\\
\hline
&(K)&&(M$_{\textup{sun}}$)&(R$_{\textup{sun}}$)&($m/s^2$)&(AU)&(M$_{\textup{jup}}$)&(R$_{\textup{jup}}$)&($m/s^2$)&(K)&&&\\
\hline

55-Cnc-e&5196&0.31&0.91&0.94&4.43&0.02&0.03&0.17&21.40&2349&5.95&\citet{Winn:2011aa}&\citet{Demory:2016aa}\\
GJ-436&3416&-0.03&0.51&0.46&4.83&0.03&0.08&0.37&13.00&669&10.68&\citet{Gillon:2007aa}&\citet{Lanotte:2014aa}\\
GJ-1214&3026&0.39&0.15&0.22&4.94&0.01&0.02&0.25&7.60&547&14.67&\citet{Charbonneau:2009aa}&\citet{Harpse:2013aa}\\
GJ-3470&3652&0.17&0.51&0.48&4.78&0.04&0.04&0.35&6.80&604&12.27&\citet{Bonfils:2012aa}&\citet{Biddle:2014aa}\\
HAT-P-1&5975&0.13&1.15&1.17&4.36&0.06&0.53&1.32&7.46&1322&10.40&\citet{Bakos:2007aa}&\citet{Nikolov:2014aa}\\
HAT-P-3&5185&0.27&0.90&0.87&4.51&0.04&0.58&0.95&16.14&1189&11.86&\citet{Torres:2007aa}&\citet{Southworth:2012aa}\\
HAT-P-4&5860&0.24&1.27&1.60&4.13&0.04&0.68&1.34&9.42&1691&11.00&\citet{Kovacs:2007aa}&\citet{Southworth:2011aa}\\
HAT-P-6&6570&-0.13&1.29&1.52&4.19&0.05&1.06&1.40&13.50&1704&10.54&\citet{Noyes:2008aa}&\citet{Southworth:2012aa}\\
HAT-P-11&4780&0.31&0.81&0.69&4.66&0.05&0.08&0.40&13.20&838&9.47&\citet{Bakos:2010aa}&\citet{Southworth:2011aa}\\
HAT-P-12&4650&-0.29&0.73&0.70&4.61&0.04&0.21&0.94&6.37&960&12.80&\citet{Hartman:2009aa}&\citet{Lee:2012aa}\\
HAT-P-13&5653&0.41&1.32&1.76&4.07&0.04&0.91&1.49&10.15&1725&10.62&\citet{Bakos:2009aa}&\citet{Southworth:2012ab}\\
HAT-P-17&5246&0.00&0.86&0.84&4.52&0.09&0.53&1.01&12.90&792&10.54&\citet{Howard:2012aa}&\citet{Howard:2012aa}\\
HAT-P-18&4870&0.10&0.77&0.72&4.61&0.06&0.20&0.95&5.42&841&12.76&\citet{Hartman:2011aa}&\citet{Esposito:2014aa}\\
HAT-P-19&4990&0.23&0.84&0.82&4.54&0.05&0.29&1.13&5.62&1010&12.90&\citet{Hartman:2011aa}&\citet{Hartman:2011aa}\\
HAT-P-25&5500&0.31&1.01&0.96&4.48&0.05&0.57&1.19&10.00&1202&13.19&\citet{Quinn:2012aa}&\citet{Quinn:2012aa}\\
HAT-P-26&5011&0.01&0.82&0.79&4.56&0.05&0.06&0.56&4.47&1001&11.74&\citet{Hartman:2011ab}&\citet{Hartman:2011ab}\\
HAT-P-30&6338&0.12&1.24&1.22&4.36&0.04&0.71&1.34&9.80&1630&10.36&\citet{Johnson:2011aa}&\citet{Johnson:2011aa}\\
HAT-P-32&6207&-0.04&1.16&1.22&4.33&0.03&0.86&1.79&6.60&1786&11.29&\citet{Hartman:2011ac}&\citet{Hartman:2011ac}\\
HAT-P-33&6446&0.07&1.38&1.64&4.15&0.05&0.76&1.69&6.60&1782&11.19&\citet{Hartman:2011ac}&\citet{Hartman:2011ac}\\
HAT-P-39&6340&0.19&1.40&1.63&4.16&0.05&0.60&1.57&5.90&1752&12.42&\citet{Hartman:2012aa}&\citet{Hartman:2012aa}\\
HAT-P-40&6080&0.22&1.51&2.21&3.93&0.06&0.61&1.73&5.13&1770&11.34&\citet{Winn:2011aa}&\citet{Winn:2011aa}\\
HAT-P-41&6390&0.21&1.42&1.68&4.14&0.04&0.80&1.69&6.90&1941&11.36&\citet{Hartman:2012aa}&\citet{Hartman:2012aa}\\
HAT-P-44&5295&0.33&0.94&0.95&4.46&0.05&0.35&1.24&5.62&1108&13.21&\citet{Hartman:2014aa}&\citet{Hartman:2014aa}\\
HAT-P-45&6330&0.07&1.26&1.32&4.30&0.05&0.89&1.43&10.70&1652&12.79&\citet{Hartman:2014aa}&\citet{Hartman:2014aa}\\
HAT-P-46&6120&0.30&1.28&1.40&4.25&0.06&0.49&1.28&7.30&1458&11.94&\citet{Hartman:2014aa}&\citet{Hartman:2014aa}\\
HAT-P-47&6703&0.00&1.39&1.51&4.22&0.06&0.21&1.31&2.95&1605&10.69&\citet{Bakos:2016aa}&\citet{Bakos:2016aa}\\
HAT-P-48&5946&0.02&1.10&1.22&4.30&0.05&0.17&1.13&3.24&1361&12.16&\citet{Bakos:2016aa}&\citet{Bakos:2016aa}\\
HAT-P-51&5449&0.27&0.98&1.04&4.39&0.05&0.31&1.29&4.58&1192&13.44&\citet{Hartman:2015aa}&\citet{Hartman:2015aa}\\
HAT-P-65&5835&0.10&1.21&1.86&3.98&0.04&0.53&1.89&3.63&1930&13.15&\citet{Hartman:2016aa}&\citet{Hartman:2016aa}\\
HATS-5&5304&0.19&0.94&0.87&4.53&0.05&0.24&0.91&7.08&1025&12.63&\citet{Zhou:2014aa}&\citet{Zhou:2014aa}\\
HATS-6&3770&0.20&0.57&0.57&4.68&0.04&0.32&1.00&7.90&713&15.16&\citet{Hartman:2015ab}&\citet{Hartman:2015ab}\\
HATS-19&5896&0.24&1.30&1.75&4.07&0.06&0.43&1.66&3.80&1570&13.03&\citet{Bhatti:2016aa}&\citet{Bhatti:2016aa}\\
HATS-21&5695&0.30&1.08&1.02&4.45&0.05&0.33&1.12&6.50&1284&12.19&\citet{Bhatti:2016aa}&\citet{Bhatti:2016aa}\\
HD-097658&5170&-0.23&0.77&0.74&4.58&0.08&0.02&0.20&14.70&757&7.71&\citet{Dragomir:2013aa}&\citet{VanGrootel:2014aa}\\
HD-149026&6147&0.36&1.34&1.54&4.19&0.04&0.37&0.81&13.55&1634&8.16&\citet{Sato:2005aa}&\citet{Carter:2009aa}\\
HD-189733&5050&-0.03&0.84&0.75&4.61&0.03&1.15&1.15&21.50&1191&7.68&\citet{Bouchy:2005aa}&\citet{Southworth:2010aa}\\
HD-209458&6117&0.02&1.15&1.16&4.37&0.05&0.71&1.38&9.30&1459&7.65&\citet{Henry:2000aa}&\citet{Southworth:2010aa}\\
KELT-4&6206&-0.12&1.20&1.60&4.11&0.04&0.90&1.70&7.74&1823&10.47&\citet{Eastman:2016aa}&\citet{Eastman:2016aa}\\
KELT-6&6102&-0.28&1.08&1.58&4.07&0.08&0.43&1.19&7.40&1313&10.42&\citet{Collins:2015aa}&\citet{Collins:2015aa}\\
KELT-7&6789&0.14&1.53&1.73&4.15&0.04&1.28&1.53&13.50&2048&8.54&\citet{Bieryla:2015aa}&\citet{Bieryla:2015aa}\\
KELT-8&5754&0.27&1.21&1.67&4.08&0.05&0.87&1.86&6.20&1675&10.83&\citet{Fulton:2015aa}&\citet{Fulton:2015aa}\\
KELT-10&5948&0.09&1.11&1.21&4.32&0.05&0.68&1.40&8.57&1377&10.70&\citet{Kuhn:2016aa}&\citet{Kuhn:2016aa}\\
KELT-11&5370&0.18&1.44&2.72&3.73&0.06&0.20&1.37&2.55&1712&8.03&\citet{Pepper:2017aa}&\citet{Pepper:2017aa}\\
KELT-12&6278&0.19&1.59&2.37&3.89&0.07&0.95&1.79&7.40&1800&10.64&\citet{Stevens:2017aa}&\citet{Stevens:2017aa}\\
\hline
\end{tabular}
\end{table}
\end{landscape}

\begin{landscape}
\begin{table}
\begin{tabular}{|p{1.5cm}|p{1cm}|p{1.25cm}|p{0.75cm}|p{0.75cm}|p{0.75cm}|p{0.75cm}|p{0.75cm}|p{0.75cm}|p{0.75cm}|p{0.75cm}|p{0.75cm}|p{4cm}|p{4cm}|}
\hline
System&T$_{\textup{star}}$ &[Fe/H]$_{\textup{star}}$ &M$_{\textup{star}}$ &R$_{\textup{star}}$ &logg$_{\textup{star}}$ & a &M$_\textup{p}$&R$_\textup{p}$&g$_\textup{p}$&Teq$_\textup{p}$&V$_{\textup{mag}}$&Discovery Paper&Updated Reference\\
\hline
&(K)&&(M$_{\textup{sun}}$)&(R$_{\textup{sun}}$)&($m/s^2$)&(AU)&(M$_{\textup{jup}}$)&(R$_{\textup{jup}}$)&($m/s^2$)&(K)&&&\\
\hline
KELT-15&6003&0.05&1.18&1.48&4.23&0.05&1.20&1.52&12.80&1904&11.44&\citet{Rodriguez:2016aa}&\citet{Rodriguez:2016aa}\\
KELT-17&7454&-0.02&1.64&1.65&4.22&0.05&1.31&1.52&13.90&2087&9.29&\citet{Zhou:2016aa}&\citet{Zhou:2016aa}\\
Kepler-12&5947&0.07&1.16&1.49&4.16&0.06&0.43&1.71&3.66&1485&13.53&\citet{Fortney:2011aa}&\citet{Southworth:2012aa}\\
TrES-1&5226&0.06&0.89&0.82&4.56&0.04&0.76&1.10&15.60&1147&11.79&\citet{Alonso:2004aa}&\citet{Southworth:2010aa}\\
TrES-4&6295&0.28&1.45&1.83&4.09&0.05&0.49&1.84&2.82&1795&11.59&\citet{Mandushev:2007aa}&\citet{Sozzetti:2015aa}\\
WASP-1&6160&0.14&1.24&1.47&4.20&0.04&0.85&1.48&9.80&1830&11.31&\citet{Collier-Cameron:2007aa}&\citet{Maciejewski:2014aa}\\
WASP-2&5170&0.04&0.85&0.82&4.54&0.03&0.88&1.06&19.31&1286&11.98&\citet{Collier-Cameron:2007aa}&\citet{Southworth:2012aa}\\
WASP-4&5540&-0.03&0.93&0.91&4.49&0.02&1.25&1.36&16.64&1673&12.46&\citet{Wilson:2008aa}&\citet{Southworth:2012aa}\\
WASP-6&5375&-0.20&0.84&0.86&4.49&0.04&0.48&1.23&7.96&1184&11.90&\citet{Gillon:2009aa}&\citet{Tregloan-Reed:2015aa}\\
WASP-7&6520&0.00&1.32&1.48&4.22&0.06&0.98&1.37&12.90&1530&9.48&\citet{Hellier:2009aa}&\citet{Southworth:2012aa}\\
WASP-11&4900&0.12&0.81&0.77&4.57&0.04&0.49&0.99&12.45&992&11.89&\citet{West:2009ab}&\citet{Mancini:2015aa}\\
WASP-12&6313&0.21&1.43&1.66&4.16&0.02&1.47&1.90&10.09&2580&11.69&\citet{Hebb:2009aa}&\citet{Collins:2015aa}\\
WASP-13&6025&0.11&1.22&1.66&4.09&0.06&0.51&1.53&5.44&1531&10.51&\citet{Skillen:2009aa}&\citet{Southworth:2012aa}\\
WASP-15&6573&0.09&1.30&1.52&4.19&0.05&0.59&1.41&7.39&1676&10.92&\citet{West:2009aa}&\citet{Southworth:2013aa}\\
WASP-16&5630&0.07&0.98&1.09&4.36&0.04&0.83&1.22&13.92&1389&11.31&\citet{Lister:2009aa}&\citet{Southworth:2013aa}\\
WASP-17&6550&-0.25&1.29&1.58&4.15&0.05&0.48&1.93&3.16&1755&11.50&\citet{Anderson:2010aa}&\citet{Southworth:2012ac}\\
WASP-19&5460&0.14&0.94&1.02&4.39&0.02&1.14&1.41&14.21&2077&12.31&\citet{Hebb:2010aa}&\citet{Mancini:2013aa}\\
WASP-20&6000&-0.01&1.09&1.14&4.36&0.06&0.38&1.28&5.80&1282&10.68&\citet{Anderson:2015aa}&\citet{Evans:2016aa}\\
WASP-21&5924&-0.22&0.89&1.14&4.28&0.05&0.28&1.16&5.07&1333&11.50&\citet{Bouchy:2010aa}&\citet{Ciceri:2013aa}\\
WASP-25&5736&0.06&1.05&0.92&4.53&0.05&0.60&1.25&9.54&1210&11.85&\citet{Enoch:2011aa}&\citet{Southworth:2014aa}\\
WASP-29&4875&0.11&0.82&0.81&4.54&0.05&0.24&0.78&10.00&970&11.21&\citet{Hellier:2010aa}&\citet{Gibson:2013aa}\\
WASP-31&6175&-0.20&1.16&1.25&4.31&0.05&0.48&1.55&4.56&1575&11.94&\citet{Anderson:2011aa}&\citet{Anderson:2011aa}\\
WASP-33&7430&0.10&1.56&1.51&4.27&0.03&2.16&1.68&19.00&2710&8.30&\citet{Collier-Cameron:2010aa}&\citet{Lehmann:2015aa}\\
WASP-34&5704&0.08&1.01&0.93&4.50&0.05&0.59&1.22&9.10&1250&10.37&\citet{Smalley:2011aa}&\citet{Smalley:2011aa}\\
WASP-35&6072&-0.05&1.07&1.09&4.40&0.04&0.72&1.32&9.50&1450&10.95&\citet{Enoch:2011ab}&\citet{Enoch:2011ab}\\
WASP-39&5460&-0.12&0.93&0.90&4.50&0.05&0.28&1.27&4.07&1116&12.10&\citet{Faedi:2011aa}&\citet{Faedi:2011aa}\\
WASP-41&5546&0.06&0.99&0.89&4.54&0.04&0.98&1.18&17.45&1242&11.64&\citet{Maxted:2011aa}&\citet{Southworth:2016aa}\\
WASP-42&5315&0.29&0.95&0.89&4.51&0.06&0.53&1.12&10.38&1021&12.57&\citet{Lendl:2012aa}&\citet{Southworth:2016aa}\\
WASP-43&4520&-0.01&0.72&0.67&4.64&0.02&2.03&1.04&47.00&1440&12.37&\citet{Hellier:2011aa}&\citet{Gillon:2012aa}\\
WASP-49&5600&-0.23&1.00&1.04&4.41&0.04&0.40&1.20&7.13&1399&11.36&\citet{Lendl:2012aa}&\citet{Lendl:2016aa}\\
WASP-52&5000&0.03&0.80&0.79&4.55&0.03&0.43&1.25&6.85&1315&12.20&\citet{Hebrard:2013aa}&\citet{Mancini:2017aa}\\
WASP-54&6296&0.00&1.21&1.83&4.00&0.05&0.64&1.65&5.32&1759&10.42&\citet{Faedi:2013aa}&\citet{Faedi:2013aa}\\
WASP-55&6070&0.09&1.16&1.10&4.42&0.06&0.63&1.33&8.73&1300&11.76&\citet{Hellier:2012aa}&\citet{Southworth:2016aa}\\
WASP-58&5800&-0.45&0.94&1.17&4.27&0.06&0.89&1.37&10.70&1270&11.66&\citet{Hebrard:2013aa}&\citet{Hebrard:2013aa}\\
WASP-62&6230&0.04&1.25&1.28&4.32&0.06&0.57&1.39&6.76&1440&10.22&\citet{Hellier:2012aa}&\citet{Hellier:2012aa}\\
WASP-63&5715&0.28&1.32&1.88&4.01&0.06&0.38&1.43&4.17&1540&11.16&\citet{Hellier:2012aa}&\citet{Hellier:2012aa}\\
WASP-67&5417&0.18&0.83&0.82&4.53&0.05&0.41&1.09&8.45&1003&12.54&\citet{Hellier:2012aa}&\citet{Mancini:2014aa}\\
WASP-69&4700&0.15&0.83&0.81&4.54&0.05&0.26&1.06&5.32&963&9.87&\citet{Anderson:2014ac}&\citet{Anderson:2014ac}\\
WASP-70&5700&-0.01&1.11&1.22&4.31&0.05&0.59&1.16&10.00&1387&10.79&\citet{Anderson:2014ac}&\citet{Anderson:2014ac}\\
WASP-74&5990&0.39&1.48&1.64&4.18&0.04&0.95&1.56&8.91&1910&9.76&\citet{Hellier:2015aa}&\citet{Hellier:2015aa}\\
WASP-76&6250&0.23&1.46&1.73&4.13&0.03&0.92&1.83&6.31&2160&9.53&\citet{West:2016aa}&\citet{West:2016aa}\\
WASP-79&6600&0.03&1.52&1.91&4.06&0.05&0.90&2.09&4.70&1900&10.04&\citet{Smalley:2012aa}&\citet{Smalley:2012aa}\\
WASP-80&4145&-0.14&0.60&0.59&4.67&0.03&0.56&0.99&14.34&825&11.87&\citet{Triaud:2013}&\citet{Mancini:2014ab}\\
WASP-82&6500&0.12&1.64&2.22&3.96&0.04&1.25&1.71&9.75&2202&10.08&\citet{West:2016aa}&\citet{Smith:2015aa}\\
\hline
\end{tabular}
\end{table}
\end{landscape}

\begin{landscape}
\begin{table}
\begin{tabular}{|p{1.5cm}|p{1cm}|p{1.25cm}|p{0.75cm}|p{0.75cm}|p{0.75cm}|p{0.75cm}|p{0.75cm}|p{0.75cm}|p{0.75cm}|p{0.75cm}|p{0.75cm}|p{4cm}|p{4cm}|}
\hline
System&T$_{\textup{star}}$ &[Fe/H]$_{\textup{star}}$ &M$_{\textup{star}}$ &R$_{\textup{star}}$ &logg$_{\textup{star}}$ & a &M$_\textup{p}$&R$_\textup{p}$&g$_\textup{p}$&Teq$_\textup{p}$&V$_{\textup{mag}}$&Discovery Paper&Updated Reference\\
\hline
&(K)&&(M$_{\textup{sun}}$)&(R$_{\textup{sun}}$)&($m/s^2$)&(AU)&(M$_{\textup{jup}}$)&(R$_{\textup{jup}}$)&($m/s^2$)&(K)&&&\\
\hline
WASP-83&5480&0.29&1.11&1.05&4.44&0.06&0.30&1.04&6.17&1120&12.87&\citet{Hellier:2015aa}&\citet{Hellier:2015aa}\\
WASP-84&5280&0.09&0.85&0.77&4.60&0.08&0.69&0.98&16.52&833&10.83&\citet{Anderson:2014ac}&\citet{Anderson:2015ab}\\
WASP-88&6430&-0.08&1.45&2.08&3.96&0.06&0.56&1.70&4.68&1772&11.39&\citet{Delrez:2014aa}&\citet{Delrez:2014aa}\\
WASP-90&6440&0.11&1.55&1.98&4.03&0.06&0.63&1.63&5.37&1840&11.69&\citet{West:2016aa}&\citet{West:2016aa}\\
WASP-93&6700&0.07&1.33&1.52&4.20&0.04&1.47&1.60&13.20&1942&10.97&\citet{Hay:2016aa}&\citet{Hay:2016aa}\\
WASP-94&6170&0.26&1.45&1.62&4.18&0.06&0.45&1.72&3.48&1604&10.06&\citet{Neveu-VanMalle:2014aa}&\citet{Neveu-VanMalle:2014aa}\\
WASP-95&5830&0.14&1.11&1.13&4.38&0.03&1.13&1.21&21.80&1570&10.09&\citet{Hellier:2014aa}&\citet{Hellier:2014aa}\\
WASP-96&5500&0.14&1.06&1.05&4.42&0.05&0.48&1.20&7.59&1285&12.19&\citet{Hellier:2014aa}&\citet{Hellier:2014aa}\\
WASP-97&5670&0.23&1.12&1.06&4.43&0.03&1.32&1.13&23.40&1555&10.58&\citet{Hellier:2014aa}&\citet{Hellier:2014aa}\\
WASP-101&6380&0.20&1.34&1.29&4.34&0.05&0.50&1.41&5.75&1560&10.34&\citet{Hellier:2014aa}&\citet{Hellier:2014aa}\\
WASP-103&6110&0.06&1.21&1.41&4.22&0.02&1.47&1.65&14.34&2489&12.50&\citet{Gillon:2014aa}&\citet{Southworth:2016ab}\\
WASP-108&6000&0.05&1.17&1.22&4.34&0.04&0.89&1.28&12.39&1590&11.22&\citet{Anderson:2014ab}&\citet{Anderson:2014ab}\\
WASP-109&6520&-0.22&1.20&1.35&4.26&0.05&0.91&1.44&10.00&1685&11.44&\citet{Anderson:2014ab}&\citet{Anderson:2014ab}\\
WASP-110&5400&-0.06&0.89&0.88&4.50&0.05&0.51&1.24&7.60&1134&12.27&\citet{Anderson:2014ab}&\citet{Anderson:2014ab}\\
WASP-113&5890&0.10&1.32&1.61&4.20&0.06&0.47&1.41&5.50&1496&11.77&\citet{Barros:2016aa}&\citet{Barros:2016aa}\\
WASP-117&6040&-0.11&1.13&1.17&4.28&0.09&0.28&1.02&6.56&1024&10.15&\citet{Lendl:2014aa}&\citet{Lendl:2014aa}\\
WASP-118&6410&0.16&1.32&1.70&4.10&0.05&0.51&1.44&5.71&1729&11.02&\citet{Hay:2016aa}&\citet{Hay:2016aa}\\
WASP-121&6460&0.13&1.35&1.46&4.24&0.03&1.18&1.86&9.40&2358&10.52&\citet{Delrez:2016aa}&\citet{Delrez:2016aa}\\
WASP-122&5720&0.32&1.24&1.52&4.17&0.03&1.28&1.74&9.66&1970&11.00&\citet{Turner:2016aa}&\citet{Turner:2016aa}\\
WASP-123&5740&0.18&1.17&1.28&4.29&0.04&0.90&1.32&11.70&1520&11.03&\citet{Turner:2016aa}&\citet{Turner:2016aa}\\
WASP-124&6050&-0.02&1.07&1.02&4.44&0.04&0.60&1.24&8.90&1400&12.70&\citet{Maxted:2016aa}&\citet{Maxted:2016aa}\\
WASP-126&5800&0.17&1.12&1.27&4.28&0.04&0.28&0.96&6.80&1480&10.80&\citet{Maxted:2016aa}&\citet{Maxted:2016aa}\\
WASP-127&5750&-0.18&1.08&1.39&4.18&0.05&0.18&1.37&2.14&1400&10.16&\citet{Lam:2017aa}&\citet{Lam:2017aa}\\
WASP-131&5950&-0.18&1.06&1.53&4.09&0.06&0.27&1.22&4.17&1460&10.08&\citet{Hellier:2017aa}&\citet{Hellier:2017aa}\\
WASP-132&4750&0.22&0.80&0.74&4.61&0.07&0.41&0.87&12.60&763&12.40&\citet{Hellier:2017aa}&\citet{Hellier:2017aa}\\
WASP-139&5300&0.20&0.92&0.80&4.59&0.06&0.12&0.80&4.17&910&12.39&\citet{Hellier:2017aa}&\citet{Hellier:2017aa}\\
WASP-140&5300&0.12&0.90&0.87&4.51&0.03&2.44&1.44&25.00&1320&11.13&\citet{Hellier:2017aa}&\citet{Hellier:2017aa}\\
XO-1&5750&0.02&1.04&0.94&4.51&0.05&0.92&1.21&15.80&1210&11.14&\citet{McCullough:2006aa}&\citet{Southworth:2010aa}\\
XO-2&5332&0.43&0.96&1.00&4.44&0.04&0.60&1.02&14.13&1328&11.25&\citet{Burke:2007aa}&\citet{Damasso:2015aa}\\
\hline
\end{tabular}
\caption{All the stellar and planetary parameters adopted from TEPCAT \citep{Tepcat2011} database, for the model simulations of 117 exoplanets in the grid are listed here. First column shows planet names with 'b' omitted indicating first planet of the stellar system as in TEPCAT database. Subsequent columns show, stellar temperature (T$_{\textup{star}}$) in Kelvin, stellar metallicity ([Fe/H]$_{\textup{star}}$ ), stellar mass (M$_{\textup{star}}$ ) in units of solar mass,  stellar radius (R$_{\textup{star}}$ ) in units of solar radius, logarithmic (base 10) stellar gravity (logg$_{\textup{star}}$ ) in $m/s^2$, semi-major axis (a) in AU, planetary mass (M$_\textup{p}$) in units of Jupiter mass,  planetary radius (R$_\textup{p}$) in units of Jupiter radius, planetary surface gravity (g$_\textup{p}$) in $m/s^2$, planetary equilibrium temperature (Teq$_\textup{p}$) in Kelvin assuming 0 albedo and efficient redistribution, V magnitude (V$_{\textup{mag}}$) of the host star, discovery paper reference (Discovery Paper) and finally the most updated reference.}
\end{table}
\end{landscape}

%%%%%%%%%%%%%%%%%%%%%%%%%%%%%%%%%%%%%%%%%%%%%%%%%%

% Don't change these lines
\bsp	% typesetting comment
\label{lastpage}
\end{document}